\documentclass[a4paper,fleqn,usenatbib]{mnras}

\usepackage[T1]{fontenc}
\usepackage{ae,aecompl}

\usepackage{graphicx}	
\usepackage{amsmath}	
\usepackage{amssymb}	

\usepackage[normalem]{ulem}
\usepackage[x11names]{xcolor}

\newcommand{\Eq}[1]{Eq.~(\ref{#1})}

\newcommand{\Fig}[1]{Fig.~\ref{#1}}

\newcommand{\Sec}[1]{\S\ref{#1}}
\newcommand{\Tab}[1]{Table~\ref{#1}}

\ifdefined\editorial

	\newcommand{\jz}[1]{{\color{cyan}\bf Comment JZ:#1}}
	\newcommand{\task}[1]{{}}
	\newcommand{\jzc}[1]{{\color{red} \bf CRITICAL: #1}}
	\newcommand{\jztd}[1]{{\bf [ .. #1 ..]}}
        
  \newcommand{\exinput}[1]{{\color{red} \bf #1}}

	\newcommand{\nb}[1]{{\color{red}TODO:#1}}
	\newcommand{\nbc}[1]{{\color{blue}\bf Comment NB:#1}}
	
  \newcommand{\jztonb}[1]{{\color{blue} JZ->NB: #1}}

\else

	\newcommand{\jz}[1]{{}}
	\newcommand{\task}[1]{{}}
	\newcommand{\jzc}[1]{{}}
	\newcommand{\jztd}[1]{{}}
  \newcommand{\exinput}[1]{{}}

	\newcommand{\nb}[1]{{}} 
	\newcommand{\nbc}[1]{{}}
	\newcommand{\jztonb}[1]{}
\fi

\newcommand{\jzd}[1]{}

\newcommand{\rewmgii}{\hbox{$EW_0^{\lambda 2796}$}}%

\newcommand{\incl}{i}
\newcommand{\tarcs}{\mathrm{arcsec}}

\newcommand{\NeV}{\textsc{[{\rm Ne}\kern 0.1em{\sc v}}]}
\newcommand{\NII}{\textsc{[{\rm N}\kern 0.1em{\sc ii}}]}
\newcommand{\OIII}{\textsc{[{\rm O}\kern 0.1em{\sc iii}}]}
\newcommand{\OII}{\textsc{[{\rm O}\kern 0.1em{\sc ii}}]}
\newcommand{\OI}{\textsc{{\rm O}\kern 0.1em{\sc i}}}

\newcommand{\MgI}{\textsc{{\rm Mg}\kern 0.1em{\sc i}}}
\newcommand{\MgII}{\textsc{{\rm Mg}\kern 0.1em{\sc ii}}}
\newcommand{\FeII}{\textsc{{\rm Fe}\kern 0.1em{\sc ii}}}
\newcommand{\MnII}{\textsc{{\rm Mn}\kern 0.1em{\sc ii}}}
\newcommand{\ZnII}{\textsc{{\rm Zn}\kern 0.1em{\sc ii}}}
\newcommand{\CrII}{\textsc{{\rm Cr}\kern 0.1em{\sc ii}}}
\newcommand{\NaI}{\textsc{{\rm Na}\kern 0.1em{\sc i}}}

\newcommand{\HI}{\textsc{{\rm H}\kern 0.1em{\sc i}}}
\newcommand{\NeIII}{\textsc[{{\rm Ne}\kern 0.1em{\sc iii}}]}
\newcommand{\CIII}{\textsc{{\rm C}\kern 0.1em{\sc iii}}}

\newcommand{\HII}{\textsc{{\rm H}\kern 0.1em{\sc ii}}}
\newcommand{\lya}{\textsc{{\rm Ly}\kern 0.1em$\alpha$}}
\newcommand{\Ly}{\textsc{{\rm Ly}\kern 0.1em$\alpha$}}
\newcommand{\Ha}{\textsc{{\rm H}\kern 0.1em$\alpha$}}
\newcommand{\Hb}{\textsc{{\rm H}\kern 0.1em$\beta$}}
\newcommand{\Hg}{\textsc{{\rm H}\kern 0.1em$\gamma$}}
\newcommand{\SII}{\hbox{[{\rm S}\kern 0.1em{\sc ii}}]}
\newcommand{\Ne}{\hbox{[{\rm Ne}\kern 0.1em{\sc v}}]}
\newcommand{\kms}{\hbox{km~s$^{-1}$}}
\newcommand{\kpc}{\hbox{kpc}}
\newcommand{\Mpc}{\hbox{Mpc}}
\newcommand{\cm}{\hbox{cm}}
\newcommand{\uerglf}{\mathrm{erg}\,\mathrm{s}^{-1}\,\mathrm{cm}^{-2}}
\newcommand{\uergcont}{\mathrm{erg}\,\mathrm{s}^{-1}\,\mathrm{cm}^{-2}\,{\text \AA}^{-1}}
\newcommand{\uergsb}{\mathrm{erg}\,\mathrm{s}^{-1}\,\mathrm{cm}^{-2}\,\mathrm{arcsec}^{-2}}

\newcommand{\mpy}{\mathrm{M}_{\odot}\,\mathrm{yr}^{-1}}

\newcommand{\tarcsec}{\mathrm{arcsec}}
\newcommand{\sqarcsec}{\tarcsec{}^2}

\newcommand{\dex}{\mathrm{dex}}

\newcommand{\dens}{n} 

\newcommand{\vvir}{v_\mathrm{vir}}

\newcommand{\rvir}{r_\mathrm{vir}}

\newcommand{\vlos}{v_\mathrm{los}}

\newcommand{\vmax}{v_\mathrm{max}}
\newcommand{\ebv}{\mathrm{E(B\mbox{--}V)}}

\newcommand{\Mvir}{M_\mathrm{vir}}
\newcommand{\Mstar}{M_*}

\newcommand{\rhalf}{R_{\rm e}} 

\newcommand{\msun}{\mathrm{M_{\odot}}}

\newcommand{\mdotout}{\dot{M}_\mathrm{out}}

\newcommand{\thetaout}{\theta_\mathrm{out}}

\newcommand{\vout}{v_\mathrm{out}}

\def\sersic{{S{\'e}rsic}}

\newcommand{\mfl}{MEGAFLOW}{}

{}
\newcommand{\galfit}{{\sc Galfit}}{}
\newcommand{\gpk}{{\sc galpak$^{\rm 3D}$}}{}
\newcommand{\camel}{{\sc camel}}{}
{}
\newcommand{\cgmpy}{{\sc cgmpy}}{}
\newcommand{\pampelmuse}{{\sc pampelmuse}}{}
\newcommand{\todse}{{\sc tdose}}{}
\newcommand{\ppxf}{{\sc ppxf}}{}
\newcommand{\coniecto}{{\sc coniecto}}{}
\newcommand{\mappingsv}{{\sc mappings v}}{}

\newcommand{\sfsf}{$\mathrm{SF}^2$}
\newcommand{\texptot}{11.2}

\newcommand{\quasarid}{SDSSJ0937+0656}

\newcommand{\rewmgiiqso}[0]{1.8}

\newcommand{\nameprime}[0]{\emph{main}}
\newcommand{\bkpcprimeqso}{38.7} 
\newcommand{\alphaprimeqso}[0]{87\pm1} 

\newcommand{\namesecond}[0]{\emph{secondary}}
\newcommand{\bkpcsecondqso}{69.0} 
\newcommand{\alphasecondqso}[0]{58\pm1}

\newcommand{\bkpcthirdqso}[0]{245} 

\newcommand{\namebzeropnine}[0]{back\_0p9}
\newcommand{\namebonepfive}[0]{back\_1p5}
\newcommand{\namebonepseven}[0]{back\_1p7}

{}
{}

{}
{}
{}
{}
{}
{}

\newcommand{\papone}{paper\,I}{}
\newcommand{\paptwo}{paper\,II}{}
\newcommand{\papthree}{paper\,III}{}
\newcommand{\papfour}{paper\,IV}{}
\newcommand{\papfive}{paper\,V}{}
\newcommand{\papsix}{paper\,VI}{}
\newcommand{\papseven}{paper\,VII}{}

\defcitealias{Zabl:2019a}{\paptwo}
\defcitealias{Schroetter:2019a}{\papthree}
\defcitealias{Zabl:2020a}{\papfour}
\defcitealias{Wendt:2021a}{\papfive}
\defcitealias{Schroetter:2021a}{\papsix}
\defcitealias{Freundlich:2021a}{\papseven}

\newcommand{\mainXpropXsedXmass}[0]{10.05_{-0.11}^{+0.15}}
\newcommand{\mainXpropXsedXlogtau}[0]{7.9_{-0.0}^{+1.9}}
\newcommand{\mainXpropXsedXage}[0]{8.4_{-0.0}^{+0.7}}
\newcommand{\mainXpropXsedXcurrsfr}[0]{22.8_{-19.0}^{+15.4}}
\newcommand{\mainXpropXsedXebv}[0]{0.42_{-0.26}^{+0.05}}
\newcommand{\secondXpropXsedXmass}[0]{9.29_{-0.04}^{+0.22}}

\newcommand{\mainXpropXgpkXflux}[0]{17.6_{-0.1}^{+0.1}\times10^{-17}}
\newcommand{\mainXpropXderivedXebvmass}[0]{0.29_{-0.09}^{+0.10}}
\newcommand{\mainXpropXderivedXsfroiimass}[0]{8{\scriptstyle \pm4}}

\newcommand{\mainXpropXderivedXdeltams}[0]{0.4_{-0.5}^{+0.3}}

\newcommand{\bzeropnineXpropXrastr}[0]{09:37:49.42}
\newcommand{\bzeropnineXpropXdecstr}[0]{+06:56:55.8}
\newcommand{\bzeropnineXpropXalphamain}[0]{57}
\newcommand{\bzeropnineXpropXbmainkpc}[0]{30}

\newcommand{\bonepfiveXpropXrastr}[0]{09:37:49.77}
\newcommand{\bonepfiveXpropXdecstr}[0]{+06:56:49.6}
\newcommand{\bonepfiveXpropXalphamain}[0]{16}
\newcommand{\bonepfiveXpropXbmainkpc}[0]{41}
\newcommand{\bonepfiveXpropXredshift}[0]{1.5}
\newcommand{\bonepsevenXpropXrastr}[0]{09:37:49.34}
\newcommand{\bonepsevenXpropXdecstr}[0]{+06:56:54.3}
\newcommand{\bonepsevenXpropXalphamain}[0]{34}
\newcommand{\bonepsevenXpropXbmainkpc}[0]{21}
\newcommand{\bonepsevenXpropXredshift}[0]{1.7}
\newcommand{\quasarXpropXrastr}[0]{09:37:49.59}
\newcommand{\quasarXpropXdecstr}[0]{+06:56:56.3}
\newcommand{\quasarXpropXalphamain}[0]{84}
\newcommand{\quasarXpropXbmainkpc}[0]{39}
\newcommand{\quasarXpropXredshift}[0]{1.8}
\newcommand{\secondXpropXrastr}[0]{09:37:49.38}
\newcommand{\secondXpropXdecstr}[0]{+06:56:47.1}
\newcommand{\secondXpropXalphamain}[0]{59}
\newcommand{\secondXpropXbmainkpc}[0]{32}

\newcommand{\bzeropnineXpropXmabatmgiimain}[0]{24.4}

\newcommand{\bonepfiveXpropXmabatmgiimain}[0]{23.6}

\newcommand{\bonepsevenXpropXmabatmgiimain}[0]{23.3}
\newcommand{\absXmetiriXstartvelmgii}[0]{-96}
\newcommand{\absXmetiriXendvelmgii}[0]{125}
\newcommand{\absXfitXMgIXvelcompa}[0]{-44}
\newcommand{\absXfitXMgIXvelcompb}[0]{23}
\newcommand{\absXfitXMgIXvelcompc}[0]{95}

\newcommand{\windmodelXthetaout}[0]{35}
\newcommand{\windmodelXvout}[0]{130}

\newcommand{\mainXpropXgalfitXn}[0]{0.50} %
\newcommand{\mainXpropXgalfitXbtoa}[0]{0.25} %
\newcommand{\mainXpropXgalfitXincl}[0]{75.7} %
\newcommand{\mainXpropXgalfitXre}[0]{4.6} 
 
\newcommand{\mainXincl}[0]{75}

\newcommand{\mainXgpkXsigma}[0]{57_{-1}^{+1}}
\newcommand{\mainXgpkXvmax}[0]{190_{-8}^{+8}}
\newcommand{\mainXderivedXvvirfromkin}[0]{180_{-14}^{+14}}
\newcommand{\mainXderivedXmvirfromkin}[0]{12.19_{-0.09}^{+0.09}}
\newcommand{\mainXderivedXrvirfromkin}[0]{208_{-16}^{+16}}	
\renewcommand{\mainXpropXgpkXflux}[0]{19.6_{-0.1}^{+0.1}\times10^{-17}}
\renewcommand{\mainXpropXderivedXsfroiimass}[0]{9_{-4}^{+4}}

\newcommand{\propXabsXmout}[0]{4}
\newcommand{\propXabsXnhi}[0]{19.5}
\newcommand{\propXeta}[0]{0.5}

\renewcommand{\mainXpropXgalfitXre}[0]{4.6\pm0.1}

\newcommand{\mainXpropXbabsmag}[0]{-20.4}

\newcommand{\bzeropnineXpropXredshift}[0]{0.857} 
\renewcommand{\bonepfiveXpropXredshift}[0]{1.495} 
\renewcommand{\bonepsevenXpropXredshift}[0]{1.713} 
\renewcommand{\quasarXpropXredshift}[0]{1.82}

\newcommand{\mainXpropXrastr}[0]{09:37:49.41}  
\newcommand{\mainXpropXdecstr}[0]{+06:56:51.5} 

\renewcommand{\bonepfiveXpropXmabatmgiimain}[0]{24.3}
\renewcommand{\bonepsevenXpropXmabatmgiimain}[0]{24.1}
\renewcommand{\bzeropnineXpropXmabatmgiimain}[0]{24.8}
\newcommand{\quasarXpropXmabatmgiimain}[0]{19.4} 

\renewcommand{\secondXpropXalphamain}[0]{60}
\renewcommand{\secondXpropXbmainkpc}[0]{32}

\renewcommand{\quasarXpropXalphamain}[0]{82}
\renewcommand{\quasarXpropXbmainkpc}[0]{39}

\renewcommand{\bonepfiveXpropXalphamain}[0]{17}
\renewcommand{\bonepfiveXpropXbmainkpc}[0]{40}

\renewcommand{\bonepsevenXpropXalphamain}[0]{32}
\renewcommand{\bonepsevenXpropXbmainkpc}[0]{22}

\renewcommand{\bzeropnineXpropXalphamain}[0]{56}
\renewcommand{\bzeropnineXpropXbmainkpc}[0]{31}

\renewcommand{\mainXpropXrastr}[0]{09:37:49.42}
\renewcommand{\mainXpropXdecstr}[0]{+06:56:51.5}

\title[Extended \MgII{} emission]{MusE GAs FLOw and Wind (MEGAFLOW) VIII. Discovery of a \MgII{} emission halo probed by a quasar sightline 
 }

\author[J. Zabl et al.]{
Johannes Zabl,$^{1,2}$\thanks{E-mail: johanneszabl@gmail.com}
Nicolas F. Bouch\'e,$^{1}$
Lutz Wisotzki,$^{3}$
Joop Schaye,$^{4}$
\newauthor
Floriane Leclercq,$^{5}$
Thibault Garel,$^{5,1}$
Martin Wendt,$^{6,3}$
Ilane Schroetter,$^{7}$ 
\newauthor
Sowgat Muzahid,$^{8,3}$
Sebastiano Cantalupo,$^{9}$ 
Thierry Contini,$^{10}$
Roland Bacon,$^{1}$
\newauthor
Jarle Brinchmann,$^{4,11}$
Johan Richard$^{1}$ \\
$^{1}$ Univ Lyon, Univ Lyon1, Ens de Lyon, CNRS, Centre de Recherche Astrophysique de Lyon UMR5574, F-69230 Saint-Genis-Laval, France\\
$^{2}$ Institute for Computational Astrophysics and Department of Astronomy \& Physics,  Saint Mary's University, 923 Robie Street,
Halifax,\\ Nova Scotia, B3H 3C3, Canada\\
$^{3}$ Leibniz-Institut f\"ur Astrophysik Potsdam (AIP), An der Sternwarte 16, 14482 Potsdam, Germany \\
$^{4}$ Leiden Observatory, Leiden University, PO Box 9513, NL-2300 RA Leiden, the Netherlands\\
$^{5}$ Observatoire de Gen\`eve, Universit\'e de Gen\`eve, 51 Ch. des Maillettes, 1290 Versoix, Switzerland \\
$^{6}$ Institut f\"ur Physik und Astronomie, Universit\"at Potsdam, Karl-Liebknecht-Str. 24/25, 14476 Golm, Germany \\
$^{7}$ GEPI, Observatoire de Paris, CNRS-UMR8111, PSL Research University, Univ. Paris Diderot, 5 place Jules Janssen, 92195 Meudon, France \\
$^{8}$ IUCAA, Post Bag-04, Ganeshkhind, Pune, India - 411007 \\
$^{9}$ Department of Physics, ETH Z\"urich,Wolfgang-Pauli-Strasse 27, 8093 Z\"urich, Switzerland \\
$^{10}$ Institut de Recherche en Astrophysique et Plan\'etologie (IRAP), Universit\'e de Toulouse, CNRS, UPS, F-31400 Toulouse, France\\
$^{11}$ Instituto de Astrof{\'i}sica e Ci{\^e}ncias do Espa\c{c}o, Universidade
do Porto, CAUP, Rua das Estrelas, PT4150-762 Porto, Portugal \\
}

\date{Accepted XXX. Received YYY; in original form ZZZ}

\pubyear{2020}

\begin{document}

\label{firstpage}
\pagerange{\pageref{firstpage}--\pageref{lastpage}}
\maketitle

\begin{abstract}
Using deep ($\texptot\,\mathrm{hr}$) VLT/MUSE data from the MEGAFLOW survey, we report the first detection of extended \MgII{} emission from a galaxy's halo that is probed by a quasar sightline. The $\MgII\,\lambda\lambda\,2796, 2803$ emission around the $z=0.702$ galaxy
($\log(M_*/\mathrm{M_\odot})=\mainXpropXsedXmass{}$) is detected out to 
$\approx25\,\kpc$ from the central galaxy
and covers $1.0\times10^3\,\kpc^2$  above  a surface brightness of
$14\times10^{-19}\,\uergsb{}$ ($2\,\sigma$; integrated over
$1200\,\kms = 19\textnormal{\AA}$ and averaged over
$1.5\,\tarcsec^{2}$). The \MgII{} emission around this highly inclined galaxy
($i\simeq\mainXincl$deg) is strongest along the galaxy's projected minor axis,
consistent with the \MgII{} gas having been ejected from the galaxy into a
bi-conical structure. The quasar sightline, which is aligned with the galaxy's
minor axis, shows strong \MgII{} absorption
($\rewmgii{}=\rewmgiiqso{}\,\textnormal{\AA}$) at an impact parameter of
$\quasarXpropXbmainkpc{}\,\kpc$ from the galaxy. Comparing the kinematics of
both the emission and the absorption - probed with VLT/UVES -, to the
expectation from a simple toy model of a bi-conical outflow, we find good
consistency when assuming a relatively slow outflow
($\vout=\windmodelXvout{}\,\kms$).
We investigate potential origins of the extended \MgII{} emission using
  simple toy models. With continuum scattering models we encounter serious difficulties in explaining the luminosity of the \MgII{} halo and in  reconciling density estimates from emission and absorption. Instead, we find that shocks might be a more viable source to power the extended \MgII{} (and non-resonant \OII{}) emission.

\end{abstract}

\begin{keywords}
galaxies: evolution -- galaxies: haloes -- intergalactic medium -- quasars: absorption lines -- quasars: individual: SDSSJ0937$+$0656
\end{keywords}



\section{Introduction}
\label{sec:intro}

There is a substantial body of indirect observational evidence that galaxy evolution is controlled by the interplay between galaxies and their surrounding circum-galactic medium (CGM). 
Galaxies need to accrete gas - the fuel for star formation - from the
circumgalactic medium, as the amount of gas inside a galaxy at a given time is not sufficient to explain the build up of their total stellar mass \citep[e.g.][]{Daddi:2010a,Tacconi:2010a,Tacconi:2013a,Freundlich:2013a,Saintonge:2013a}. Once formed, massive stars will end up as supernovae and power large-scale outflows.

The low density CGM is traditionally studied through the absorption it imprints on background sources. 
The background sources can be either the associated galaxies themselves (down-the-barrel) or unassociated background sources (transverse sightlines). 
Through collecting large samples of sightlines, \emph{statistical} insights into the distribution, physical state, and kinematics of the CGM gas can be gained.
These samples have revealed a complex multi-phase CGM (see \citealt{Tumlinson:2017a} for a recent review). 

The resonant $\MgII{}\,\lambda\lambda 2796, 2803$ doublet is an especially useful tracer of the cool ($10^{4\mbox{--}5}\,\mathrm{K}$) and metal enriched CGM due to its strength and rest-wavelength, which allows to study it with ground-based optical spectroscopy for redshifts between $0.3\lesssim z \lesssim 2.5$.

Studies with transverse quasar sightlines have revealed that \MgII{} is not
isotropically distributed around galaxies, but is instead found preferentially
along the galaxy minor or major axis
\citep[e.g.][]{Bordoloi:2011a,Bouche:2012a,Kacprzak:2012a,Lan:2014a,Nielsen:2015a,
  Martin:2019a,Schroetter:2019a,Zabl:2019a}. Down-the-barrel observations also
support this conclusion \citep[e.g.][]{Bordoloi:2014c}.
This bi-modality is consistent with a simple picture where galaxies are surrounded by extended gas disks, from which they likely accrete gas, and supernova driven winds that are ejected into a bi-conical outflow launched perpendicular to the disk. The kinematics of both the presumed disk sightlines \citep{Steidel:2002a, Chen:2005a, Kacprzak:2010a, Kacprzak:2011a, Bouche:2013a, Bouche:2016a, Ho:2017a, Ho:2020a, Rahmani:2018a, Zabl:2019a} and the presumed outflow sightlines \citep[e.g.][]{Bouche:2012a, Kacprzak:2014a, Muzahid:2015a, Schroetter:2015a,
  Schroetter:2016a, Schroetter:2019a, Rahmani:2018b, Martin:2019a, Zabl:2020a} are in good agreement with this simple picture.

Compared to transverse sightlines, down-the-barrel observations have the advantage that the sign of the velocity is known, but they have the major disadvantage that the location of the \MgII{} absorbing gas is not directly constrained, and could thus be ISM (inter-stellar medium), CGM or even IGM (inter-galactic medium) gas.
Blue-shifted \MgII{} absorption, which definitely represents outflows, is common in down-the-barrel observations and is strongest for galaxies that are observed face-on \citep[e.g.][]{Weiner:2009a, Rubin:2010b}, but the observed absorption strength does not only depend on inclination, but also on other - non-geometric - properties such as stellar mass, star-formation density, and redshift \citep[e.g.][]{Martin:2012a, Rubin:2014a}.

Another complication with down-the-barrel absorption studies is infilling by ISM \MgII{} \emph{emission}. While more massive star-forming galaxies ($\log(\Mstar{}/\msun{})\gtrapprox10.0$) appear to show mainly \MgII{} absorption with little or no emission, lower-mass galaxies ($\log(\Mstar{}/\msun{})\lessapprox 9.0$) appear to show only emission, with intermediate-mass galaxies between these masses showing P-Cygni profiles \citep[e.g.][]{Finley:2017b,Feltre:2018a}.
These variations in \MgII{} are possibly  a consequence of the complex radiative transfer that the resonant \MgII{} doublet experiences in the ISM gas \citep[][]{Prochaska:2011a}, similar to the radiative effects responsible for shaping the spectral profile and escape of \HI{} \lya{} \citep[e.g.][]{Dijkstra:2012a,Verhamme:2006a,Laursen:2009a}.

For \lya{} it is now well established that emission is not confined to
the ISM, but extends far out into the CGM. This halo-scale emission is
not limited to the rare extreme \lya{} halos around UV-bright quasars
\citep[e.g.][]{Borisova:2016a}, radio-loud AGNs, and \lya{}
blobs that are found in highly overdense regions  and are likely powered by
obscured AGNs
\citep[e.g.][]{Steidel:2000a,Francis:2001a,Yang:2009a,Prescott:2009a,Prescott:2015a,Cai:2017a}.
Extended emission is - at a weaker level - also typical around individual star
forming galaxies
\citep[e.g.][]{Steidel:2011a,Feldmeier:2013a,Momose:2014a,Wisotzki:2016a,Wisotzki:2018a,Leclercq:2017a}.
Recent studies start to focus on spatially and kinematically resolving these typical
\lya{} nebulae \citep[e.g.][]{Erb:2018a,Claeyssens:2019a,Leclercq:2020a}.

In extension of other similarities between \lya{} and \MgII{}, one can expect
that galaxies are also surrounded by \MgII{} emission halos. These would represent a reservoir of cool metal enriched gas, as revealed by the quasar absorption studies mentioned before.
However, despite this expectation, there are only very few detections of extended \MgII{} CGM emission \citep{Rubin:2011a, Martin:2013a}, with also non-detections being reported \citep{RickardsVaught:2019a}.
This situation is about to change thanks to the tremendous efficiency gain provided by the current panoramic IFU spectrographs: MUSE /VLT and KCWI/Keck. 
In addition to increasing the number of objects with \MgII{} halo detections, these IFU data also allow for spatially resolved analysis of the extended emission with a level of detail not feasible based on slit spectroscopy. 
In a recent pioneering study, \citet{Burchett:2021a} performed such a detailed analysis, where they used the KCWI to re-observe a \MgII{} nebula around a starburst galaxy previously found by \citet{Rubin:2011a}.

The  MUSE GAs FLOw and
Wind (MEGAFLOW) survey (\papone{}: \citealt{Schroetter:2016a};
						\paptwo{}: \citealt{Zabl:2019a};
						\papthree{}: \citealt{Schroetter:2019a};
						\papfour{}: \citealt{Zabl:2020a};
						\papfive{}: \citealt{Wendt:2021a};
            \papsix{}: \citealt{Schroetter:2020a},
            \papseven{}: \citealt{Freundlich:2021a})
is aimed at studying the  properties of  gas flows around star-forming galaxies with   low-ionization absorption by \MgII{} ($\lambda\lambda$ 2796,2803) in 22 quasar fields.
This survey, which combines VLT/UVES and VLT/MUSE observations, has enabled us to bring the sample size of galaxy-quasar pairs from a dozen \citep{Bouche:2012a, Schroetter:2015a} to almost $80$ pairs of which about 2/3 are suitable for studying outflows  \citep[][hereafter, paper~III]{Schroetter:2019a}.
However, given that  typical exposure times in the MEGAFLOW survey are 2--4hr,
we acquired deeper MUSE observations with $>10$hr integration on two fields
(SDSSJ0937$+$0656; SDSSJ0014$-$0028) 
in 2015-2018 with the aim to test whether the outflows detected in absorption against background quasars    would be visible in emission with metal lines, such as \MgII{}.
Here, we present the discovery of extended \MgII{} emission around a $z=0.7$ galaxy in the field of SDSSJ0937$+$0656.

The paper is organised as follows. We present our observations in \Sec{sec:obs},
the \nameprime{} galaxy and other objects in the field in \Sec{sec:object}, and
our results for the CGM emission and absorption in \Sec{sec:halo} and
\Sec{sec:back_absorption}, respectively. Then, in \Sec{sec:toy_model}, we use a
simple toy model to discuss the \MgII{} halo's properties. Finally, in
\Sec{sec:discussion}, we conclude with a comparison to other objects.
Throughout, we use a 737 cosmology ($H_0=70\,\kms\,\Mpc^{-1}$,
$\Omega_{\rm m}=0.3$, and $\Omega_{\Lambda} = 0.7$) and a \citet{Chabrier:2003a}
stellar Initial Mass Function (IMF). All distances are proper distances.
All wavelengths and redshifts are in vacuum and are corrected to a heliocentric
velocity standard. All
fluxes are corrected for Galactic extinction \citep{Schlegel:1998a}.
All stated errors are $1\sigma$, unless otherwise noted.

\section{Observations}
\label{sec:obs}

\begin{table}
\caption{\label{tab:obs} Total exposure times and average seeing (Moffat FWHM at
  $7050\,\textnormal{\AA}$) split into observing nights.  Only data with good
  quality that was used for the final cube are included. Also listed are the observing modes. NOAO
  and AO are short for WFM-NOAO-N or WFM-AO-N, respectively.}
\centering
\begin{tabular}{crrrr}
\hline
DATE-OBS & Mode & $\mathrm{t}_\mathrm{exp}$ & Seeing  & Prog. ID. \\
&  & [hr] & \arcsec{} &  \\
\hline
2015-04-15	 & NOAO	 & 1.00	 & 0.55	 & 095.A-0365(A) \\
2015-04-17	 & NOAO	 & 1.00	 & 0.90	 & 095.A-0365(A) \\
2017-11-16	 & AO	   & 0.28	 & 0.55	 & 0100.A-0089(B) \\
2018-02-14	 & AO	   & 1.56	 & 0.85	 & 0100.A-0089(A) \\
2018-02-15	 & AO	   & 3.11	 & 0.70	 & 0100.A-0089(A) \\
2018-03-18	 & AO	   & 0.78	 & 0.65	 & 0100.A-0089(A) \\
2018-04-11	 & AO	   & 1.56	 & 0.59	 & 0101.A-0287(A) \\
2018-04-14	 & AO	   & 1.56	 & 0.83	 & 0101.A-0287(A) \\
2018-04-17	 & AO	   & 0.39	 & 0.80	 & 0101.A-0287(A) \\
\hline
\end{tabular}
\end{table}

The field centered on the $z=1.82$ quasar SDSSJ0937$+$0656 was observed for a total of 13.6 hr with MUSE in 2015-2018.
This on-source integration is slightly deeper than for the 3
\arcmin{}$\times$3\arcmin{}  mosaic MUSE observations in the UDF
\citep{Bacon:2017a} and allows us to reach a \MgII{} surface-brightness limit of
$14.0\times10^{-19}\,\uergsb{}$ (2$\sigma$) (see \Sec{sec:mgii_map}).

The observations are described in Table~\ref{tab:obs}.
While the median seeing was  $0\arcsec.8$, some observations were rejected due to either technical issues or poor seeing
(FWHM$>1\arcsec.0$).
The total usable science exposure is $\texptot\,\mathrm{hr}$. We created our main cube from this dataset.
In addition to this full set of usable exposures, we also selected a sub-set
including only the 3.36\,hr with the best image quality (FWHM < 0\arcsec.6), in order to better constrain
morphology and spatial variations of the foreground galaxy's line fluxes.

We reduced the data similarly as in papers II\mbox{--}IV, but using the newer
Data Reduction Software (DRS) v$\geq2.4$ \citep{Weilbacher:2016a, Weilbacher:2020a} which includes the auto-calibration of the fluxes in the slices \citep[e.g.][]{Bacon:2015a,Bacon:2017a}.
Compared to our initial data reduction, where we masked regions with strong flat-fielding imperfections at the edges of slices, here we used the method advocated by \citet{Bacon:2021a} to improve the flat-fielding through applying a `super sky-flat`. This consists of median combining all the sky-subtracted exposures (masking the sources) to produce a sky-flat, which can then be subtracted from the individual exposures.
We create this master flat from the pixtables rather than the cubes to enhance computation efficiency\footnote{Alternatively, each of the typically 30 frames used in a master flat must be resampled 30 times to put each of them on the same WCS grid.}, and refer to this sky-flat in the following as `super-fast super-flat' (\sfsf{}).

In practice, we used here a \sfsf{} constructed from 36 exposures from another MUSE GTO program (the MUSE eXtreme Deep Field [MXDF]; PI: Bacon) taken during observing runs overlapping with our observations (in August 2018).
After testing, we applied this \sfsf{} to our AO (adaptive optics) data taken from Nov 2017 to April 2018 as it does not only improve the flat-fielding at the slice edges, but also reduces sky-line residuals. For the non-AO exposures, we applied simpler masks, as in papers II\mbox{--}IV.
Further details about the new reduction chain will be given in the survey paper (Bouch{\'e} et al. in prep.). 

As part of the MEGAFLOW survey, we also observed the quasar SDSSJ0937+0656 with the VLT high-resolution spectrograph UVES (Ultraviolet and Visual 
Echelle Spectrograph; \citealt{Dekker:2000a}) for a total integration time of 9~ks in the nights of 2015-12-20 and 2016-01-11.
Further details about the observation, reduction, and continuum normalisation can be found in \citetalias{Zabl:2019a}.

\section{The object}
\label{sec:object}

\begin{figure*}
  \centering
  \includegraphics[height=7cm]{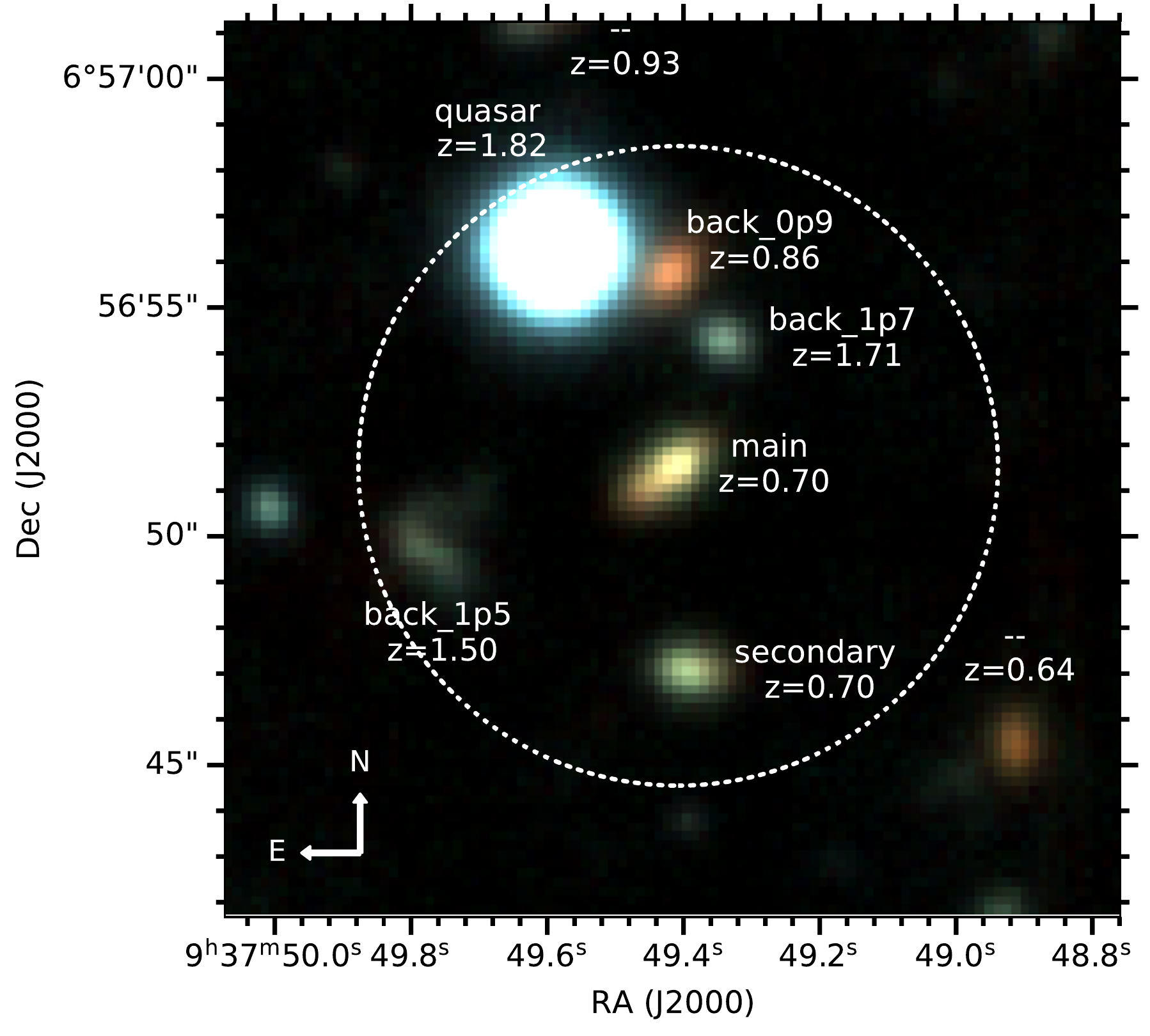} 
  \includegraphics[height=7cm]{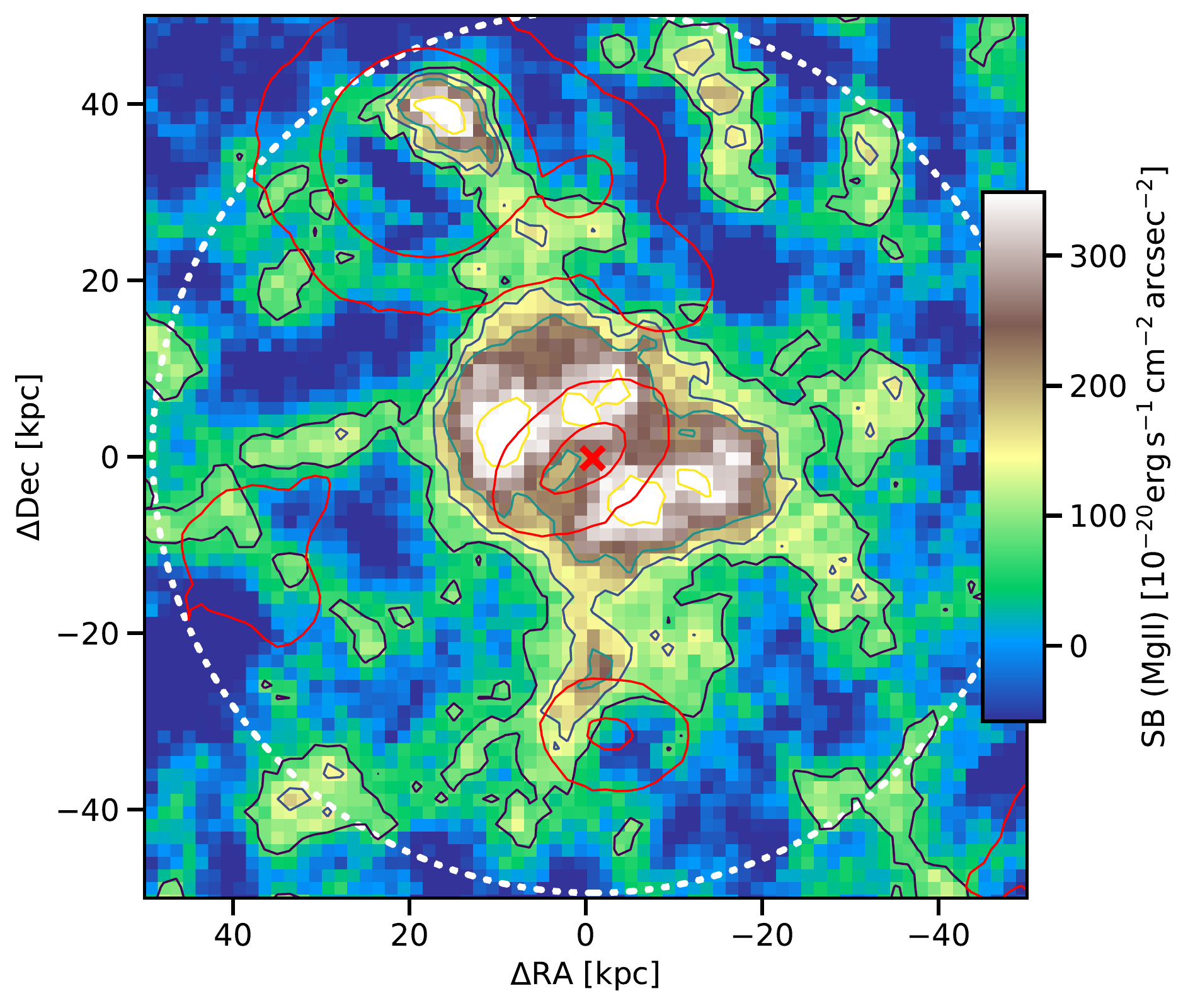}
  \\
  
  \caption{\label{fig:main_maps} \textbf{Left:} $20\,\tarcs \times 20\,\tarcs$ field around the \nameprime{} galaxy with (pseudo-) i', r', and V filters in
  the red, green, and blue channels of the colour image, respectively. The surrounding galaxies are labelled
  and the redshifts are stated (see also \Tab{tab:backsightlineinfo} and
  \Sec{sec:abs_galaxy_obs}).  The dotted white ring is centred on the
  \nameprime{} galaxy and has a radius of $50\,\kpc$ at $z=0.7021$. The centre, which is
    used throughout the paper, was determined from a \galfit{} fit to a pseudo
    r' image (see \Fig{fig:galfit_rp}).
\textbf{Right:} SB emission map created with a $600\,\kms$ double NB filter (see Fig. \ref{fig:mgii_spec_full_extract}). The map was smoothed with a tophat filter with a diameter of $10\,\kpc$ (1.4\arcsec{}). The dotted white ring has an identical radius as the colour image in the left panel.
The black, grey, green, 
and yellow contours
correspond to surface brightness levels of 7, 14, 21 and
$35\times10^{-19}\,\uergsb{}$ (1, 2, 3, and $5\sigma$). For the \MgII{} SB map all (projected-) neighbours, including the quasar, have been subtracted in the cube (see text \Sec{sec:mgii_map}). The red contours (23.9 and $26.2\,\mathrm{mag}\,\tarcsec^{-2}$; arbitrarily chosen) are for a MUSE white-light continuum image convolved with the same kernel as the SB image. These contours show both the position of the \nameprime{} galaxy and those of the neighbours (c.f. left). The continuum peak of the \nameprime{} galaxy is indicated as red cross.
  }
\end{figure*}

The quasar \quasarid{} sightline has three strong \MgII{} absorbers at redshifts of 0.702, 0.857, 0.933 with rest-frame equivalent widths $\rewmgii=\rewmgiiqso$, 1.0\, and $1.5\,\textnormal{\AA}$, respectively. In all three cases we have been able to identify at least one galaxy associated with the absorber within $100\,\kpc$ of the quasar.
The galaxies associated with the $z=0.702$ and $z=0.933$ absorbers were first discussed in \citetalias{Schroetter:2019a}{}. The galaxy associated with the $z=0.857$ absorber was not included in the previous \mfl{} papers due to its low \OII{} flux.

In the present paper, we focus on the absorber at $z=0.702$, for two
reasons: first, because it has the lowest possible \MgII{} redshift
within the MUSE wavelength coverage, it favors our ability to detect
low surface brightness emission. Second, because this absorber has (at least) one galaxy whose position angle (PA) is such that the quasar is approximately along the galaxy projected minor axis, which favors, as motivated in the
introduction, outflows causing \MgII{} absorption. 
As discussed in \citetalias{Schroetter:2019a}{}, this $z=0.702$ absorber actually  has two galaxies at the redshift of the absorber within $100\,\kpc$ of the quasar.\footnote{For completeness, in addition to these two
  galaxies within $100\,\kpc$ of the quasar, we have identified within a search
  window of $\pm 400\,\kms$ around the absorber redshift two further galaxies at $224$ and
  $\bkpcthirdqso\,\kpc$, respectively.
In addition to their much larger impact parameter, these are also 
 $>$10 times fainter than the $\nameprime$ galaxy.}
 In addition, we searched for a putative fainter companion located under the quasar PSF and found no evidence for such a companion to a flux limit 10$\times$ fainter than the \nameprime{} galaxy. The quasar PSF can be well subtracted around the observed wavelength of the \OII{} doublet ($6343\,\textnormal{\AA}$). By placing the \nameprime{} galaxy at the quasar location and rescaling its flux by 1/5 or 1/10, such companion would have been detected in both cases.

The \nameprime{} galaxy is at an impact parameter 
$b_1=\bkpcprimeqso\,\kpc(\approx 0.2\,\rvir)$, while the \namesecond{} galaxy's impact parameter is approximately twice as large with
$b_2=\bkpcsecondqso\,\kpc(\approx 0.6\,\rvir)$, or even three times as large when normalized with the respective galaxy's virial radius. \Fig{fig:main_maps} (left) shows the
location of the \nameprime{} and \namesecond{} galaxies in relation to
the quasar location and three background galaxies,
\namebonepseven{} at $z=1.71$, \namebonepfive{} at $z=1.50$, and
\namebzeropnine{} at $z=0.86$. Two further background galaxies within $50\,\kpc$ of the \nameprime{} galaxy are too faint to be visible in \Fig{fig:main_maps} (left), and we will not discuss them further.\footnote{A $z=1.06$ galaxy (09:37:49.67 +06:56:54.2) is close to the quasar and a $z=1.27$ galaxy (09:37:49.69 +06:56:51.0) is in close proximity of the \namebonepfive{} galaxy.}

Using the azimuthal angle
 $\alpha$,  defined as the projected angle between the galaxy's major axis and the apparent quasar location (see Fig.~1 of \citetalias{Zabl:2019a}), the sightline is suitable as a `wind' probe for both  \nameprime{} and \namesecond{} galaxies, as the  azimuthal angle is $\alphaprimeqso\,\deg$ ($\alphasecondqso\,\deg$) for the \nameprime{} (\namesecond{}) galaxy.
 But, given the strong anti-correlation between $\rewmgii$ and
$b$ \citep[e.g.][]{Lanzetta:1990a,Chen:2010a,Nielsen:2013a}, we will assume that the majority of the \MgII{}
absorption is caused by gas flows associated with the closest of the two galaxies,
\nameprime{} ($b_1=\bkpcprimeqso\,\kpc$). The much lower
$M_*$ and $SFR$ of the \namesecond{} galaxy, which has less than 1/5 of the
\nameprime{} galaxy's $M_*$ and about half its $SFR$, makes it a  minor
  companion, which gives further credibility to
associating the absorption primarily with the \nameprime{} galaxy.

As discussed in \citetalias{Schroetter:2019a}{}, the \nameprime\ galaxy is a main-sequence galaxy which has a SFR of $10-20$ $\mpy$ and a stellar mass of $\log M_\star/\msun\simeq10.5$, which was estimated in \citetalias{Schroetter:2019a}{} from the Tully-Fisher like scaling relation between $S05$ ($\equiv \sqrt{0.5\,\vmax^2 + \sigma_0^2}$) and stellar mass \citep[e.g.][]{Kassin:2007a}. In addition, \citetalias{Freundlich:2021a} recently constrained the $SFR$ of this galaxy using dust continuum observations to $<23.8\,\mpy$  ($3\,\sigma$), ruling out a dust-obscured starburst. Here, we perform a 
spectral energy distribution (SED) fit directly to our MUSE data  and found that
$\log M_\star/\msun=\mainXpropXsedXmass$  using our custom  code \coniecto{}
\citep[as in \citetalias{Zabl:2019a}, \citetalias{Zabl:2020a}, and][]{Zabl:2016a}. Using strong-line
calibrations from \citet{Maiolino:2008a} we infer for the \nameprime{} galaxy a
gas-phase metallicity of about solar, which is a typical value for its mass and
redshift. Further details about stellar mass and metallicity can be found in
Appendix~\ref{appendix:mstar} and \ref{appendix:linefluxes}, respectively.

Perhaps the most relevant galaxy property for this paper is the galaxy's inclination. However, we find that the \nameprime\ galaxy contains a significant  clump of SFR offset from the kinematic centre (see Appendix~\ref{appendix:morpho}) which impacts the morphology and  the determination of the inclination $i$ from \OII{}.
In addition, as we will show in \Sec{section:halo:morpho}, the \OII\ flux map
contains extended emission (towards the minor axis), which will bias the
determination of $i$. Thus, the most reliable estimate of the inclination (and
PA) of the \nameprime{} galaxy can only be obtained from fitting a model to the
continuum.
 
Using a pseudo-broadband image for the $r'$ passband, the best-fit flux model obtained  with   \galfit{} \citep{Peng:2010a}
 (see Appendix~\ref{appendix:morpho}  for further details)
 has a \sersic{} index of $\mainXpropXgalfitXn$,  half-light radius
 $\rhalf=\mainXpropXgalfitXre\,\kpc$, and an inclination of $\incl =
 \mainXpropXgalfitXincl\,\deg$, using a Moffat PSF with the parameters as
 measured from the quasar. We will adopt this inclination,
 $i\simeq\mainXincl\,\deg$ for the rest of this work.

We measure the kinematics of the \nameprime{} galaxy using \camel{}
\citep{Epinat:2012a} and \gpk{} \citep{Bouche:2015b} as described in
Appendix~\ref{appendix:kinematics} and find that the maximum velocity is
$V_{\rm max}=\mainXgpkXvmax\,\kms$, and the disk dispersion is
$\sigma_0=\mainXgpkXsigma\,\kms$. These parameters were inferred using a fixed
inclination ($\mainXpropXgalfitXincl\,\deg$) and a deconvolved \OII{} flux
model that we obtained with \galfit{}.\footnote{We expect systematics of about
  $5\mbox{--}10 \deg$ (see, e.g., Appendix A of \citetalias{Zabl:2019a}) to
  dominate uncertainties on the \incl{}.} The estimated virial velocity and halo
mass from the kinematics are $\vvir=\mainXderivedXvvirfromkin\,\kms$ and $\log
M_{\rm h}/\mathrm{M}_\odot= \mainXderivedXmvirfromkin$, respectively.
For the galaxy's redshift we find both from the kinematic fitting and from visual inspection of a position-velocity-diagram a best fit redshift of $z=0.7021$. We estimate the uncertainty of the \OII{} redshift to be about $20\,\kms$.
\task{Reference to coordinates of the galaxy}

\Tab{tab:prop_gals} lists all properties of the \nameprime{} galaxy associated
with the $z=0.702$ absorber. Coordinates of both the \nameprime{}{} and the
\namesecond{} galaxy are listed in \Tab{tab:backsightlineinfo}.

\begin{table}
  \caption{\label{tab:prop_gals}
 Physical properties of the \nameprime{} galaxy associated with the $z=0.70$ absorber.
  For further details, see \citetalias{Zabl:2019a}.
  (1) \OII{} flux;
  (2) nebular extinction from $\ebv{}-M_*$ relation;
  (3) nebular extinction from SED fit;
  (4) instantaneous SFR from \OII{} assuming E(B-V) from (2);
  (5) instantaneous SFR from (3; SED fit);
  (6) stellar mass from SED fit;
  (7) rest-frame B absolute magnitude from best-fit SED model;
  (8) distance from the main-sequence of star-forming galaxies (assuming main-sequence from \citealt{Boogaard:2018a});
  (9) age of galaxy from SED fit (time since onset of star-formation);
  (10) decay time in delayed $\tau$ SFH from SED fit;
  (11) half-light radius $\rhalf$ from \galfit{} continuum fit;
  (12) rotation velocity from \gpk{} fit;
  (13) velocity dispersion from \gpk{} fit;
   (14) adopted inclination $i$ (deg);
  (15) virial velocity;
  (16) virial radius from $\vvir$;
  (17) virial mass from $\vvir$;
  (18) redshift from \OII{}.
  }
  \centering
  \begin{tabular}{rrrrr}
    \hline
  Row  & Property            & \nameprime{}                    & Unit                \\
  \hline
  (1)  & $f_{\OII}$          & $\mainXpropXgpkXflux$           & $\uerglf$           \\
  (2)  & E(B-V) ($M_*$)      & $\mainXpropXderivedXebvmass$    & mag                 \\
  (3)  & E(B-V) (SED)        & $\mainXpropXsedXebv$            & mag                 \\
  (4)  & SFR ($f_{\OII}$)    & $\mainXpropXderivedXsfroiimass$ & $\mpy$              \\
  (5)  & SFR (SED)           & $\mainXpropXsedXcurrsfr$        & $\mpy$              \\
  (6)  & $M_*$ (SED)         & $\mainXpropXsedXmass$           & $\log(\msun)$       \\
  (7)  & $B$                 & $\mainXpropXbabsmag$                          & mag                 \\
  (8)  & $\delta(MS)$        & $\mainXpropXderivedXdeltams$    & dex                 \\
  (9)  & age              	 & $\mainXpropXsedXage$            & $\log(\mathrm{yr})$ \\
  (10) & $\tau$              & $\mainXpropXsedXlogtau$         & $\log(\mathrm{yr})$ \\
  (11) & $\rhalf$            & $\mainXpropXgalfitXre$  & kpc                 \\
  (12) & $\vmax$             & $\mainXgpkXvmax$                & $\kms$              \\
  (13) & $\sigma_{0}$        & $\mainXgpkXsigma$               & $\kms$              \\
  (14) & $i$                 & $\mainXpropXgalfitXincl$        & deg                 \\
  (15) & $\vvir$             & $\mainXderivedXvvirfromkin$     & $\kms$              \\
  (16) & $\rvir$             & $\mainXderivedXrvirfromkin$     & $\kpc$              \\
  (17) & $\Mvir$ (from kin.) & $\mainXderivedXmvirfromkin$     & $\log(\msun)$       \\
  (18) & $z$                 & $0.7021\pm0.0001$  & \\
      \hline
    \end{tabular}
\end{table}

\begin{table*}
\caption{\label{tab:backsightlineinfo} Information about position and
  orientation of the \namesecond{} galaxy and the four background sightlines w.r.t. the \nameprime{} galaxy, including impact
  parameter, $b$, azimuthal angle, $\alpha$ ($0\deg \leq \alpha \leq 90\deg$), redshift, and observed
  magnitude at the wavelength corresponding to \MgII{} at the redshift
  of the \nameprime{} galaxy.
}
\begin{tabular}{l l|rrrrrr}
Name                                         & [unit]             & main              & \namesecond{}              & quasar                     & \namebonepfive{}                & \namebzeropnine{}                & \namebonepseven{}                \\
\hline
  R.A. (J2000)                                 & hms                &
                                                                      \mainXpropXrastr  & \secondXpropXrastr       & \quasarXpropXrastr{}       & \bonepfiveXpropXrastr{}         & \bzeropnineXpropXrastr{}         & \bonepsevenXpropXrastr{}         \\
Dec (J2000)                                  & dms                & \mainXpropXdecstr & \secondXpropXdecstr      & \quasarXpropXdecstr{}      & \bonepfiveXpropXdecstr{}        & \bzeropnineXpropXdecstr{}        & \bonepsevenXpropXdecstr{}        \\
$b$  (from \nameprime{})                     & $\kpc{}$           & \mbox{--}         & \secondXpropXbmainkpc{}  & \quasarXpropXbmainkpc{}    & \bonepfiveXpropXbmainkpc{}      & \bzeropnineXpropXbmainkpc{}      & \bonepsevenXpropXbmainkpc{}     \\
$\alpha$ (from \nameprime{})                 & $\deg{}$           & \mbox{--}         & \secondXpropXalphamain{} & \quasarXpropXalphamain{}   & \bonepfiveXpropXalphamain{}     & \bzeropnineXpropXalphamain{}     & \bonepsevenXpropXalphamain{}     \\
redshift (of background galaxy)              &                    & 0.702             & 0.702                    & \quasarXpropXredshift{}    & \bonepfiveXpropXredshift{}      & \bzeropnineXpropXredshift{}      & \bonepsevenXpropXredshift{}      \\
m @ $4760\,\textnormal{\AA}$                 & mag                & 23.5              & 23.8                     & \quasarXpropXmabatmgiimain & $\bonepfiveXpropXmabatmgiimain$ & $\bzeropnineXpropXmabatmgiimain$ & $\bonepsevenXpropXmabatmgiimain$ \\
\end{tabular}
\end{table*}

\section{Halo in emission}
\label{sec:halo}

The main purpose of the present study is to use our \mfl{}-deep MUSE cube ($\texptot\,\mathrm{hr}$) to test whether we can detect the circumgalactic \MgII{} in \emph{emission} around galaxies with known \MgII{} \emph{absorption} in transverse quasar sightlines.
We focus here on the most promising candidate, which is  the \nameprime{} galaxy (see \Sec{sec:object}) associated with the strong $z=0.70$ \MgII{} absorber. 
We defer a statistical analysis of the presence or absence of \MgII{} emission halos in the full \mfl{} sample to a later paper. We note that we do not detect extended \MgII{} emission around the galaxies that are associated to the other two absorbers ($z=0.86$ and $0.93$) in the \quasarid{} sightline.

\subsection{\MgII{} emission map}
\label{sec:mgii_map}

We searched for extended \MgII{} emission around the \nameprime{} galaxy by creating a double pseudo-narrowband filter centred on the expected observed frame wavelength of \MgII{}. The transmittance curve of this double filter was centred on the $2796$ and $2803$ lines of the \MgII{} doublet, respectively. Each of the two passbands was chosen to be $600\,\kms$ wide.\footnote{The separation of the two lines is $770\,\kms$. This means that the filter has a gap of $170\,\kms$.}
After converting to units of surface-brightness, we refer to this narrowband (NB) image in the following as the \MgII{} surface-brightness map (SB-map).

We subtracted the continuum to first order by taking the median in the spectral direction in a $\pm10000\,\kms$ window around \MgII{}.\footnote{To the blue of \MgII{}, not the full $10000\,\kms$ was
available.} For non-resonant lines, such as e.g. \OII{}, this standard approach of continuum subtraction is usually sufficient to remove the continuum of both the targeted galaxy and other galaxies in the field.
However, for resonant lines, such as \MgII{}, the same gas that we aim to probe in emission will also imprint itself as absorption on the background sources. This absorption will show up in an off-band subtracted NB image (or cube) as negative flux.
To mitigate this effect, we subtracted the background sources from the layers that contribute to the NB images using a 3D model.

For subtracting the quasar (including the \MgII{} absorption), we tested both using an empirical, non-parametric, PSF model and a wavelength-dependent parametric model. 
In the latter case, we determined the best-fit Moffat PSF parameters
\citep{Moffat:1969a} as a function of wavelength using the \pampelmuse{} code
\citep{Kamann:2013a}. The fitted FWHM varies monotonically from 0.81\arcsec{} in the blue
($4750\,\textnormal{\AA}$) to 0.56\arcsec{} in the red ($9300\,\textnormal{\AA}$) which will be used for the morphology and kinematics analysis of the \nameprime{} galaxy.
While overall a good representation of the PSF,
 this Moffat model leaves residuals for the relatively bright quasar  that  can  impact \MgII{} emission measurements around the
\nameprime{} galaxy.
  Therefore, for the quasar subtraction over the
wavelength range relevant for the NB filter, we constructed an
empirical PSF from an image with a relatively large velocity range
($\pm 10000\,\kms$) around the considered line, excluding the line
itself.
To further increase the signal in the outer part of this empirical PSF image and to avoid being impacted by neighbouring sources, we replaced the empirical PSF in the outer part ($>1.4\arcsec{}$ away from the quasar) with the median flux in annuli.

Similarly, for subtracting  the relevant background
galaxies (\namebonepseven{}, \namebonepfive{}, \namebzeropnine{}) and the \namesecond{} galaxy, we first created a model with \galfit{} using an $r'$ (pseudo-)broadband image created from the MUSE cube and fit this model to each layer of the cube taking into account the appropriate PSF.  The fitting was done in a conceptually identical way to that used in codes such as \pampelmuse{} \citep{Kamann:2013a} or \todse{} \citep{Schmidt:2019a}. Subsequently, we removed the best fit from each layer.
These layer-by-layer fits also provide optimally extracted spectra.

The resulting \MgII{} SB-map  is shown in Fig.~\ref{fig:main_maps} (right). We smoothed it with a tophat with a diameter of $1.4\arcsec$ ($10\,\kpc$ at $z=0.70$). 
Contours corresponding to the five (yellow), three (green), two (grey), and one (black) sigma significance levels of the smoothed map are shown on the map. These significance levels correspond to surface-brightnesses of 35, 21, 14, and $7\times10^{-19}\,\uergsb{}$, respectively.\footnote{Values are the average SB within the spatial smoothing filter, which has an area of $1.5\,\mathrm{arcsec}^2$.} The significance levels were determined based on the standard deviation of 80 independent pixels in object-free regions over the full MUSE FOV that were far enough apart from each other not to be correlated after smoothing.
Also shown (in red) are the contours of the white light-image with the same smoothing applied.

\begin{figure}
  \centering
  \includegraphics[width=0.8\columnwidth]{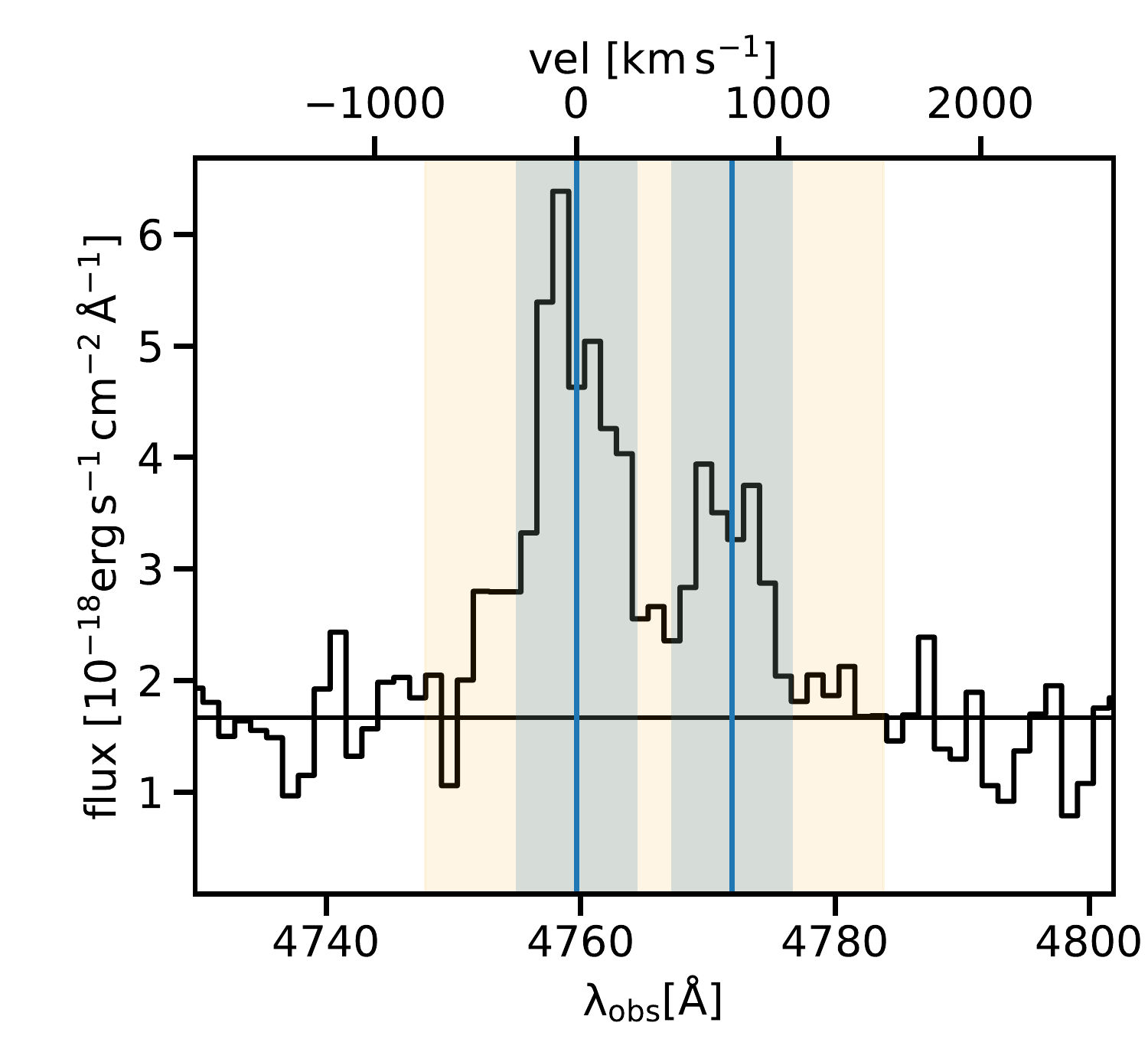}
  \caption{Spectrum around the observed wavelength of the \MgII{} doublet extracted from a large spatial aperture corresponding to the $2\sigma$ contour in \Fig{fig:main_maps} (right). The black horizontal line indicates the local continuum level estimated by excluding the wavelength range covered by the yellow region. The blue shaded regions indicate the width ($600\,\kms$ each) of the double NB filter used for creating the SB map in \Fig{fig:main_maps}. By contrast, the yellow shaded region extents continuously over the full range from $-750\,\kms$ bluewards of the 2796 line to $750\,\kms$ redwards of the 2803 line.
   A SB map using this more inclusive filter is discussed in Appendix~\ref{app:wide_mgii_sb_map}. }
  \label{fig:mgii_spec_full_extract}
\end{figure}

In \Fig{fig:mgii_spec_full_extract}, we show the \MgII{} spectrum of the full halo. This spectrum was extracted from the $2\,\sigma$ contour of the \MgII{} SB map (grey contour in right panel of \Fig{fig:main_maps}). 
The blue vertical bands in \Fig{fig:mgii_spec_full_extract} indicate the passband of the double NB filter used to create the SB map in \Fig{fig:main_maps} (right). This passband encompasses the majority of the \MgII{} flux and is hence a well motivated choice. While it misses some flux at higher velocities, we show in Appendix~\ref{app:wide_mgii_sb_map} that this extra flux does not change the morphology of the halo.
The total \MgII{} flux (2796+2803) within the blue ($\pm300\,\kms$) passband is $41\pm3\times10^{-18}\,\uerglf{}$, with a 2796/2803 ratio of $1.9\pm0.3$. This ratio agrees with the value of 2.0
expected for optically thin emission. The flux corresponds to a luminosity of $9.0\pm0.7\times10^{40}\,\mathrm{erg}\,\mathrm{s}^{-1}$.

\subsection{Morphology of the \MgII{} halo}
\label{section:halo:morpho}

\begin{figure*}
	\centering
	\includegraphics[width=\textwidth]{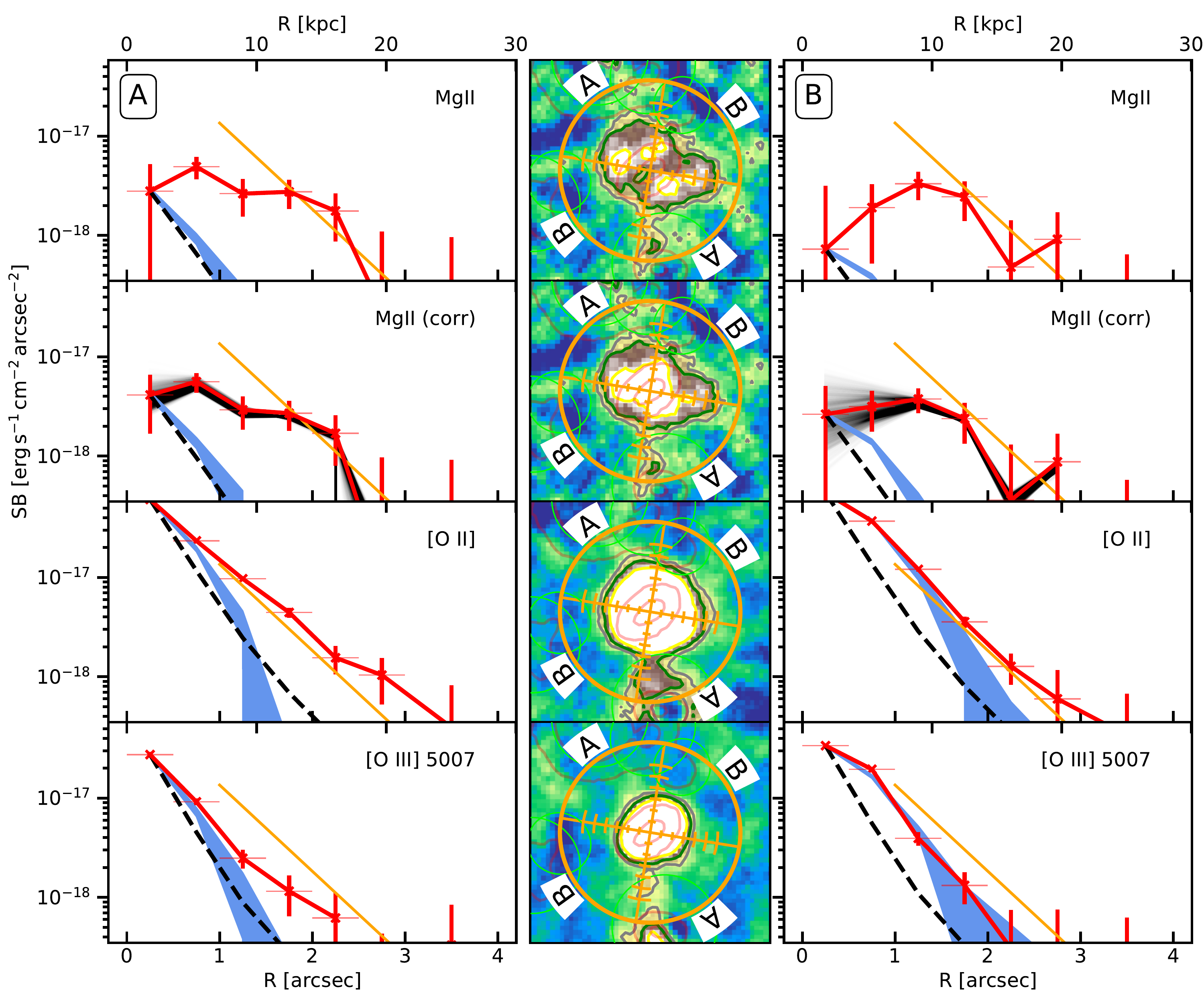}
	\caption{\label{fig:sb_profiles_pa} 
	Radial surface brightness profiles (median) are shown for \MgII{} (1st row), \OII{} (3rd row), and
  $\OIII{}\,\lambda 5007$ (4th row) along both the minor (left, A) and major axis (right, B) of
  the \nameprime{} galaxy. 
  The red solid curves show the respective emission line surface brightness (SB).
  The black dashed line shows the PSF profile as determined from the quasar and the blue curve is the profile of the galaxy's continuum.
  The arbitrary orange exponential profile is identical in all six panels and is
  meant as a visual aid for comparison of the different profiles.
  The respective cone-like regions used for the minor and major axis profiles
  are indicated (orange) in the line SB maps (middle column). The radial
    bin edges are indicated by little fins, with the first bin starting in the centre. All radial
    bins have a width of 0\arcsec{}.5, except the outermost, which has
    double the width. The outer edge of this last bin is the full circle, which
    is at a radius of 4\arcsec{} ($=28.6\kpc$)    For comparability, the \OII{} and \OIII{} SB maps were determined with the same filter width as that used for \MgII{} ($600\kms$; for \OII{} extended by $224\kms$ to take account for the doublet). Regions potentially contaminated by residuals from neighbouring galaxies have been excluded from the profile extraction both for the line and continuum profiles, and are indicated as green ellipses.
  The errors for both the line (red vertical bars) and the continuum (extent of blue shading) have been determined based on empty apertures (see text in \Sec{section:halo:morpho}) and are $2 \sigma$. The thin red horizontal bars indicate the radial bins.
  Both PSF and continuum profiles have been determined from a $\pm10000\,\kms$ wide window around the respective emission line, excluding the emission itself. Further, these two curves have been arbitrarily scaled to match the line surface brightness profile in the innermost bin.
  Finally, a version of the \MgII{} profile after correcting for likely weak continuum absorption is shown in the 2nd row. The grey curves are based on random realizations of absorption profiles taken from the MCMC chain obtained, and indicate the uncertainty of this correction (see \Sec{section:halo:morpho}).
	} 
\end{figure*}

\Fig{fig:main_maps} (right) shows that the \MgII{} emission extends out to at least $20\,\kpc$ from the galaxy. Above the $2 \sigma$ SB threshold of $14.0\times10^{-19}\,\uergsb{}$ it encompasses an area of $19\,\sqarcsec$, corresponding to $9.8\times10^2\,\kpc^2$.
The morphology of the \MgII{} emission is very different from that of the continuum. The strongest \MgII{} emission appears along the minor axis of the galaxy.

In order to analyze the morphology of the \MgII{} emission, we show in
\Fig{fig:sb_profiles_pa} the radial surface brightness profile extracted along the minor and major axes of the \nameprime{} galaxy respectively.
The \MgII{} profile along the minor axis is relatively flat out to
about $20\,\kpc$, after which it drops significantly.  While not extending as far out along the major axis, the profile 
is still substantially flatter than the continuum and even increasing in the inner part.

\begin{figure}
  \centering
  \includegraphics[width=0.9\columnwidth]{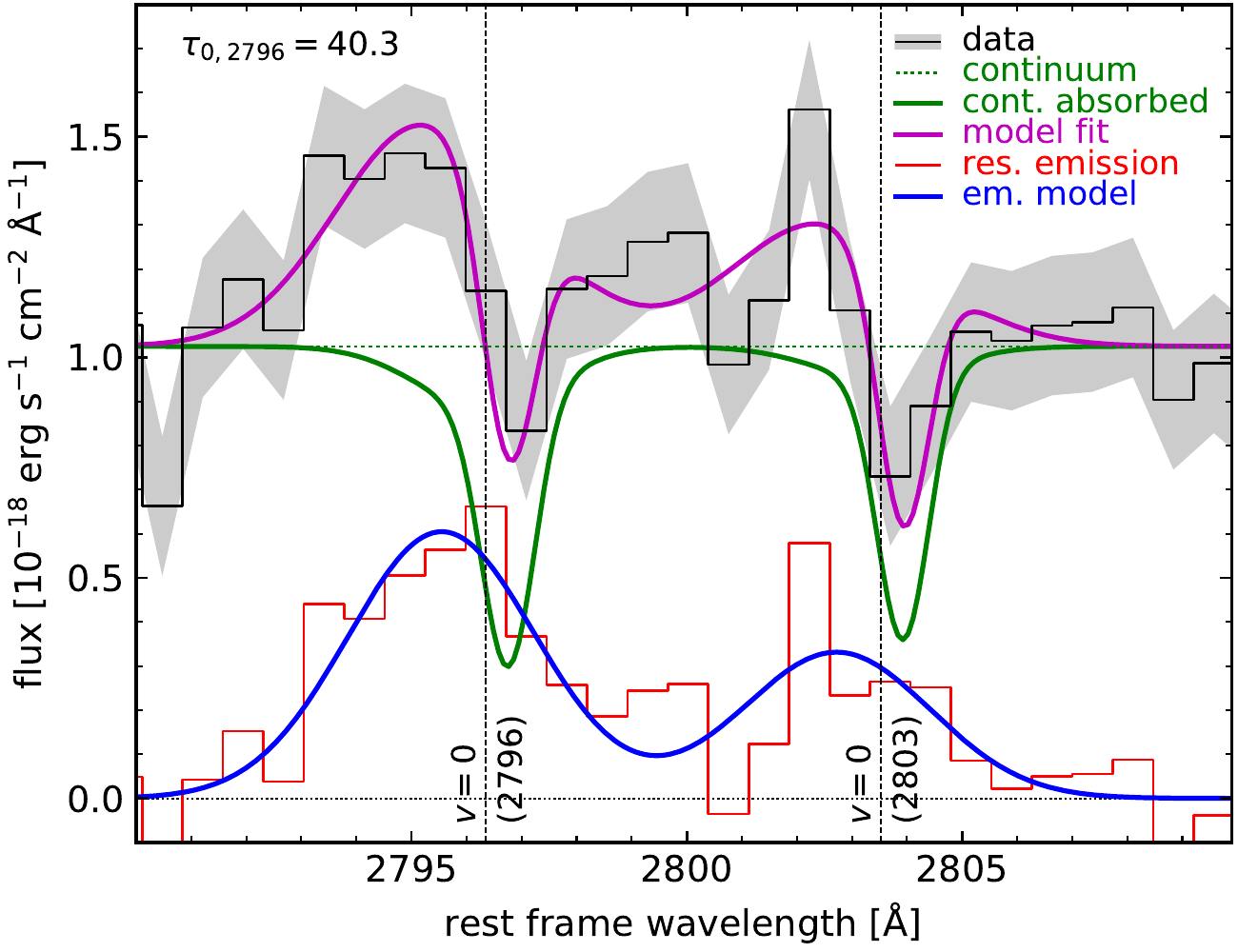}
  \caption{\label{fig:infillcorrfit} Decomposition of the \nameprime{} galaxy's
    \MgII{} spectrum into absorption and emission. The observed spectrum (black
    steps with grey errorband) was extracted from an elliptical aperture with a
    semi-major axis of $1\arcsec$. The purple line is the best-fit model, which
    is a linear combination of a continuum with doublet absorption (green solid
    line) and doublet emission (blue solid line). A direct subtraction of the
    absorption spectrum from the observed spectrum (red steps) is as expected
    consistent with the emission model. The observed wavelengths of \MgII{}
      $\lambda 2796$ and $2803$ at the systemic redshift of the
      \nameprime{} galaxy are indicated by vertical lines. Further details are
    given in \Sec{section:halo:morpho}. }
\end{figure}

One complication is that the lack of positive flux at the position of the central galaxy might not be an actual lack of emission, but can be caused by \MgII{} absorption in the galaxy's down-the-barrel spectrum.
Indeed, \MgII{} spectra centred on galaxies often show absorption, with part of the absorption infilled by redshifted emission (P-Cygni like profiles; e.g., \citealt{Weiner:2009a, Martin:2009a, Erb:2012a, Finley:2017b, Feltre:2018a}). 
In order to recover the full emission, one can attempt to simultaneously fit emission and absorption \citep[e.g.][]{Martin:2012a}. 

\Fig{fig:infillcorrfit} (black curve) shows the central \MgII{} spectrum of the \nameprime{} galaxy extracted from an elliptical aperture with a semi-major axis of 1\arcsec{}. 
While no strong absorption and little emission is directly visible in this spectrum, it is still possible that emission and absorption almost exactly cancel. To investigate this possibility, we performed a formal decomposition using a model and an MCMC algorithm as described in Wisotzki et al. in prep. For details about the code we refer the reader to this publication, but in short: The model assumes Gaussians for both the optical depth distribution of the absorption and for the flux distribution of the emission. 
The velocity offsets and widths of these Gaussians are free parameters and
  are independent between
absorption and emission, but were assumed to be identical for the 2796 and 2803
lines in both cases. The optical depth ratio between the 2796 and 2803 absorption was fixed to 2:1 and unit covering was assumed. The flux ratio between the two emission lines was a free parameter within the 2796/2803 range from 0.1 to 2.5.
Finally, the model was convolved with the spectral resolution of the MUSE data. 

\Fig{fig:infillcorrfit} (purple line) shows the resulting best fit.\footnote{Best fit means here a model with parameters corresponding to the 50\% percentile in the marginalized 1D distributions.} This best-fit model describes the data very well and is a super-position of the (green) model absorption spectrum and the (blue) model emission spectrum.
This means that the actual \MgII{} emission from within the elliptical aperture is likely a factor of about 2 higher than what one would measure from a simple NB image.

In \Fig{fig:sb_profiles_pa} (second row), we show the impact of this correction on the SB emission map (created as in \Sec{sec:mgii_map}). The full, corrected emission is estimated by rescaling the offband continuum cube by the absorption spectrum, as inferred from the decomposition fit described above, before subtracting it from the cube.
An implicit assumption in this process is that the normalized absorption profile is identical over the galaxy's full extent. As expected, the correction mainly removes the central flux suppression. However, the radial profile still remains relatively flat without a central peak.

\subsection{Kinematics of the \MgII{} halo}
\label{section:halo:kinematics}

\begin{figure}
\centering
  \includegraphics[width=\columnwidth]{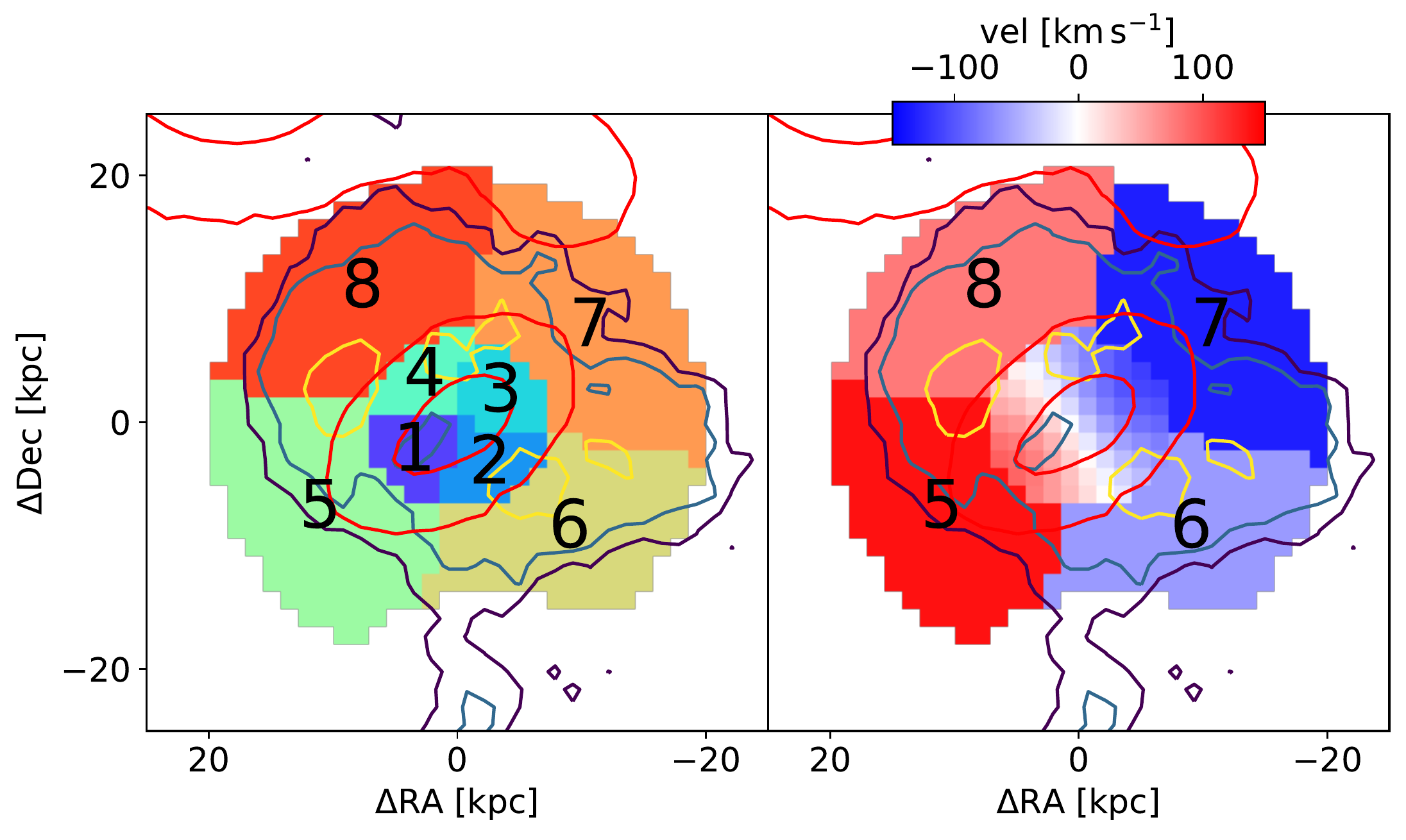} \\
  \includegraphics[width=0.9\columnwidth]{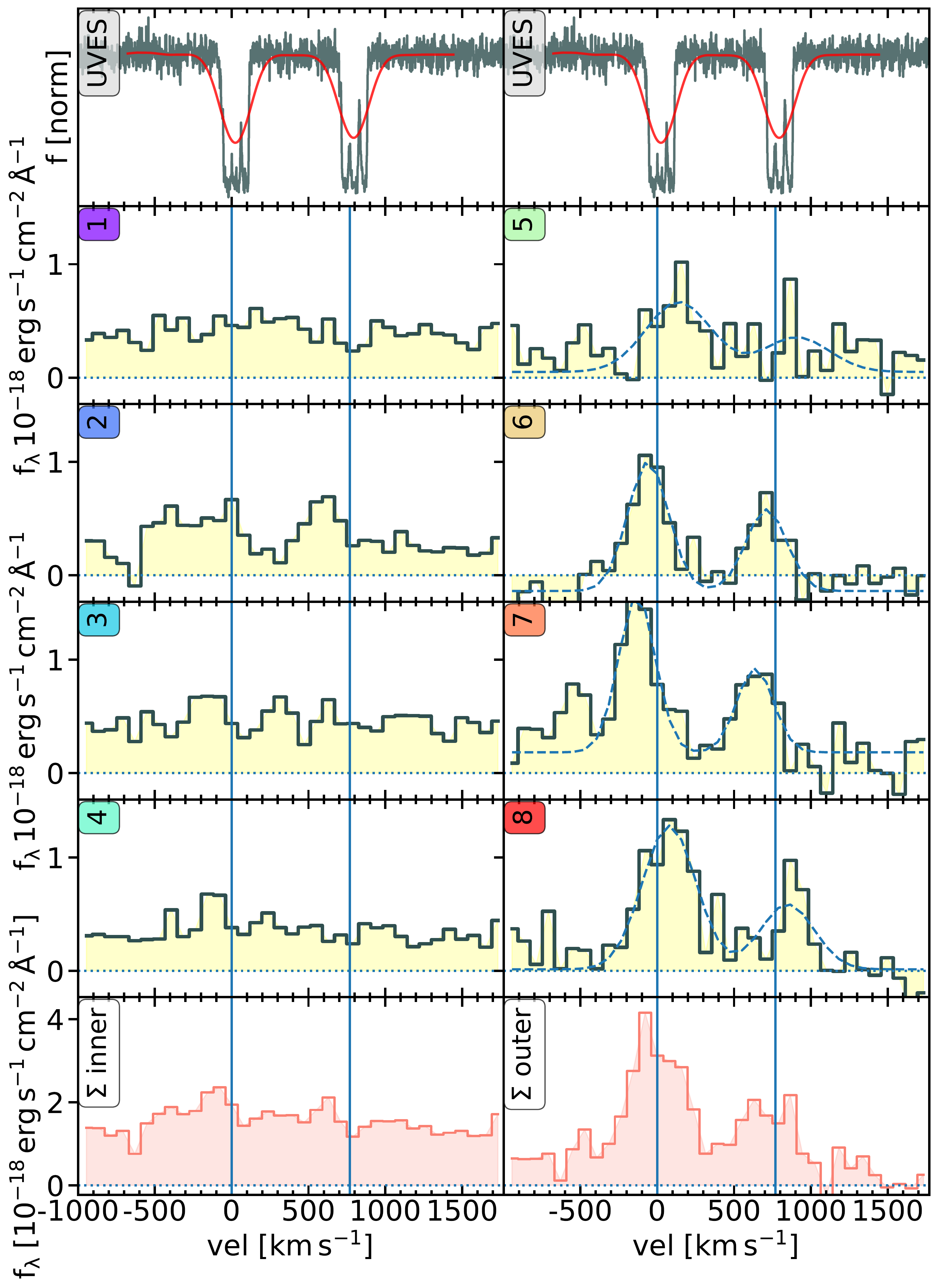}
  \caption{\label{fig:quadrant_spectra} Spatial variations of the
    \MgII{} emission spectrum.  We extracted spectra in 8 different
    regions, as defined in the map on the top-left. The color-coding in the top-left panel is
      arbitrary. The respective extracted spectra, 
    are shown in yellow below.
    Regions 1 to 4 (left column)
    extend out to a radius $r\leq 1\arcsec{}\;(7.2\,\kpc)$, while
    regions 5 to 8 (right column) cover the radial range
    $1\arcsec{}<r \leq 2.8\arcsec{} (7.2 \mbox{--}20\,\kpc)$. The odd
    numbered regions are along the \nameprime{} galaxy's major axis,
    while the even numbered regions are along the minor axis. 
     The
    two red spectra in the bottom row correspond to the sum of the four yellow spectra in the corresponding columns. This is identical to an extraction from an inner aperture (left) and an outer annulus (right), respectively.
    The two panels in the top row are identical and show the UVES quasar spectrum
    at the wavelength corresponding to \MgII{} at the redshift of the
    \nameprime{} galaxy (see \Sec{sec:abs_quasar_obs}). The UVES spectrum (grey) is also shown
    convolved with the resolution of MUSE (red).
    The velocity scale of the spectra is for $\MgII{}\,2796$
      and is w.r.t. the systemic redshift of the galaxy. The two blue vertical
      lines indicate the observed wavelengths of the 2796 and 2803 lines at this systemic
      redshift, respectively.
      Gaussian
      doublet fits are shown for the four outer regions as blue dashed line.
      The velocities obtained from these fits are shown in the
      upper-right map, where for comparison also the galaxy velocity field as measured from
      \OII{} with \gpk{} is shown.}
\end{figure}

\begin{figure*}
  \centering
  \begin{minipage}{\textwidth}
    \centering
    \raisebox{-0.4\height}{\includegraphics[width=0.3\textwidth]{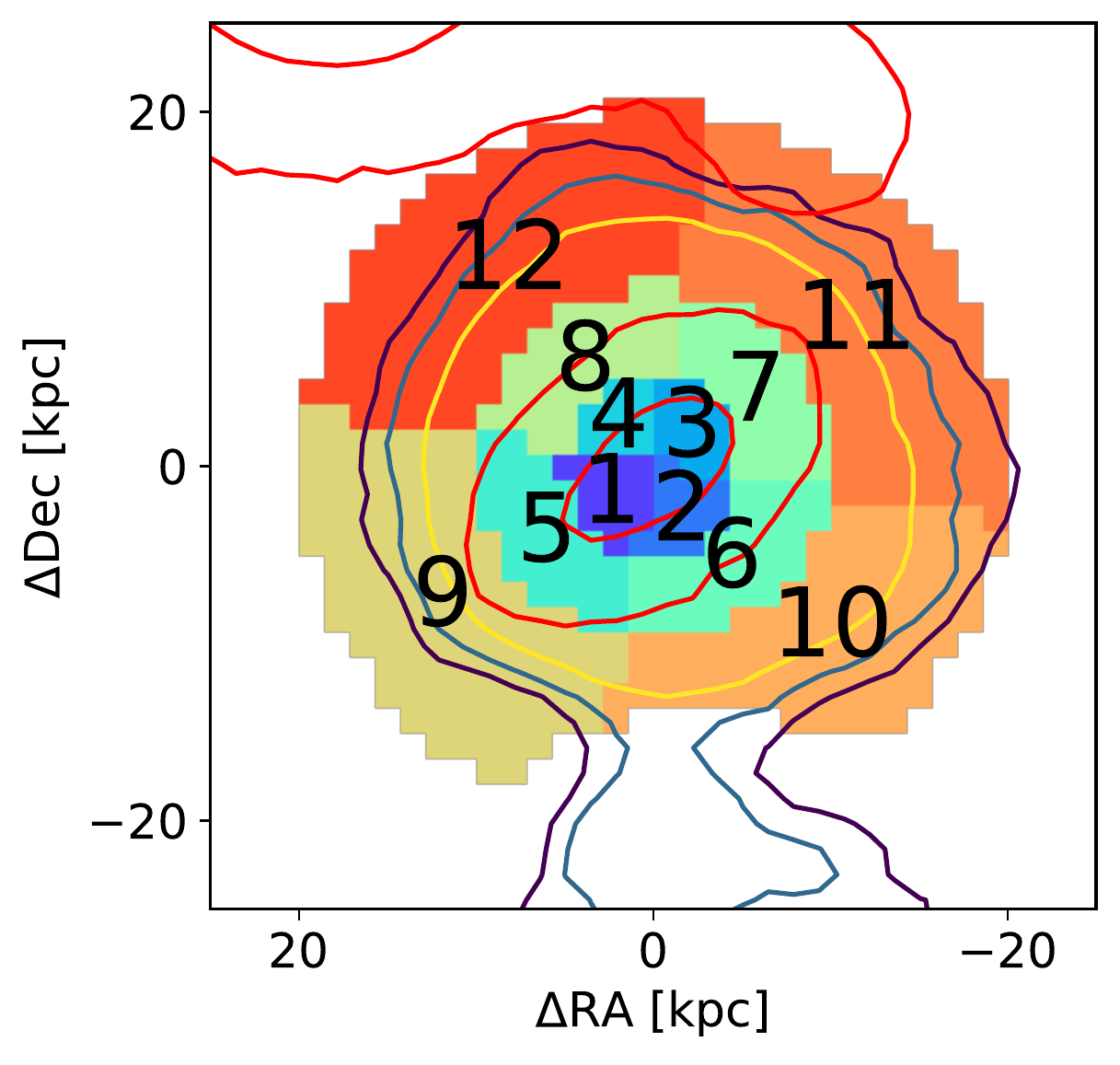}}
    \raisebox{-0.5\height}{\includegraphics[width=0.64\textwidth]{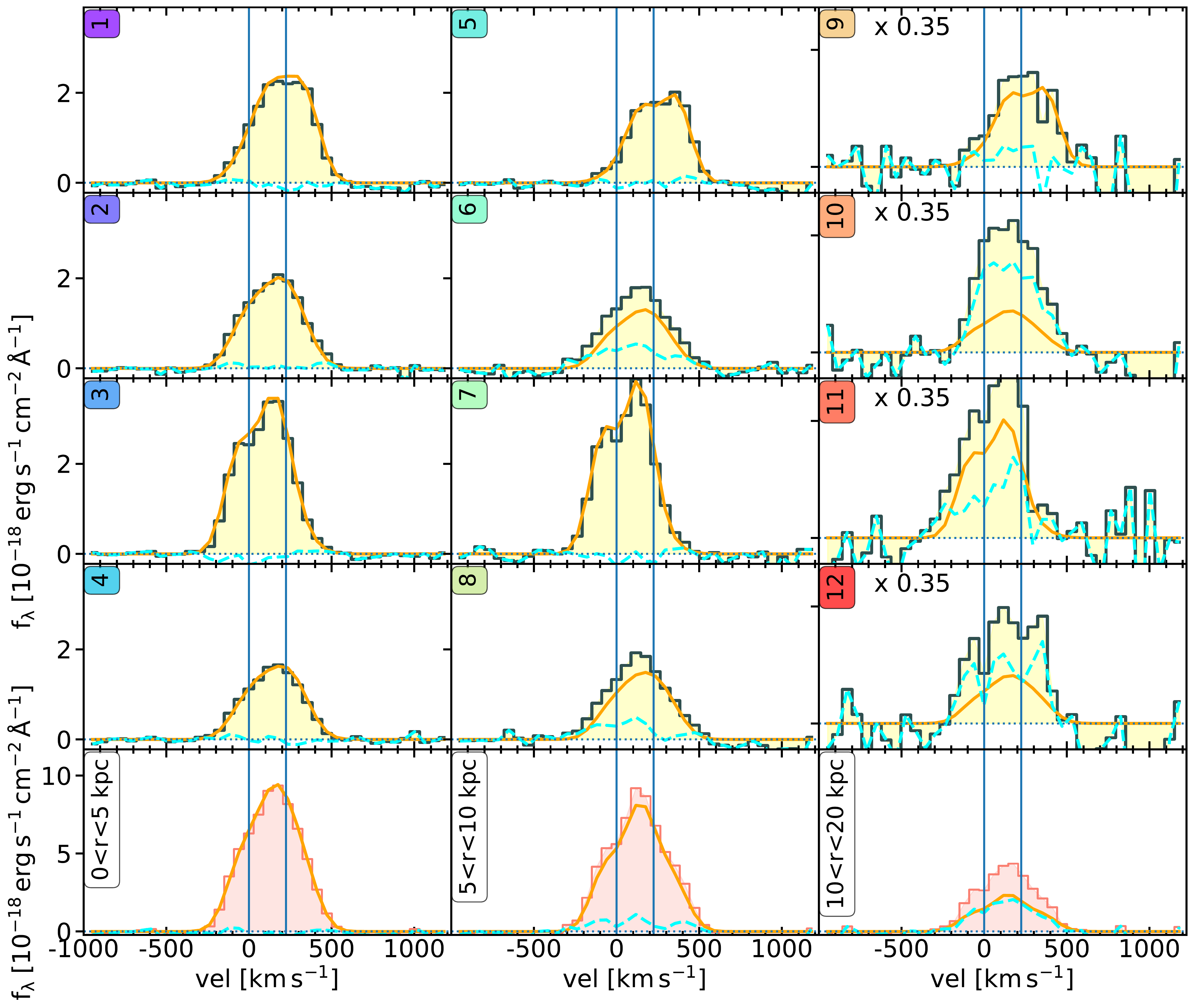}} \\
  \end{minipage}
    \caption{\label{fig:quadrant_spect_oii} 
      \OII{} spectra from multiple spatial regions around the \nameprime{}
      galaxy. Spectral extractions for the 12 numbered regions indicated in the
      map (left) are shown on the right (black line and yellow shading). A
      region around the neighbouring \namesecond{} galaxy has been excluded to
      mostly avoid contribution from residual \OII{} emission of this galaxy, which
      reduces the size of regions 9 and 10 slightly compared to regions 11 and
      12. The velocity scale is for $\OII{}\,3727$
        and is w.r.t. the systemic redshift of the galaxy. The two blue vertical
        lines indicate the observed wavelengths of the 3727 and 3729 lines at this systemic
        redshift, respectively.
 The y-scale for the spectra of regions $9\mbox{--}12$ has been
      reduced by a factor 0.35 compared to that used in regions $1\mbox{--}8$.
      The sum of the spectra from the four inner regions (regions $1\mbox{--}4$;
      $r<5\,\kpc\;(0.7\arcsec{})$), the four middle regions (regions
      $5\mbox{--}8$; $5\,\kpc\;(0.7\arcsec{})<r<10\,\kpc\;(1.4\arcsec{})$), and
      the four outer regions (regions $9\mbox{--}12$;
      $10\,\kpc\;(1.4\arcsec{})<r<20\,\kpc\;(2.8\arcsec{})$) are in the bottom
      row. The orange solid lines are spectra extracted from the best-fit \gpk{}
      model cube (see \Sec{sec:object}) in an identical way as from the data.
      The cyan dashed line is the residual after subtracting the \gpk{} model
      from the data and can be interpreted as excess flux above the ISM emission.
    }
\end{figure*}

While the SB map is illustrative (shown in \Fig{fig:main_maps}
right), the MUSE data cube allows to go further
and analyse the spatial variations of the \MgII{} kinematics and
spectral shape.  \Fig{fig:quadrant_spectra} shows the \MgII{} spectra
extracted in different regions. The red spectra in the left and right bottom panels of \Fig{fig:quadrant_spectra} are spectra extracted in
annuli with a radial range of $0\mbox{--}1\arcsec{}$
($0.0\mbox{--}7.2\,\kpc$) and $1\mbox{--} 2.8\arcsec{}$
($7.2\mbox{--}20\,\kpc$).  

Going one step beyond simple annuli, we split the two annuli into four
quadrants each (Regions 1\mbox{--}8 of \Fig{fig:quadrant_spectra}).
The four regions comprising the inner annulus (1\mbox{--}4), which
cover the ISM of the \nameprime{} galaxy rather than its CGM, show
very little \MgII{} emission individually. The outer regions 5
and especially 7 (along the \nameprime{} galaxy's major axis)
appear to have less \MgII{} emission than the outer regions 6 and 8 (along the minor axis), consistent with the visual impression from
\Fig{fig:main_maps} (right). 

Interestingly, the south-east region (5) is redshifted w.r.t. the \nameprime{}
galaxy's systemic redshift ($+139\,\kms$)\footnote{Best-fit value from Gaussian
  doublet fit.}, while
the north-west region (7) is blueshifted ($-132\,\kms$), consistent with an extension
of ISM rotation field (see velocity coded region map in \Fig{fig:quadrant_spectra}). This
  could indicate that the \MgII{} emission seen along the major axis
traces the co-rotating extended gaseous disks that have been seen so far in absorption (e.g.~\citealt{Bouche:2016a}; \citealt{Ho:2017a}; \citetalias{Zabl:2019a}).

Furthermore, \Fig{fig:quadrant_spectra} also reveals that the two
  regions along the minor axis (6 and 8) have a clear velocity shift
  with respect to each other. The region towards the south-west (6) is
  blue-shifted ($-59\,\kms$) and the opposite region (8) is red-shifted ($+78\,\kms$).  Given that
  these two regions are approximately coincident with the two bipolar
  peaks seen in the \MgII{} emission map (\Fig{fig:main_maps}
  right), this provides us with crucial kinematic information to characterize a
  putative bi-conical outflow (see also \Sec{sec:dis_em_kinematics}).
  In Appendix \ref{sec:app:rotouter}, we rule out that the kinematics in these two regions might be
    caused by a combination of rotation and the observed slightly in-homogenous flux
    distribution.

In summary, the \MgII{} emission appears to have complex, but coherent
kinematics connected to both the disk kinematics along the major-axis and to the outflow kinematics along the minor-axis.
Therefore, the line shape of a spectrum obtained by summing over the full halo
(\Fig{fig:mgii_spec_full_extract}) or over annuli (\Fig{fig:quadrant_spectra};
bottom row) might be
shaped more by a superposition of different kinematic components than
by radiative effects. In \Sec{sec:toy_model}, we discuss a possible explanation for the
coherent motions.

\subsection{Halo emission from other lines}
\label{sec:em-from-other-lines}

For comparison, we also show the \OII{} and \OIII{} profiles in
\Fig{fig:sb_profiles_pa}. The profiles along the major axis are mostly consistent  with
the slope of the continuum profile. 
Along the minor axis, however, the \OII{} emission is significantly more extended and the \OIII{} emission potentially slightly
more extended than the continuum. 
The observed $\OII{}/\MgII{}$ ratio\footnote{We determined the flux ratios using maps with their PSF matched to the map with the lowest resolution ($\MgII$). In practice, we determined the matching kernels empirically using Gaussians modified by polynomials \citep{Alard:1998a}.} decreases along the minor axis from $20$  in the central $0\arcsec.5$ to $\approx1 \mbox{--} 2$ at $2\arcsec\pm0.5$. Even when correcting for potential down-the-barrel \MgII{} absorption in the centre (see \Sec{section:halo:morpho}), the ratio changes still by an order of magnitude between the centre ($\OII{}/\MgII{}=10$) and the outer parts.
The $\OIII{}/\OII{}$ ratio seems to decrease over the same range from 1/2 in the centre to $\approx1/3$ at $2\arcsec$.

For \OII{}, we performed a spatially resolved kinematics analysis in multiple regions along the minor and major axes, similar to what we did for \MgII{} in \Fig{fig:quadrant_spectra}. Given the more centrally peaked radial SB of \OII{} compared to \MgII{}, we decided to put the regions at different radii than in the case of \MgII{} ($r\leq 5\,\kpc{}$, $5\,\kpc{}<r<10\,\kpc{}$, and  $10\,\kpc{}<r<20\,\kpc{}$).
Spectra for the twelve regions defined in the left panel of
\Fig{fig:quadrant_spect_oii} are shown in the right of the same figure (black
line/solid yellow shading). 
It is very revealing to compare each of these spectra to spectra obtained in an identical way from the best-fit \gpk \OII{} model (see \Sec{sec:object}). The difference between these model spectra (orange) and the data allows us to assess excess halo emission (cyan).
Little excess flux is needed in the regions along the major axis even at large radii. Except in the region 11, the excess flux is essentially zero, and even in region 11 the excess flux is subdominant. 
By contrast, already in the radial range from $5\,\kpc{}<r<10\,\kpc{}$  significant excess flux is needed for the two minor axis regions (regions 6 and 8). The excess flux dominates in the two outer minor axis regions. The total \OII{} excess flux out to $20\,\kpc$ is $26\pm2\times10^{-18}\uerglf$.
For comparison, the total \MgII{} flux (2796+2803) out to $20\,\kpc$ is $39\pm5\times10^{-18}\uerglf$. Making the assumption that all \MgII{} flux but only the excess portion of the \OII{} flux are produced outside the ISM, this implies a \MgII{}/\OII{} ratio larger than one in the CGM.

While it is difficult to interpret the kinematics of the excess flux due to the unknown doublet ratio and the low signal-to-noise (S/N), the kinematics appear consistent with that seen in \MgII{} along the minor axis. This is a relatively small velocity shift compared to systemic. 
The comparison between the \MgII{} halo morpho-kinematics and the  morpho-kinematics of the \OII{} halo excess flux is very suggestive of a similar spatial distribution of the \OII{} emitting and the \MgII{} emitting/scattering gas.

\section{Halo in absorption}
\label{sec:back_absorption}

While \MgII{} emission maps provide a novel way to \emph{image} the
cool metal-enriched CGM, as in, e.g. \citet[][]{Rubin:2011a,Martin:2013a,Burchett:2021a}, Wisotzki et al. (in prep), Leclercq et al. (in prep), the total luminosity of the halo has a natural limit in the number of photons injected by the central galaxy that are in resonance with $\MgII$ ions in the CGM. By contrast, the absorption that the gas imprints on the spectra of background sightlines is independent of the brightness of the central source, and solely depends on the column density along the line-of-sight.

By selection we have a quasar sightline piercing through the halo of the \nameprime{} galaxy, at an impact parameter of $\bkpcprimeqso{}\,\kpc$. We describe the observed absorption in the quasar sightline in \Sec{sec:abs_quasar_obs}. In \Sec{sec:abs_quasar_mod}, we apply our generic outflow toy model \citep{Bouche:2012a} to the observed quasar absorption. In addition to the quasar sightline, there are three background galaxies detected within $50\,\kpc$ of the \nameprime{} galaxy (see \Fig{fig:main_maps} (left)).  
The continuum of these galaxies is detected, but despite the depth of
the MUSE observation, with relatively low S/N. We briefly summarize the limited information we can extract at the current depth from these sightlines in \Sec{sec:abs_galaxy_obs}.

All four sightlines are just outside the region where we significantly detect \MgII{} emission, and hence allow us in principle to extend the mapping provided by the emission in an absorption tomographic way \citep[e.g.][]{Bowen:2016a,Lopez:2018a,Lopez:2019a}, especially if in the future observations of the background sources could be obtained with the ELTs.

\subsection{Observed absorption in the quasar sightline}
\label{sec:abs_quasar_obs}

\begin{figure}
\begin{center}
	\includegraphics[width=0.8\columnwidth]{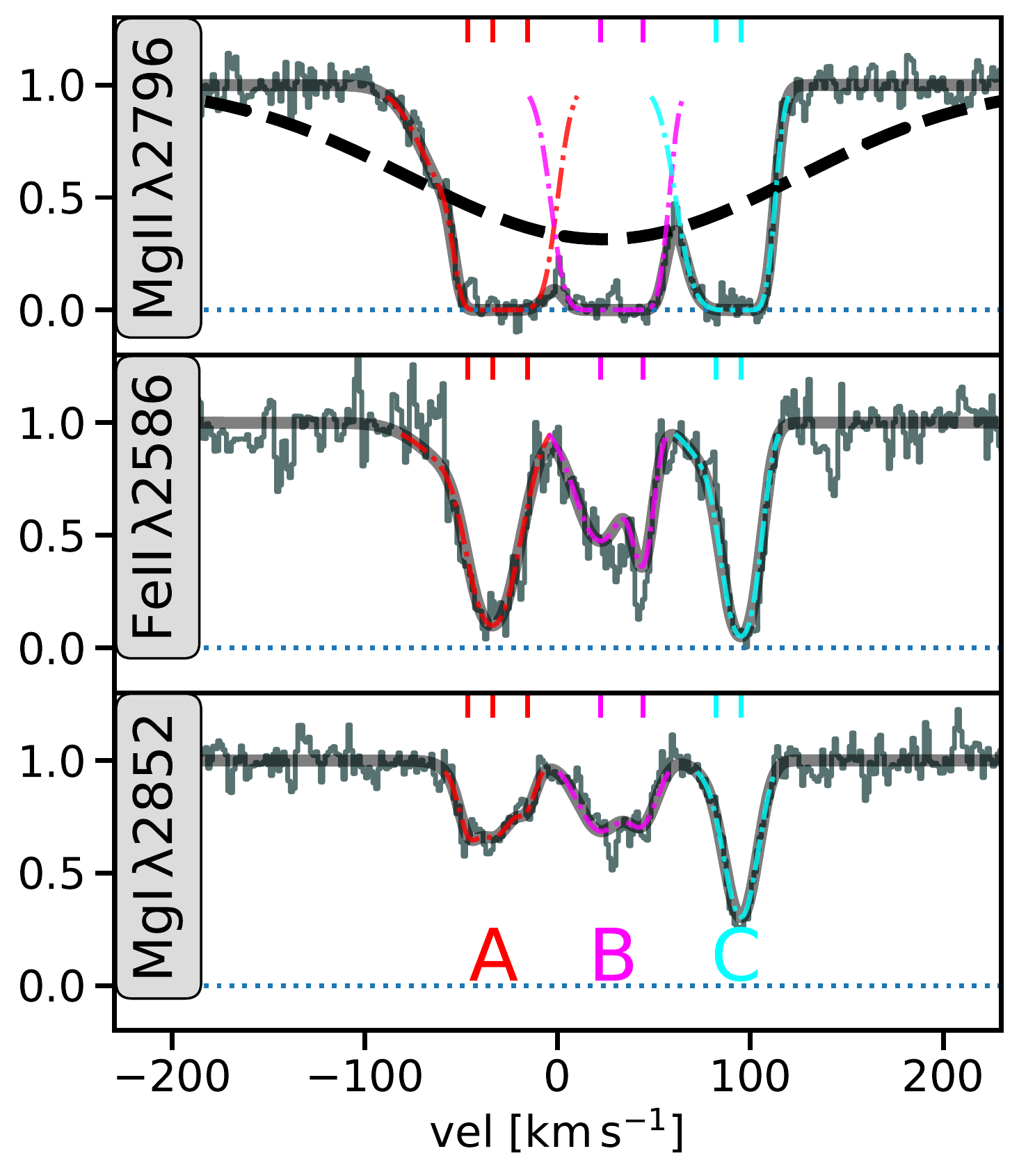}
\end{center}
\caption{\label{fig:uves_lines}
Absorption seen in the quasar UVES spectrum at the redshift of the \nameprime{} galaxy.
Detected lines are transitions of \MgII{}, \MgI{}, and \FeII{}, for each of which one transition is shown.
The thin grey line are the observed UVES data, while the thick line is a model fit to the data. This model consists of the seven components at velocities as indicated on the top. Three main kinematic components (A, B, C) can be visually identified, the contributions of which to the model are indicated by the red, magenta, and cyan lines, respectively.
The black dashed line in the $\MgII{}\,\lambda2796$ panel is the UVES spectrum artificially degraded to the resolution of MUSE at this wavelength.
}
\end{figure}

The UVES data for the quasar sightline allows us to study the
absorption strength and kinematics not only for \MgII{}, based on
which the system was selected, but also for several other transitions
(\Fig{fig:uves_lines}).  In addition to \MgII{}, we detect absorption
in multiple \FeII{} lines and $\MgI{}\,\lambda2852$.  Weaker lines
from $\ZnII{}$, $\MnII{}$, and $\CrII{}$, which would haved allowed us to constrain
the dust content (e.g.~\citealt{DeCia:2016a}; \citet{Wendt:2021a}[\papfive{}]), are covered but not
significantly detected.

The \MgII{} absorption covers a velocity range from
$\absXmetiriXstartvelmgii{}$ to $\absXmetiriXendvelmgii{}\,\kms$
around the \nameprime{} galaxy's systemic redshift.  As can be seen
from the less saturated lines, the highest column densities appear to
be in three main kinematic components. These three components are
centered around $\absXfitXMgIXvelcompa{}$, $\absXfitXMgIXvelcompb{}$,
and $\absXfitXMgIXvelcompc{}\,\kms$ from the \nameprime{} galaxy's
systemic redshift.\footnote{Velocities are based on \MgI{}.}

We determined the \FeII{} and \MgI{} column densities using a
multicomponent fitting approach as described in \citetalias{Wendt:2021a}, which uses the evolutionary algorithm of
\citet{Quast:2005a} as used by \citet{Wendt:2012a}. We find
$\log (N_\FeII{}/\mathrm{cm}^{-2}) = 14.5$ and
$\log(N_\MgI{}/\mathrm{cm}^{-2}) = 12.6$.  While we cannot directly
constrain the \MgII{} column density, as \MgII{} is strongly saturated,
we can use two different approaches to obtain a crude estimate.  One
option is to assume a ratio between \MgII{} and \MgI{} that is typical for \MgII{}
absorbers, which was found to be around
600:1 by \citet{Lan:2017a}. This would result in
$\log(N_\MgII{}/\mathrm{cm}^{-2}) = 15.4$.  Another approach is to
assume a typical dust depletion strength and pattern, which is
necessary to estimate Mg from Fe. As found in \citetalias{Wendt:2021a}{},
sightlines along the galaxy's minor axis typically have [Zn/Fe] 
$\approx0.8$.
For this [Zn/Fe] \citet{DeCia:2016a} find a difference between the Mg and Fe
depletion strengths ($\delta_\mathrm{Fe} - \delta_\mathrm{Mg}$) of
about -0.5. Combined with the difference between the Mg and Fe (solar)
abundances of $0.1\,\dex{}$, this results in
$\log(N_\MgII{}/\mathrm{cm}^{-2}) = 15.1$, consistent with the result
starting from \MgI{}. While clearly both methods are very uncertain,
the similarity of the estimates provides some confidence.

\begin{figure*}
  \includegraphics[width=\textwidth]{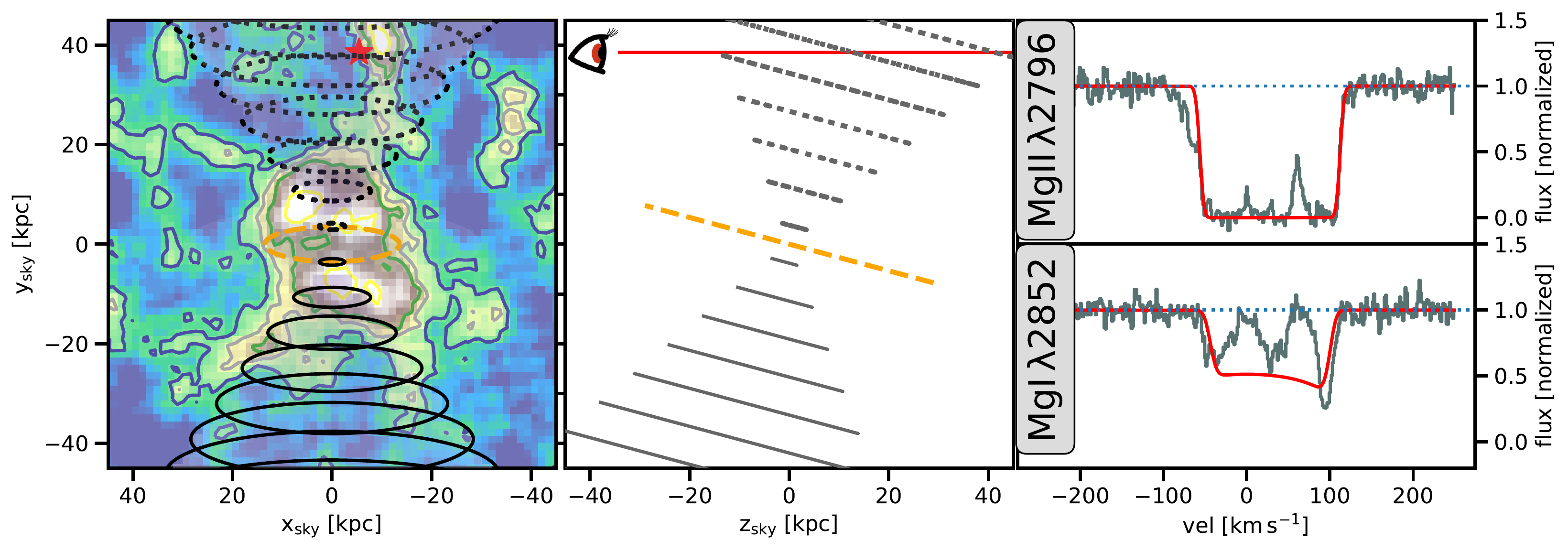}
  \caption{\label{fig:toy_model} Toy model to explain the Mg absorption kinematics seen in the quasar sightline.
  \textbf{Left:} A front view of the toy bi-conical wind model overlayed on the
  \MgII{} emission surface brightness map from \Fig{fig:main_maps} (right) with
  the \nameprime{} galaxy's major axis aligned with the plot's x-axis. The disk
  of the central galaxy is indicated by a dashed orange ellipse. The black ellipses represent the orientation and extent of the bi-conical outflow model (see \Sec{sec:toy_model}), where dotted ellipses represent the far cone, while solid ellipses show the near cone. The position of the quasar is indicated with a red star.
  \textbf{Center:} a side view of the bi-conical wind model, where the x-axis is parallel to the line-of-sight. The quasar sightline is shown in red, where the observer direction is indicated by the eye.
  \textbf{Right:} The absorption predicted by this model is overplotted in red on the Mg absorption observed in the UVES spectrum (grey) (Top: $\MgII{}\,\lambda 2796$; Bottom: $\MgI{}\,\lambda 2582$).
   \task{need to decide where exactly to put the centre of the model}}
  \end{figure*}

\subsection{Modelled absorption in quasar sightline}
\label{sec:abs_quasar_mod}

  The distribution of \MgII{} around galaxies is anisotropic, which can be explained with a two component picture consisting of an extended gas disk and a bi-conical outflow, as motivated in the introduction.
  The quasar sightline is along the \nameprime{} galaxy's minor axis ($\alpha=\alphaprimeqso{}\deg$), and as such the absorption seen in this sightline is plausibly caused by outflowing gas.
  
  Following \citet{Bouche:2012a}, we try to match the velocity profile of the
  observed Mg absorption using a conical outflow with constant outflow velocity,
  $\vout$, and (half-) opening angle $\thetaout$. In \citetalias{Schroetter:2019a}{}, we find typical values
  of $20\mbox{--}40\,\deg$ and $\approx 100 \mbox{--}300\,\kms$ for $\thetaout{}$
  and $\vout$, respectively. In order to ensure mass conservation, the volume density of the considered ion,
  $\dens$, scales with $r^{-2}$, and is normalized at $r=20\kpc$ ($\dens_{20}$). Here,
  $r$ is the distance from the galaxy center.
  The orientation of the cones w.r.t. the quasar sightline is fully constrained
  by the orientation of the galaxy on the sky, except for the sign of
  the inclination. However, as the mean absorption velocity is - at least
  sightly - redshifted from the systemic redshift of the \nameprime{} galaxy,
  it seems more plausible that the sightline crosses the far cone.

  Using our code \cgmpy{} (see \citetalias{Zabl:2020a}{} for a description), we calculated
  model absorption profiles for \MgII{} and \MgI{} and tried to find a
  parameter set ($\vout$, $\thetaout$, $\dens_{20}$) that qualitatively matches the
  observed profile. As \MgII{} is strongly saturated, we primarily compared model
  and data for the unsaturated
  \MgI{}. A model with $\theta=\windmodelXthetaout\deg$,
  $\vout=\windmodelXvout\,\kms$, $\dens_{20;\MgI{}}=2.1\times10^{-10}\mathrm{cm}^{-2}$ captures the overall velocity spread of the observed \MgI{} absorption
  (\Fig{fig:toy_model}; lower right). As a consistency check, we also evaluated
  this model for \MgII{} (\Fig{fig:toy_model}; upper right; red solid curve), where we assumed
  $\dens_{20;\MgII{}}$ to be 600 times  $\dens_{20;\MgI{}}$
  (c.f. \Sec{sec:abs_quasar_obs}; $\dens_{20;\MgII{}}=1.3\times10^{-7}\mathrm{cm}^{-3}$), and found good agreement between
  model and data.

  It is important to emphasize that we generally do not expect such
  model profiles to be a perfect match to the data, as the real
  distribution of \MgII{} is without doubt more complex than the smooth
  and coherent gas distribution and kinematics in our toy model.
  A first order correction that can often improve the similarity between
    observed and modelled outflow profiles (see \papthree{}) is to assume a hollow
    cone, an assumption that is corroborated by observations of local
    galaxies \citep[e.g.][]{McKeith:1995a}. However, the presence of absorption component B (see
    \Fig{fig:uves_lines}) suggests that the outflow cone around the \nameprime{}
    galaxy is not
    entirely hollow.

   We note
    that a different model for the same system was presented in \papthree{}.
    There, an additional outflow originating from the distant \namesecond {}
    galaxy was employed to explain the full absorption profile. However, with the
    refined galaxy redshift and inclination measurements in this present work,
    obtained with more accurate methodology possible thanks to the much deeper
    data now available (see
    \Sec{sec:object}), the outflow from the \nameprime{} galaxy alone appears sufficient to
    explain all the absorption.  
  
  Following \citet{Bouche:2012a} and \citet{Schroetter:2015a}, we can use
  the model
  to calculate a mass outflow rate $\mdotout$. Using eq.~5 from
  \citetalias{Schroetter:2019a}{} we find $\mdotout=\propXabsXmout\,\mpy$ with the parameters
  inferred above and assuming an \HI{} column density
  $\log(N_\HI / \cm^{-2})=\propXabsXnhi\pm0.3$. The latter was obtained by using the relation
  between $\rewmgii$ and $N_\mathrm{\HI}$ from \citet{Lan:2017a}, with the
    uncertainty on the $N_\mathrm{\HI}$ being due to scatter in the relation. This
  yields a mass-loading factor $\eta \equiv \mdotout / SFR = \propXeta$.
  This is a value at the lower end of the typical range at this redshift
  \citep[e.g.][\citetalias{Schroetter:2019a}{}]{Schroetter:2015a,Sugahara:2017a}.

As a consistency check, we compared the density estimate from \cgmpy{} to an analytical calculation.
	We estimated the $\MgII$ column density, $\log(N_{\MgII} / \cm^{-2})\approx 15.0-15.5$,
in \Sec{sec:abs_quasar_obs} from the measured $\MgI$ column density, $N_{\MgI}$. The $N_{\MgII}$
 can then be converted for the assumed toy model geometry into a
volume density using geometrical arguments, using eq. B5 from \citet{Bouche:2012a}:
\footnote{The equation
  assumes a background sightline crossing a conical outflow at 
  $i=90\deg$ and $\alpha=90\deg$, which is quite close to the
  configuration of the quasar sightline probing the halo of the
  \nameprime{} galaxy. }

  \begin{equation}
  \label{eq:b5_bouche2012}
  N_\MgII{} = \frac{\dens_{20} r_{20}^{2}}{b} 2 \thetaout \Rightarrow \dens_{20;\MgII{}} =
  \frac{N_\MgII}{2\thetaout}\frac{b}{r_{20}^2}
\end{equation} 
We find  $\dens_{20;\MgII{}}\approx5\times10^{-8}\,\cm^{-3}$, where we assumed $\thetaout=35\deg$.
This is as expected about a factor two smaller than than the \cgmpy{} estimate, as the \cgmpy{} model does not take account of the sub-structure in the \MgI{} profile{} (\Fig{fig:toy_model}  right).

\subsection{Observed absorption in galaxy sightlines}
\label{sec:abs_galaxy_obs}

\begin{figure*}
\includegraphics[width=\textwidth]{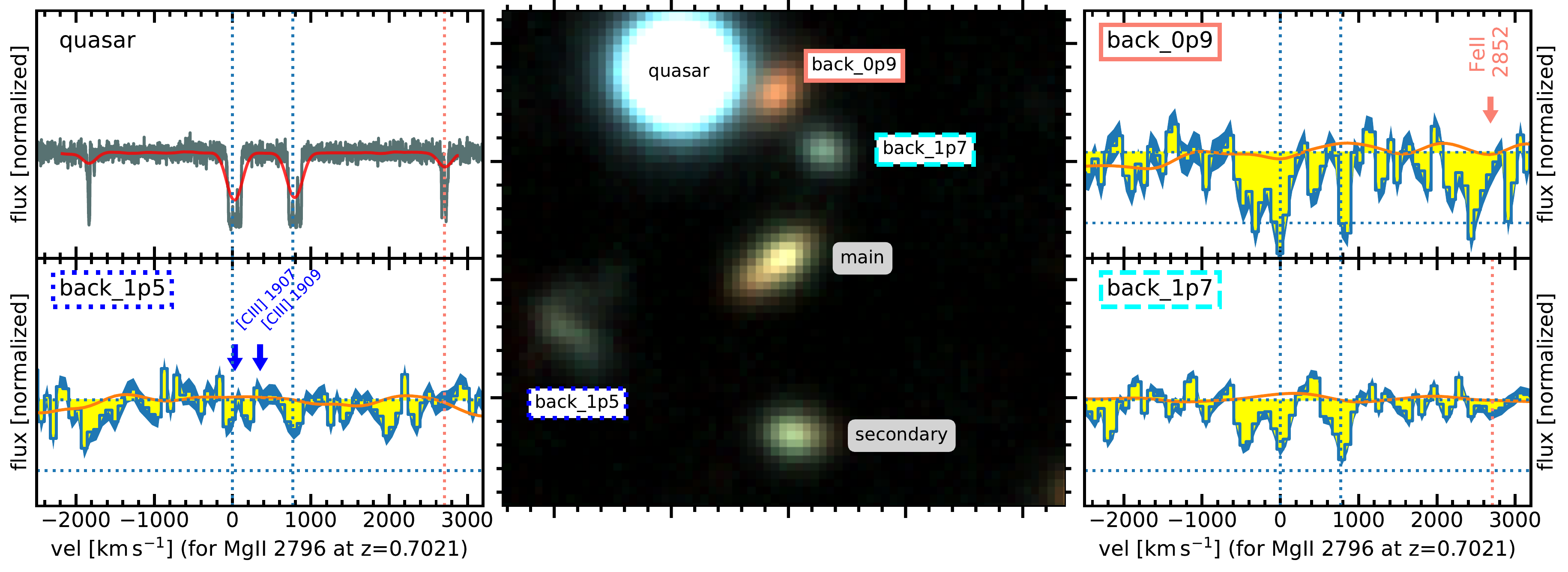}
\caption{
\label{fig:absorption_galaxy_sightlines}
\MgII{} absorption from four background sightlines that probe the halo of the \nameprime{} galaxy.
Shown is the wavelength region from $4720\,\textnormal{\AA}$ to $4810\,\textnormal{\AA}$, which corresponds to a velocity range from $-2500\,\kms$ to  $3200\,\kms$ around $\MgII{}\,\lambda2796$ at $z=0.7021$. \textbf{Center:} Zoom-in of \Fig{fig:main_maps} (left), which shows the position of the four background sources w.r.t. the \nameprime{} galaxy.
\textbf{Upper Left:} UVES spectrum of the quasar sightline (grey). The red line also shows the UVES spectrum, but downgraded to the resolution of MUSE.
\textbf{Lower Left, Lower \& Upper Right:} The three background galaxy sightlines, \namebonepfive{}, \namebzeropnine{}, \namebonepseven{} as seen by MUSE. The orange line is a model obtained from a \ppxf{} fit to the full spectral range covered by MUSE using the UV-extended E-Miles SSPs \citep{Vazdekis:2016a}. The spectral resolution of these models is by factors of 1.8, 2.5, and 2.7 lower than the data for \namebzeropnine{}, \namebonepfive{}, and \namebonepseven{}, respectively.
The blue dotted vertical lines in each of the four spectral panels indicate the positions of $\MgII{}\,\lambda 2796$ and $\MgII{} \lambda 2803$ at the redshift of the \nameprime{} galaxy. The \namebonepfive{} galaxy spectrum has its \ion{C}{III}] emission coincident with the wavelength of $\MgII{}\,\lambda2796$ in the \nameprime{} foreground galaxy, potentially masking the absorption. Furthermore, galaxy \namebzeropnine{} has at the edge of the shown wavelength range its $\FeII{}\,\lambda2852$ absorption. As galaxy \namebzeropnine{} is at a lower redshift than the three others sightlines, CGM $\FeII{}\,\lambda2852$ absorption at the redshift of $\namebzeropnine{}$ is possible in the other sightlines. The corresponding wavelength is indicated as the orange dotted-line.
 }
\end{figure*}

The three background galaxies within $50\,\kpc$ of the \nameprime{}
galaxy (\namebzeropnine{}, \namebonepfive{}, \namebonepseven{}) have
at $4760\,\textnormal{\AA}$, which is the wavelength of \MgII{} at the
redshift of the \nameprime{} galaxy, observed magnitudes of
$\bzeropnineXpropXmabatmgiimain$, $\bonepfiveXpropXmabatmgiimain$, and
$\bonepsevenXpropXmabatmgiimain$, respectively. The coordinates and
alignment of these galaxies with respect to the \nameprime{} galaxy
are listed in \Tab{tab:backsightlineinfo}.

The spectra of all three background sources are shown in
\Fig{fig:absorption_galaxy_sightlines}, in addition to the quasar
spectrum.
Potential complications with using background galaxy spectra are
photospheric and ISM/CGM absorption features intrinsic to the
background galaxy, which by accident might coincide in the observed-frame
wavelength with the absorption of interest in the foreground
CGM. To test the presence of photospheric absorption features, we
performed a full spectral fit with \ppxf{} \citep{Cappellari:2017a} to
each of the three background galaxies using the UV-extended E-Miles
SSP \citep{Vazdekis:2016a}. The best-fit model, after normalising in
the same way as the actual data, is shown as orange lines in
\Fig{fig:absorption_galaxy_sightlines}. It is clear that the
background galaxies' photospheric features are  not
important for the noise level of our data. ISM features, by contrast, are more problematic. In the
case of \namebonepseven{}, there is $\CIII ] \,\lambda\lambda 1907,1909$
exactly at the observed wavelength of $\MgII\,\lambda 2796$, making it
impossible to fit this component of the \MgII{} doublet.  For
\namebzeropnine{} there is ISM/CGM absorption due to $\FeII\,\lambda 2852$ at
its redshift, but it is far enough in velocity
from $\MgII{}$ at the redshift of the \nameprime{} galaxy to not be
problematic. In addition, both in \namebzeropnine{} and \namebonepseven{} there is a component bluewards of $\MgII\,\lambda2796$. However, this features seems absent in $\MgII{}\,\lambda 2803$. Therefore, we could not convincingly identify this blueshifted absorption to be due to \MgII{}.

Due to the low S/N and the complications mentioned above, we refrained
from quantitatively measuring the $\rewmgii$ and velocity
centroid. Qualitatively, the data in all three sightlines, both the
two more aligned with the major axis ($\alpha \lesssim 30\deg$) and
the sightline more aligned with minor axis (\namebzeropnine{};
$\alpha=\bzeropnineXpropXalphamain\,\deg$), appear to have \MgII{}
absorption consistent with being at least as strong as in the
quasar sightline. This is especially the case for the closest of the
three sightlines (\namebonepseven{}), which seems to have the strongest
$\MgII\,\lambda 2803$ absorption among the three sightlines. Given the
known anti-correlation between $\rewmgii$ and $b$, this is not unexpected.

Kinematically, the $\MgII\,\lambda 2803$ absorption seems to be
relatively close to zero velocity. Depending on whether
the blueshifted feature seen for $\MgII{}\,\lambda 2796$ in the
$\namebonepseven$ and $\namebzeropnine$ sightlines is real, there
might also be an additional blueshifted component. Interestingly, blueshifted absorption
  in sightlines $\namebonepseven$ and $\namebzeropnine$ would be consistent with an
  extrapolation of the \nameprime{} galaxy's rotation field, and hence would provide
  further support for an extended rotating gas disk on CGM scales. However, given the low S/N, even
deeper data will be necessary to confirm the blueshifted component and accurately measure the kinematics in these
  sightlines.

\section{Discussion of \MgII{} emission morpho-kinematics}
\label{sec:toy_model}

In this section, we discuss potential origins of the extended \MgII{} emission. One main focus of our discussion is to explore to what extent the generic toy model of a bi-conical outflow, which we used to interpret the absorption in \Sec{sec:abs_quasar_mod}, can also explain the observed emission morphology and kinematics. After an initial comparison of observed and predicted emission kinematics (\Sec{sec:dis_em_kinematics}), we discuss the feasibility of resonantly scattered continuum photons as a source of the \MgII{} emission (\Sec{sec:em_cont_scattering}). For this assessment, we compare densities inferred from emission and absorption (\Sec{sec:density_sobolev}) and test to what extent models with scattered continuum photons can reproduce the observed morphology (\Sec{sec:em_toy_model_kinmorph}) and luminosity (\Sec{sec:mgii-luminosity}). Finally, we discuss alternative mechanisms to produce the extended \MgII{} (and \OII{}) emission (\Sec{sec:em_alt_mechanisms}).

\subsection{\MgII{} emission kinematics}
\label{sec:dis_em_kinematics}

\begin{figure*}
  \begin{tabular}{cc}
    \includegraphics[width=0.7\textwidth]{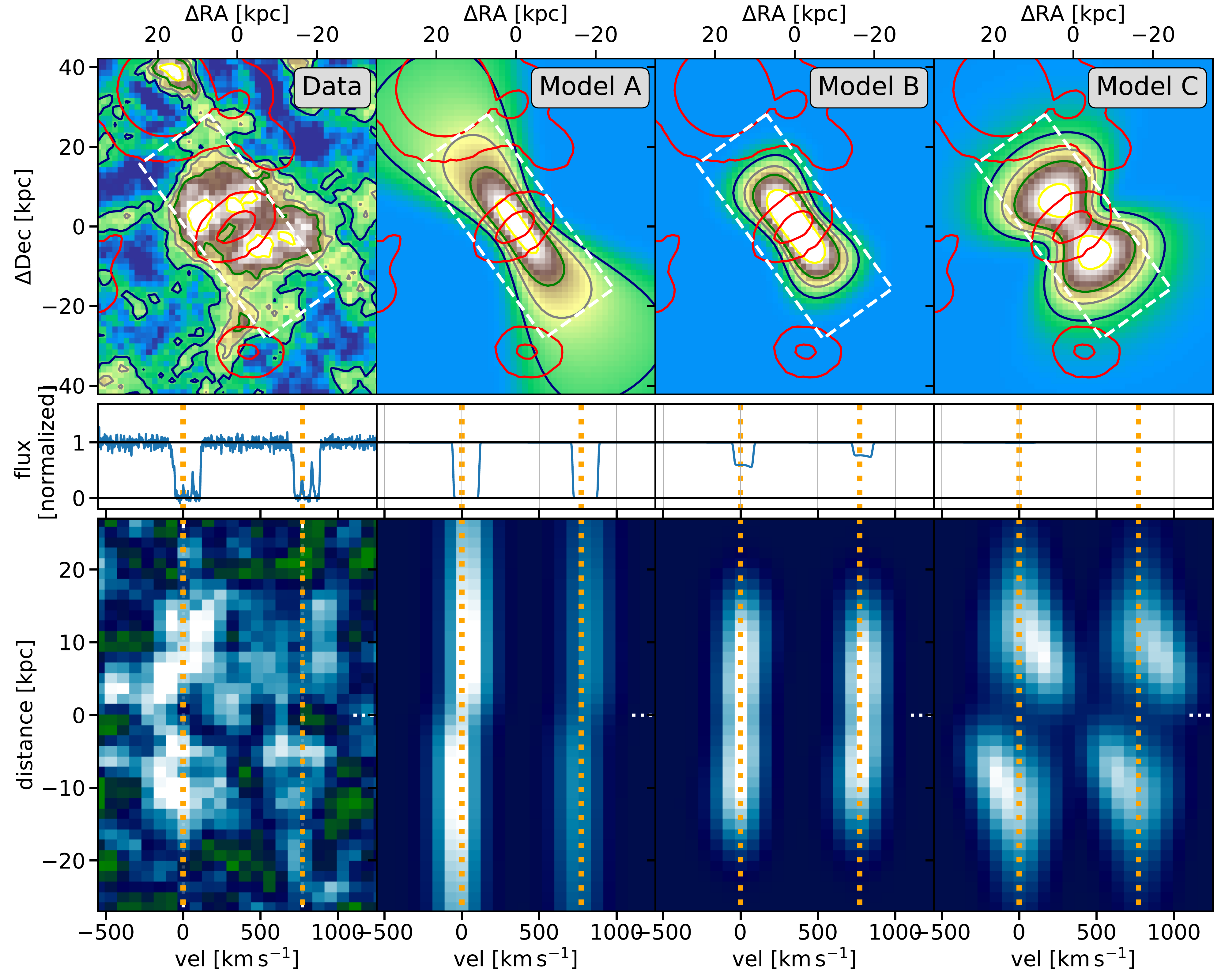} &
    \includegraphics[width=0.29\textwidth]{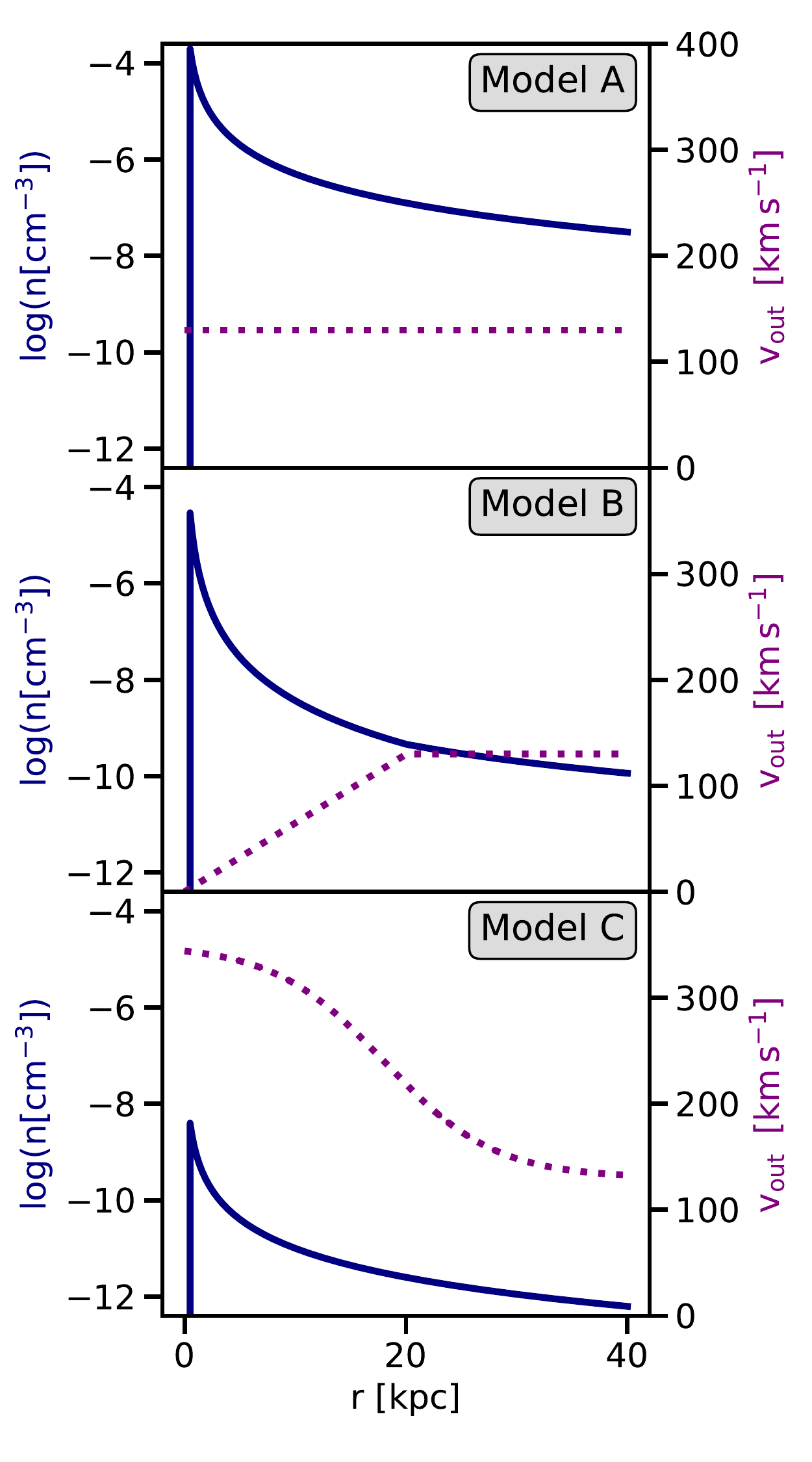}
  \end{tabular}

  \caption{\label{fig:kinematics_slit2d}
Kinematics of the \MgII{} emission along the
\nameprime{} galaxy's minor axis.  
{\bf Left:} The top panels show the spatial SB distribution of the observed
(`Data') and modelled (`Model A-C') \MgII{} emission. The bottom panels show
position-velocity diagrams (PVD) along the
\nameprime{} galaxy's minor axis. These 2D spectra were extracted from the
pseudo-slit shown with dashed-lines in the SB maps and have been smoothed with
a tophat filter (r=1 pixel). Positive distances are towards the quasar.
The zero point of the velocity scale in the plots is set to the wavelength of
$\MgII{}\,2796$ at the systemic redshift of the \nameprime{} galaxy.
The two vertical orange dotted lines indicate the wavelengths expected for
$\MgII{}\,2796$ and $\MgII{}\,\lambda 2803$, respectively. Both in the data and the models the \MgII{}
emission appears blueshifted in the direction
away from the quasar ($distance < 0\,\kpc$), while it is redshifted towards
the quasar, consistent with the redshift seen in the quasar absorption. This
absorption in the quasar sightline is shown in the {middle} panels on the same
velocity scale as for the PVDs below. \textbf{Right:} The velocity and density
profiles assumed for the three toy models shown in the left. `Model A'
(\Sec{sec:dis_em_kinematics}) assumes emissivity proportional to
\MgII{} volume density and its parameters have been tuned to explain the
absorption in the quasar sightline.  `Model B' (accelerating) and `Model C' (decelerating)
(\Sec{sec:em_toy_model_kinmorph}), which assume physically motivated
emissivity from continuum scattering, better reproduce the surface brightness
distribution than `Model A', but fail to reproduce the quasar absorption. The normalization
of the maps has been been arbitrarily chosen (see \Sec{sec:mgii-luminosity}).
}
\end{figure*}

As discussed in \Sec{section:halo:morpho}, the morphology of the CGM \MgII{}
emission is dominated by two regions of strong emission along the galaxy's
minor axis. Such a morphology is suggestive of a bi-conical outflow. To test the
viability of this interpretation, some insight can
be gained by comparing the observed emission kinematics
(see~\Sec{section:halo:kinematics}) to that predicted from an idealized toy
model.

In order to get an impression of the \MgII{} emission kinematics expected from our toy model, we used an extremely simple model: we assumed that each spatial point ``emits'' isotropically $\MgII{}$ photons with an emissivity proportional to the $\MgII$ density, $\dens_{\MgII}$.
Further, the photons escape from that point freely with the 2796/2803 ratio of 2:1, and are assumed to be emitted in the rest-frame of the outflowing $\MgII$ ions. For a  first model (Model A) we assumed a biconical outflow with the geometry, density, and velocity parameters set to those inferred from the \MgII{} absorption seen in the quasar sightline (see \Sec{sec:abs_quasar_mod}).

With these assumptions we created 	a simulated \MgII{} emission data cube, where we accounted for the appropriate MUSE PSF and line spread function, LSF, at the observed wavelength of \MgII{}.
Subsequently, we created from this model cube a surface-brightness map in an identical way as described in \Sec{section:halo:morpho} for the data.
SB maps for both the data and this "Model A" are shown in the respective columns of Fig.~\ref{fig:kinematics_slit2d}'s upper row.
The overall flux normalization in the model was chosen arbitrarily so that the modelled emission approximately matches the observed SB.
While the model SB map cannot reproduce the observations in detail, as expected from the extreme simplifications of the model, there are nevertheless similarities with the observations. E.g., the data SB map shows clear indications of a bi-conical structure as predicted by the model.

In the lower row of Fig.~\ref{fig:kinematics_slit2d} ``Data'' (``Model A'') we
show a pseudo-longslit spectrum of the observed (modelled) \MgII{} emission for
a slit aligned with the bi-conical structure. This slit is overplotted as a
white dashed line on the observed (modeled) SB maps. The pseudo-longslit
  spectrum was created from a cube with the continuum of the galaxy and the
  quasar subtracted, as described in \ref{sec:mgii_map}. The emission
kinematics in the position-velocity-diagram, PVD, extracted from this
pseudo-slit appear in approximate agreement between data and model. In
particular, both data and model emission show a similar amount of red- and
blueshift towards and away from the quasar, respectively. Further, the amount of
the observed (and modelled) redshift of the \MgII{} emission in the direction of
the quasar is consistent with the velocity of the \MgII{} absorption seen in the
quasar (middle panel). 
Importantly, this means that the geometrical model can explain simultaneously the absorption kinematics in the quasar sightline and the emission kinematics.

\subsection{Continuum scattering}
\label{sec:em_cont_scattering}

\subsubsection{Density estimate}
\label{sec:density_sobolev}
We now turn to estimating the density of $\MgII$ in the outflow from the extent of the emission.
This requires a realistic treatment of the scattering optical depth as a function of the radius.
Such an estimate can be obtained using the Sobolev approximation (\citealt{Sobolev:1960b}; for applications in the context of CGM winds see e.g.~\citealt{Rubin:2011a,Martin:2013a,Prochaska:2011a,Scarlata:2015a,Carr:2018a}).

The Sobolev approximation requires the presence of a velocity gradient, $\vert dv/dr \vert$, in the outflowing gas. This is important, as having a velocity gradient ensures
that photons are only resonant within a small radial range.  The
characteristic thickness of this interaction region, the ``Sobolev
length'', is then simply proportional to
$s_0 = b_\mathrm{th} / {\vert dv_\mathrm{out}/dr \vert_{r}}$, where $b_\mathrm{th}$ is the thermal broadening of the line.  If the ``Sobolev'' length is small compared to the extent of the wind, the ``Sobolev'' approximation holds. If
the wind is furthermore only moving radially, as assumed here, and if the
source at the centre can be approximated to be an isotropically
emitting point source, the optical depth at  a given radius can be
approximated as:

\begin{equation}
  \label{eq:tau_sobolev}
	\tau^{s} = 
	\frac{\pi e^2}{m_{e}c}f_\mathrm{osc} \lambda_0 \dens_{\MgII} \left \vert \frac{dv}{dr} \right \vert ^{-1} _{r} =
        4.54 \times 10^{-7}\,\mathrm{cm}^{3}\,\mathrm{s^{-1}} \dens_{\MgII} \left \vert \frac{dv}{dr} \right \vert^{-1} _{r}
\end{equation}

Consequently, it is possible to estimate $\dens_\MgII{}$ at a certain radius by inverting eq.~\eqref{eq:tau_sobolev},  if $\tau^s$ and $\vert dv/dr \vert$ are known at this radius. \citet{Martin:2013a} argued that $\tau^s$ drops below one at the maximum extent of the observed emission. This assumes that the extent - $\approx 20\,\kpc$ in our case - is limited by the probability for scattering ($\propto \mathrm{e}^{-\tau_s}$) and not by the depth of the data. The assumption of $\tau^s=1$ results in:

\begin{equation}
  \label{eq:rho_sobolev}
	\dens_{\MgII} = 2.2\times10^{6}\,\cm^{-3} \,\mathrm{s}   \left \vert \frac{dv}{dr} \right \vert  _{r} = 7.1\times10^{-11} \,\cm^{-3} 
			\left \vert 
				\frac{dv}{dr}\frac{\mathrm{kpc}}{\mathrm{km}\,\mathrm{s}^{-1}}
			\right \vert  _{r}
\end{equation}

\noindent
If we assume, as an order-of-magnitude estimate, a velocity gradient of 
$\vert dv/dr \vert = 130\,\kms / 20\,\kpc$, as motivated by the $\vout$ inferred from the absorption (\Sec{sec:abs_quasar_mod}) and the observed extent of the emission. Then, eq.~\eqref{eq:rho_sobolev} gives a $\MgII$ density at $20\,\kpc$, $\dens_{20;\MgII{}}$, of $5\times10^{-10}\,\cm^{-3}$.\footnote{Assuming solar metallicity and neglecting depletion, this would correspond to $\dens_{\textsc{\rm H}} = \dens_{\MgII{}} / f_\MgII{}\,10^{4.4} = 1.3\times10^{-5}\,cm^{-2}/f_\MgII{}$, with $f_\MgII{}$ being the $\MgII$ ionization fraction.}

Thanks to the background quasar, we can alternatively constrain the
gas density more directly from the absorption measured in the sightline. 
As described in \Sec{sec:abs_quasar_mod}, from the absorption we inferred $\dens_{20;\MgII{}}\approx5\times10^{-8}\,\cm^{-3}$.\footnote{In the absorption modelling of \Sec{sec:abs_quasar_mod} a scaling of $\dens \propto r^{-2}$ is assumed, which would be mass conserving only for an outflow with $\vout=const$. By contrast, we assumed for the purpose of the Sobolev approximation a velocity gradient $\vout\propto r$. This would require, if the mass in the cool phase is indeed conserved, $\dens \propto r^{-3}$. To reconcile the two assumptions, we could assume that $\vout=const$ is valid beyond $r=20\,\kpc$ only.}
This density is a factor $100$ larger than the value estimated here from the emission.
Despite the relatively large uncertainties in both estimates, it might be difficult to explain such a large discrepancy.

Assuming that the emission-based estimate is wrong, one potential reason is the assumption of $\tau^s=1$.
However, a very high $\tau^s\approx100$ would be required to consolidate
absorption and emission. This seems
inconsistent with the 2803/2796 flux ratio of
$1.9\pm0.3$ of the observed emission, as a line ratio closer to
1:1 should be expected for optically thick scattering.
Another option would be an underestimated velocity gradient. While a velocity gradient a few times larger than the one assumed here is not ruled out by the emission data (see also \ref{sec:em_toy_model_kinmorph}), the difference can certainly not explain a substantial fraction of the discrepancy.

Assuming that the absorption-based estimate is wrong,
a possible reason
might be that the `specific' sightline encounters more $\MgII$ gas
than a `typical' sightline would. To asses this possibility, we
compared the $\rewmgii$ of this sightline to those typical for
sightlines around star-forming galaxies at $b\approx40\,\kpc$. Due to
the anisotropic distribution of \MgII{} (c.f.~\Sec{sec:intro}), it is
important to restrict `typical` here to sightlines along the minor
axis. E.g., \citet{Lan:2018a} find for minor axis sightlines at this
impact parameter statistical values of about
$\rewmgii\approx1\mbox{--}2\,\textnormal{\AA}$,\footnote{\citet{Lan:2018a}
  state the sum of the 2796 and 2803 lines. Therefore, we divided
  their $EW_0$ values by 2, as for such high $\rewmgii$ both lines are
  typically saturated and the $EW$ ratio is likely 1:1.} and we conclude that the $\rewmgii$ of our sightline ($\rewmgiiqso \,\textnormal{\AA}$) is not exceptional.

\subsubsection{Emission kinematics and morphology from toy model}
\label{sec:em_toy_model_kinmorph}

In \Sec{sec:dis_em_kinematics} we explored whether the toy model
which we used to fit the absorption kinematics in the quasar
sightline (see \Sec{sec:abs_quasar_mod}) can also explain the
observed emission halo. There, we assumed that the emissivity is
proportional to the density, an assumption which resulted in an
unrealistic surface brightness profile, not surprisingly. Here, we
explore models with scattering of the central galaxy's continuum
photons by \MgII{} in the extended halo.

For the modelling of the continuum scattering we use the Sobolev
optical depth (\Eq{eq:tau_sobolev}) to obtain  an estimate of the
emissivity: a continuum photon in resonance with the \MgII{} transition
will get scattered with a probability $1-e^{-\tau^s}$. Assuming that
the photon can escape after this single scattering, the local
isotropic ``emissivity'', $\epsilon_{\text cs}$, from a small volume
element $dV=dA dr$ at radius r will be:

\begin{equation}
  \label{eq:scatter_em}
  \epsilon_{\text cs} = \frac{L_\lambda}{4 \pi r^2} \frac{\lambda_{0;r}}{c} (1 - e^{-\tau^{s}}) \left \vert \frac{dv}{dr} \right \vert _{r} \,dr\,dA
\end{equation}
where $L_\lambda$ is the luminosity density of the central point
source at the wavelength $\lambda_{0;r}$, with
$\lambda_{0;r}=\lambda_0 \left(1 + \frac{v(r)}{c} \right)$.
$\lambda_0$ is the resonant wavelength of the considered transition
in the rest-frame of the central galaxy. Under this assumption of
"emissivity" we created simulated cubes, SB maps and PVDs as
described in \Sec{sec:dis_em_kinematics}.

\begin{table}
	\begin{tabular}{rccc}
	& Model A & Model B & Model C \\
\hline
$\rho(r)$ & $c_{\text a} \left(\frac{r}{r_{0}} \right )^{-2}$  & 
 $c_{\text b}\left(\frac{r}{r_{0}} \right )^{-3}$  for $r<r_0$  &
  $c_{\text c}\left(\frac{r}{r_{0}} \right )^{-2}$ \\
 &  &  $c_{\text b}\left(\frac{r}{r_{0}} \right )^{-2}$  for $r\geq r_0$ & \\
$\vout(r)$ &  $v_{\text const}$  
		   &  $\frac{\text{d}v}{\text{d}{r}}\,r$ for $r<r_0$
		   &  $v_{1} - \frac{v_2}{1+e^{-k(r-r_t)}}$ 
		   \\
		   & 
		   & $v_{\text const}$ for $r\geq_0$
		   & \\
	\end{tabular}
	\caption{\label{tab:model_paras} Parametrizations of the density and velocity profiles shown in \Fig{fig:kinematics_slit2d} (right) and used for the models shown in  \Fig{fig:kinematics_slit2d} (left). $c_{\text a\mbox{--}c}$ are the density normalizations at $r_0$ of the respective models. The constant velocity (part) of model A (B) is $v_{const}=130\kms$. The velocity gradient in the inner part ($r\leq r_0<20\kpc$) of model B is ${\text{d}v}/{\text{d}{r}}=6.5\,\kms/\kpc$. For Model C, $v_1=350\,\kms$, $v_2=220\,\kms$, $k=0.2\,\kpc^{-1}$, and $r_t=18\,\kpc$. To avoid numerical problems, we further set in all models the density for $r<0.5\,\kpc$ to zero.}
\end{table}

As continuum scattering will only provide a viable source of \MgII{} emission
if there is a velocity gradient, we can here not use the constant velocity
profile assumed in "Model A" discussed in \Sec{sec:dis_em_kinematics}. Instead, we explore  both accelerating and decelerating
winds\footnote{Accelerating and decelerating does not necessarily mean that
individual \MgII{} clouds change their velocity. The velocity profile is here
simply the radial profile of the ensemble of clouds.}, as there is no
consensus about the velocity profiles in cool outflows. Some observations prefer radially
increasing velocities \citep[e.g.][]{Martin:2009a}, while others support
radially decreasing velocities \citep[e.g.][]{Martini:2018a,Burchett:2021a}.

Given the relatively low S/N of the emission data and the simplifications of the
model, we refrain from a futile attempt to formally constrain the geometry, density and
velocity from the data. Instead, we simply show two feasible example models, one
with increasing (`Model B') and one with decreasing velocity (`Model C')
profile. Their velocity and density profiles are listed in \Tab{tab:model_paras}
and are shown in the right panel of \Fig{fig:kinematics_slit2d}. We fixed the
geometry to a biconocial outflow with $\thetaout=35\deg$ for ('Model B'),
as estimated from the quasar absorption in \Sec{sec:abs_quasar_mod}. By
  contrast, we used for 'Model C' a wide opening angle of $\thetaout=60\deg$,
  which is motivated by the extent of the observed emission.
  Further, we chose in these models the central
  luminosity, $L_\lambda$, so that resulting simulated halo surface-brightness
  is similar to the observed one. This choice is useful for comparing kinematics
  and morphology between data and model. We critically compare these required
  $L_\lambda$ to the available $L_\lambda$
  in \Sec{sec:mgii-luminosity}.

Model B has a constant velocity gradient out to $20\,\kpc$, beyond which it has
a constant velocity. The resulting SB map and velocity profile (see
\Fig{fig:kinematics_slit2d}) are similar to the data. While the modelled SB map
does not show the central suppression as in the data, the suppression in the
data might be the consequence of \MgII{} absorption in the continuum of the
central galaxy (see \Sec{section:halo:morpho}). We note that a decelerating wind
with the same absolute velocity gradient, the same density profile, and the
  same $\thetaout{}$ as `Model B' would also have the same surface brightness
profile as `Model B`. Therefore, we used for the sake of variation in `Model C',
in addition to a large $\thetaout$(=$60\deg$), a radially changing velocity
gradient that is steepest off-centre from the galaxy. Having the steepest
gradient off-centre results even for the chosen monotonically declining density
profile in the surface brightness peaking away from the centre. This two-lobe
geometry is very reminiscent of the observed SB distribution without the
absorption correction. Further, `Model C' has a relatively high velocity of
$350\,\kms$ in the centre. Such a relatively high velocity in the inner part
seems still consistent with the data (see \Fig{fig:kinematics_slit2d} left
lower), and data and model seem to agree remarkably well, especially in the
direction away from the quasar.

In summary, the simple continuum scattering models with a biconical outflow geometry seem to be able to reproduce kinematics and, at least approximately, the SB distribution of the observations. Both models with increasing and decreasing velocity gradients seem consistent with the data.
As expected from the discussion in \Sec{sec:em_toy_model_kinmorph}, densities which allow for low enough scattering optical depths to support the observed 2:1 \MgII{} doublet ratio result for the assumed monotonously declining density profiles in too low densities at the position of the quasar sightline to explain the observed absorption (see middle row of \Fig{fig:kinematics_slit2d}).

\subsubsection{\MgII{} luminosity}
\label{sec:mgii-luminosity}

For the continuum scattering models discussed in \Sec{sec:em_toy_model_kinmorph}
and shown in \Fig{fig:kinematics_slit2d}, we arbitrarily chose the strength of
the central continuum source so that the models matched the observed MgII SB
brightness in the halo. This required continuum fluxes, $f_\lambda$, which
exceeded the observed ones by factors of $\approx20$ and $\approx10$ for models
B and C, respectively. This discrepancy between model and observations 
can be understood: The maximum flux from continuum
scattering is naturally limited by the number of continuum photons available for
scattering. This budget is determined both by the strength of the continuum,
$f_\lambda$, and the
range of radial velocities of the ouflowing \MgII{} ions ($\Delta v = 
\text{max}(\vout(r)) -  \text{min}(\vout(r))$), as scattering happens in the rest-frame of the ions. For a non-isotropic biconical outflow with (half-)opening angle $\thetaout{}$, the maximum \MgII{} flux is further reduced by a factor $1-\cos\thetaout{}$. This results in: 

\begin{align}
	 f_\mathrm{MgII} = &  2 (1 - \cos\theta_\mathrm{out}) \Delta v \frac{\lambda_{obs}}{c} f_\lambda  \label{eq:flux_cont_scattering}
\end{align}
where the factor 2 accounts for \MgII{} being a doublet\footnote{The factor 2 will get modified if $\Delta v$ is larger than the separation of the doublet ($=770\,\kms$).}. 
For our fiducial $\thetaout=35\deg$ and  $f_\lambda =2\times 10^{-18}\uergcont$ at the wavelength of \MgII{}, this results in $f_\mathrm{MgII}=1.15\times10^{-18} \uerglf{}$ for an outflow with $\Delta v = 100\,\kms$. 
For comparison, the total measured flux in the \MgII{} halo is $41\times10^{-18}\uerglf{}$, which is a factor $30\mbox{--}40$ larger than the expected one.

We briefly consider three possibilities that could reduce the tension between required and available continuum photons.
First, a $\Delta v$ of up to $\approx 500\kms$, might be consistent with the data, if the high velocity gas is only close to the galaxy (c.f. 'Model C`). However, even with $\Delta v = 500\kms$ the discrepancy would still be a factor $\approx7$.
Second, the effective cone opening angle might be larger than the assumed $\thetaout=35\deg$, especially in the inner part. In the extreme case of an isotropic wind ($\thetaout=90\deg$) the flux would be a factor 5 larger than for the assumed $\thetaout=35\deg$. Third, the $f_\lambda$ escaping the ISM in direction of the cone might be larger than the $f_\lambda$ measured from the edge on view of the galaxy, due to less dust extinction in direction perpendicular to the disk. However, it is unlikely that the difference is much more than a magnitude (factor 2.5) \citep[e.g.][]{Yip:2010a, Chevallard:2013a}. Finally, more realistic radiative transfer accounting for multiple scattering could result in the preferential escape of scattered photons towards the observer (perpendicular to the cone). An investigation of this possibility is beyond the scope of this paper.

\subsection{Alternative origins of the extended emission} 
\label{sec:em_alt_mechanisms}

As discussed in \Sec{sec:toy_model}, a biconical outflow model where \MgII{}
ions scatter continuum photons emitted from the central galaxy can, at least
approximately, reproduce the morphology of the \MgII{} nebula and
simultaneously explain the kinematics of halo gas seen in emission and
absorption. However, it results in inconsistent density estimates between
emission and absorption \Sec{sec:density_sobolev} and struggles to reproduce
the \MgII{} luminosity of the nebula \Sec{sec:mgii-luminosity}.

One way to boost the number of photons available for scattering on the
$\MgII{}$ ions in the CGM would be if the galaxy produces \MgII{} line
emission in its \HII{} regions. While \MgII{} emission from \HII{} regions is
certainly feasible \citep[e.g.][]{Erb:2012a,Feltre:2018a}, the amount of
$\MgII{}$ emission originating from the ISM of the \nameprime{} galaxy appears low, even
when correcting for potential down-the-barrel absorption (see
\Fig{fig:infillcorrfit}). Given that the statistical presence or absence of
\MgII{} emission does not depend on the galaxy's inclination
\citep{Finley:2017b}, it appears unlikely that the \MgII{} emission flux
escaping the \nameprime{} galaxy's ISM perpendicular to the disk is much
larger than suggested by our edge-on observations.

Therefore, it is unavoidable to consider, in addition to scattering, also
in-place production of \MgII{} photons in the halo. Further support for at
least a partial non-scattering origin of extended \MgII{} emission comes from
the presence of clear extended emission also in the non-resonant oxygen lines
(see \Sec{sec:em-from-other-lines}). Extended line emission from optical non-resonant lines produced
in the cool-warm ionized phase of galactic-scale winds seems indeed common, as seen around
local starbursts, such as M82 \citep[]{Lynds:1963a, Bland:1988a,
Heckman:1990a, Sharp:2010a}.

In
the outflows around local starbursts the line ratios of the various
observed lines are more consistent with shocks than with \HII{} regions photo-ionized by
stars \citep[e.g.][]{Heckman:1990a}.
The lack of access to important diagnostic lines (e.g. [\OI{}], \NII{}) does not
allow us to conclusively decide whether shocks are responsible for the extended
emission around the \nameprime{} galaxy. Nevertheless, we can still test for
feasibility by comparing the strength of the observed lines to shock model
predictions (including precursor). We used the \citet{Allen:2008a} model grid in
the updated version by \citet{Alarie:2019a} using \mappingsv{}. For a preshock
density of $n = 1\cm^{-3}$, the grid includes models with shock velocties
between 100 and $1000\,\kms$, magnetic fields between $10^{-4}$and
$10\,\mathrm{\mu G}$, and five different abundance sets.
We find that only models (shock + precursor) with low shock velocities ($\approx 100-200\kms$) can
explain the observed low \OIII{}/\OII{} ratio of $\approx 1/3$. The velocities
seem plausible as they are similar to the dispersion of the \MgII{} emission. 
Models with these velocities predict, independently of magnetic field strength, a
\MgII{}/\OII{} ratio of $\approx1/2$ (assuming LMC abundances). While the
observed ratio of $\approx1$ is somewhat higher, the predicted ratio is close
enough when considering that \MgII{} might also have a contribution from
scattering.
Further, we can compare the surface brightness predicted by such shocks to the one observed for the \MgII{} nebula. Those models with acceptable
\OIII{}/\OII{} emit \MgII{} from their shock surface with $\approx10^{-5}\mbox{--}
10^{-4}\,\mathrm{erg}\,\mathrm{s}^{-1}\,\mathrm{cm}^{-2}$. 
Assuming that shock surfaces cover the full area of the nebula, this would correspond for the $z=0.7$ halo to an observed \MgII{} SB of $\approx 10^{-18}\mbox{--}10^{-17}\uergsb{}$. Interestingly, this is of the same order as the observed SB (see \Fig{fig:sb_profiles_pa}). 
This means that emission by shocks could be a viable source of the extended \OII{} and \MgII{} emission.

Motivated by the biconical geometry of the \MgII{} halo
(\Sec{section:halo:morpho}) and the measured kinematics
(\Sec{section:halo:kinematics}),
the observations appear very consistent with an outflow.
However, the extended metal-enriched gas
could also (partially) originate from tidal stripping or from intergalactic 
  transfer via outflows from a neighboring galaxy \citep[e.g.][]{AnglesAlcazar:2017a,Mitchell:2020a}.
While the \nameprime{} galaxy does not appear to be in a dense group, in
which extended emission from non-resonant lines has been observed also
outside the local Universe \citep[][]{Epinat:2018a, Johnson:2018a, Chen:2019a}, the presence
of the \namesecond{} galaxy in the proximity of
\nameprime{} and its off-centre star-forming clump (see Appendix
\ref{appendix:morpho}) make intergalactic transfer and tidal stripping
plausible alternative 
contributions to the extended cool gas.\footnote{We note that the \MgII{} SB map (\Fig{fig:main_maps}
  right) shows potentially a weak "bridge" between the \nameprime{} and
  \namesecond{} galaxy. However, we consider it more likely to be residuals from
  the subtraction of the \namesecond{} galaxy than a real tidal feature.}
Certainly, starburst-driven outflows and tidally stripped gas will often
co-exist. As the signatures of outflows and stripped gas could be similar at the
level of our $z=0.7$ observations,  it does not yet seem
feasible to conclusively rule out that there are other contributions to the
extended emission than a pure outflow from the \nameprime{} galaxy.

\section{Conclusion \& Discussion}	
\label{sec:discussion}

The main focus of this paper was to report the first discovery of a 
\MgII{} emission halo around a galaxy near a quasar sightline, using deep ($\texptot\,\mathrm{hr}$) VLT/MUSE data from the MEGAFLOW
survey. The main findings are:

\begin{itemize}
    \item The $z=0.702$ \nameprime{} galaxy is a typical main-sequence galaxy with
      $\log(M_*/\mathrm{M_\odot})=\mainXpropXsedXmass{}$ and
      SFR=$\mainXpropXsedXcurrsfr\,\mpy$  ($\delta(MS)=
      \mainXpropXderivedXdeltams$), but has a relatively large
      inclination with $i\simeq\mainXincl$deg. The galaxy has a minor 
      companion at $\secondXpropXbmainkpc{}\,\kpc$ (less than 1/5 of $M_*$ of
      the \nameprime{} galaxy);
    \item We detect  $\MgII\,\lambda\lambda\,2796, 2803$ emission around this
      galaxy that extends to a projected radius of $25\,\kpc$ and covers $1.0\times10^3\,\kpc^2$  above  a surface brightness of $14\times10^{-19}\,\uergsb{}$ (2$\sigma$) (Fig.\ref{fig:main_maps} right);
    \item The \MgII{} emission is not isotropic, but is strongest along the galaxy's projected minor axis (\Fig{fig:sb_profiles_pa}). Similarly, we detect extended \OII{} emission along the minor axis (Figs.~\ref{fig:sb_profiles_pa} \& \ref{fig:quadrant_spect_oii});
    \item The minor axis \MgII{} emission kinematics is blue- and redshifted on opposite sides of the galaxy's major axis, respecitvely (Figs.~\ref{fig:quadrant_spectra} \& \ref{fig:kinematics_slit2d}); 
    \item The quasar sightline, which is aligned with the galaxy's minor-axis, has strong \MgII{} absorption  ($\rewmgii{}=\rewmgiiqso{}\,\textnormal{\AA}$) at an impact parameter of $\quasarXpropXbmainkpc{}\,\kpc$ whose kinematics
    is consistent with the \MgII{} emission (Fig.~\ref{fig:uves_lines}\,\& \ref{fig:kinematics_slit2d}) ;
    \item We identified three background galaxies within $50\,\kpc$ of
      the \nameprime{} galaxy. While we see indications of \MgII{}
      absorption at the redshift of our halo in these sightlines, they
      lack the S/N to infer
      kinematical information from these sightlines  (Fig.~\ref{fig:absorption_galaxy_sightlines});
      \item The kinematics of the gas seen in both the
        emission and the quasar absorption are consistent with the
        expectation from a simple toy model of a bi-conical outflow
        (\Fig{fig:toy_model} \& \ref{fig:kinematics_slit2d});
       \item We estimated the CGM $\MgII$ density both from the emission - under the assumption of continuum scattering - and the quasar absorption, and find values discrepant by two orders of magnitude. Further, continuum scattering struggles to explain the brightness of the halo.
       \item Shocks in the outflow are a viable alternative to explain the missing \MgII{} photons and the non-resonant extended \OII{} emission.
\end{itemize}

\begin{figure}
  \includegraphics[width=\columnwidth]{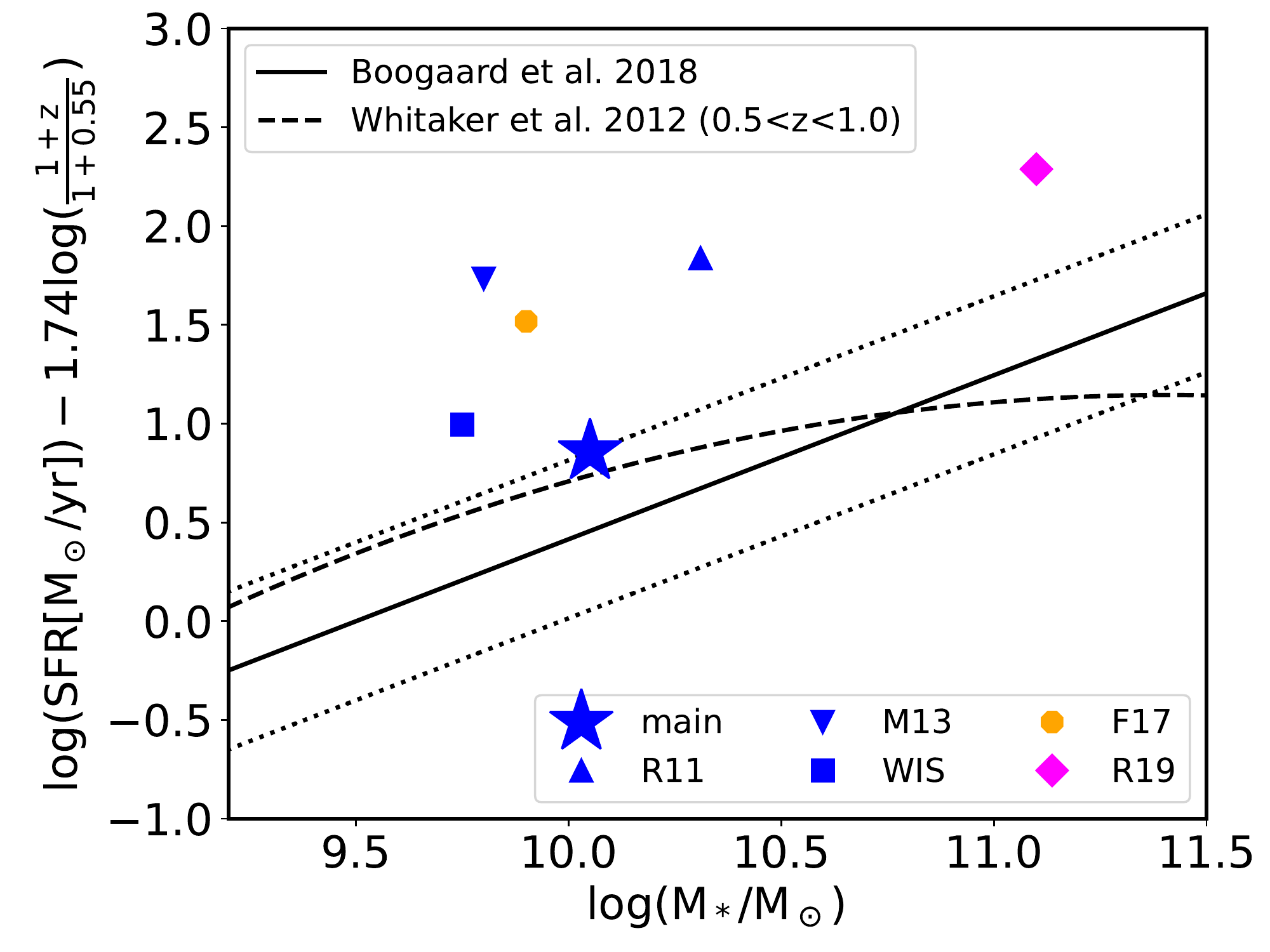}
  \caption{\label{fig:mgiihalogalsonms} Galaxies with \MgII{} halos in the $SFR\mbox{--}M_*$ plane compared to the main-sequence
    of star-formation (MS). In blue are the four \MgII{} halo objects
    found so far ($\nameprime{}$: this work, SFR from \OII; WIS: Wisotzki et al. in
    prep, SFR from \Hb; R11: \citealt{Rubin:2011a} (=\citealt{Burchett:2021a}), SFR from \Hb; M13: \citealt{Martin:2013a}, SFR from \OII).
  In addition, a galaxy with a $\FeII^*$ halo but no \MgII{}
  emission (F17: \citealt{Finley:2017a}, SFR from \OII) and a galaxy with a
  gigantic \OII{} halo and potentially slightly extended \MgII{} emission (R19, \citealt{Rupke:2019a}, SFR from SED) are indicated.
    All SFRs have for comparability been corrected to a
    reference $z=0.55$, using the evolution of the MS
    normalization as found by \citet{Boogaard:2018a}. For comparison 
    both the MS of \citealt{Boogaard:2018a} and
    \citealt{Whitaker:2012a} are shown.
    }
\end{figure}

We put the main emphasis of this paper on characterizing the kinematics, of both the
\MgII{} absorption and emission.
While we also tried to gain some insight into understanding the morphology and
luminosity of the emission nebula through comparison of simple scattering
toy models (Sobolev approximation) and shock models, further analysis should
profit from the use of full radiative transfer models for the resonant \MgII{}
\citep[e.g.][]{Prochaska:2011a, Michel-Dansac:2020a,Burchett:2021a}. Given the
complex morphology and kinematics seen in this \MgII{} halo, radiative
transfer simulations will likely need to go beyond the computationally
efficient assumption of spherical symmetry, and adopt a bi-conical structure.
After all, there is now strong evidence from \MgII{} absorption in background
sightlines that an anisotropic \MgII{} distribution in the CGM is a generic
property around star-forming galaxies (see \Sec{sec:intro}). Further, it might
be necessary to consider more realistic galaxy models than an isotropically
emitting point source.

Understanding and inferring accurate information from the
morphology and kinematics of \MgII{} halos will be one of the main
challenges in the emerging field of \MgII{} halos, with some such
efforts already made in this paper and in previous studies
\citep[e.g.][]{Rubin:2011a,Prochaska:2011a,Martin:2013a,Burchett:2021a}.  

Another
important, more phenomenological way to characterize \MgII{} halos
will be to understand around which type of galaxies they occur. In
\Fig{fig:mgiihalogalsonms}, we compare the position of the four
\MgII{} halos so far published in the literature in the
$SFR\mbox{--}M_*$ plane. While the first two detections
\citep{Rubin:2011a,Martin:2013a}, which are based on long-slit
observations,\footnote{\citet{Burchett:2021a} is a re-observation with the KCWI IFU of the \citet{Rubin:2011a} object.} were significantly above the main-sequence of
star-forming galaxies \citep[MS;
e.g.][]{Whitaker:2012a,Boogaard:2018a}
and could be considered as starbursts, 
the galaxy in Wisotzki et al.~in prep.~has only slightly enhanced star-formation
(about $2\sigma$ above the MS). The \nameprime{} galaxy presented in this
paper is even within the $1\sigma$ scatter of the MS. Therefore, it is clear
that a starburst, at least at the time of the observation, is not a
requirement for a \MgII{} halo to occur. On the other hand, even being
significantly above the MS and having spatially extended emission in
another line does not necessitate the presence of a halo. E.g. in a
starburst with extended $\FeII^*$ emission in the MUSE data no 
extended \MgII{} emission was found \citep{Finley:2017a}, and in a starburst surrounded by a 
giant \OII{} nebula \citep[][]{Rupke:2019a}, which is the largest metal line nebula around a
galaxy detected so far, only slightly extended \MgII{} was
found.

A specific inclination seems also not to be  a requirement for the presence
of \MgII{} halos. While the \nameprime{} galaxy discussed in this work
has a very high inclination ($i\approx75\deg$), the galaxy at the center
of the halo in the Wisotzki et al. in prep. is likely seen almost face on
($i \approx 0\deg$). Nevertheless, the difference in inclination can
explain a major difference between the two objects in the \MgII{}
spectrum extracted for the central galaxy. We did not detect any strong
\MgII{} absorption in the observed down-the-barrel spectrum of the \nameprime{}
galaxy, while the object of Wisotzki et al. shows a typical P-Cygni profile.
Even when correcting for potential infilling by emission (see \Sec{section:halo:morpho}),
the absorption  is not very strong ($\rewmgii{}\approx 0.8\,\textnormal{\AA}$). The lack of strong down-the-barrel absorption
for galaxies observed at high inclination, such as the \nameprime{} galaxy, is
expected for a bi-conical cool outflow launched perpendicular to the galaxy
disk and this expectation has been statistically confirmed
\citep[e.g.][]{Kornei:2012a,Rubin:2014a,Bordoloi:2014c}.

A potential similarity between the \MgII{} halo objects is that they have
non-relaxed disturbed morphologies due to clumps, potentially but not
necessarily triggered by past or present mergers. Both the
\nameprime{} galaxy and the object from \citet{Martin:2013a} lack Hubble Space Telescope data
data\task{Check that this is really the case for the Martin:2013}, which will
be necessary to gain further insight into this issue. Fortunately, it will
soon be possible to study the requirements for the presence or absence of
\MgII{} halos more systematically, thanks to the quickly increasing amount of
available wide-field IFU data from MUSE and KCWI.

\section*{Acknowledgements}

We thank the referee for a constructive report, which helped to improve the quality of the manuscript.
 This study is based on observations collected at the European Southern
Observatory under ESO programmes  095.A-0365(A), 096.A-0609(A), 0100.A-0089(A),
0101.A-0287(A).
This work has been carried out thanks to the support of the ANR FOGHAR (ANR-13-BS05-0010), the ANR 3DGasFlows (ANR-17-CE31-0017), and the OCEVU Labex (ANR-11-LABX-0060).
FL and TG acknowledge support from ERC starting grant ERC-757258-TRIPLE.
SC gratefully acknowledges support from Swiss National Science Foundation grants PP00P2\_163824 and PP00P2\_190092, and from the European Research Council (ERC) under the European Union’s Horizon 2020 research and innovation programme grant agreement No 864361.
JB acknowledges support by Funda\c{c}\~{a}o para a Ci\^{e}ncia e a Tecnologia (FCT) through the
research grants UID/FIS/04434/2019, UIDB/04434/2020, UIDP/04434/2020 and through the Investigador FCT Contract No. IF/01654/2014/CP1215/CT0003.
This  work  made  use  of  the  following  open  source
software:
\gpk{} \citep{Bouche:2015b},
\textsc{ZAP} \citep{Soto:2016b},
\textsc{MPDAF} \citep{Piqueras:2017a},
\textsc{matplotlib} \citep{Hunter:2007a},
\textsc{NumPy} \citep{vanderWalt:2011a},
\textsc{Astropy} \citep{AstropyCollaboration:2013a}.

\section*{Data availability}

The data underlying this article were accessed from the ESO archive (\url{http://archive.eso.org}; program IDs 095.A-0365(A), 096.A-0609(A), 0100.A-0089(A),
0101.A-0287(A)). 
The derived data generated in this research will be shared on reasonable request to the corresponding author.




\bibliographystyle{mnras}
\bibliography{megaflow_mgii_emission} 

\begin{thebibliography}{}
\makeatletter
\relax
\def\mn@urlcharsother{\let\do\@makeother \do\$\do\&\do\#\do\^\do\_\do\%\do\~}
\def\mn@doi{\begingroup\mn@urlcharsother \@ifnextchar [ {\mn@doi@}
  {\mn@doi@[]}}
\def\mn@doi@[#1]#2{\def\@tempa{#1}\ifx\@tempa\@empty \href
  {http://dx.doi.org/#2} {doi:#2}\else \href {http://dx.doi.org/#2} {#1}\fi
  \endgroup}
\def\mn@eprint#1#2{\mn@eprint@#1:#2::\@nil}
\def\mn@eprint@arXiv#1{\href {http://arxiv.org/abs/#1} {{\tt arXiv:#1}}}
\def\mn@eprint@dblp#1{\href {http://dblp.uni-trier.de/rec/bibtex/#1.xml}
  {dblp:#1}}
\def\mn@eprint@#1:#2:#3:#4\@nil{\def\@tempa {#1}\def\@tempb {#2}\def\@tempc
  {#3}\ifx \@tempc \@empty \let \@tempc \@tempb \let \@tempb \@tempa \fi \ifx
  \@tempb \@empty \def\@tempb {arXiv}\fi \@ifundefined
  {mn@eprint@\@tempb}{\@tempb:\@tempc}{\expandafter \expandafter \csname
  mn@eprint@\@tempb\endcsname \expandafter{\@tempc}}}

\bibitem[\protect\citeauthoryear{{Alard} \& {Lupton}}{{Alard} \&
  {Lupton}}{1998}]{Alard:1998a}
{Alard} C.,  {Lupton} R.~H.,  1998, \mn@doi [\apj] {10.1086/305984}, \href
  {https://ui.adsabs.harvard.edu/abs/1998ApJ...503..325A} {503, 325}

\bibitem[\protect\citeauthoryear{{Alarie} \& {Morisset}}{{Alarie} \&
  {Morisset}}{2019}]{Alarie:2019a}
{Alarie} A.,  {Morisset} C.,  2019, \mn@doi [\rmxaa]
  {10.22201/ia.01851101p.2019.55.02.21}, \href
  {https://ui.adsabs.harvard.edu/abs/2019RMxAA..55..377A} {55, 377}

\bibitem[\protect\citeauthoryear{{Allen}, {Groves}, {Dopita}, {Sutherland}  \&
  {Kewley}}{{Allen} et~al.}{2008}]{Allen:2008a}
{Allen} M.~G.,  {Groves} B.~A.,  {Dopita} M.~A.,  {Sutherland} R.~S.,
  {Kewley} L.~J.,  2008, \mn@doi [\apjs] {10.1086/589652}, \href
  {https://ui.adsabs.harvard.edu/abs/2008ApJS..178...20A} {178, 20}

\bibitem[\protect\citeauthoryear{{Angl{\'e}s-Alc{\'a}zar},
  {Faucher-Gigu{\`e}re}, {Kere{\v{s}}}, {Hopkins}, {Quataert}  \&
  {Murray}}{{Angl{\'e}s-Alc{\'a}zar} et~al.}{2017}]{AnglesAlcazar:2017a}
{Angl{\'e}s-Alc{\'a}zar} D.,  {Faucher-Gigu{\`e}re} C.-A.,  {Kere{\v{s}}} D.,
  {Hopkins} P.~F.,  {Quataert} E.,   {Murray} N.,  2017, \mn@doi [\mnras]
  {10.1093/mnras/stx1517}, \href
  {https://ui.adsabs.harvard.edu/abs/2017MNRAS.470.4698A} {470, 4698}

\bibitem[\protect\citeauthoryear{{Astropy Collaboration} et~al.,}{{Astropy
  Collaboration} et~al.}{2013}]{AstropyCollaboration:2013a}
{Astropy Collaboration} et~al., 2013, \mn@doi [\aap]
  {10.1051/0004-6361/201322068}, \href
  {https://ui.adsabs.harvard.edu/abs/2013A&A...558A..33A} {558, A33}

\bibitem[\protect\citeauthoryear{{Bacon} et~al.,}{{Bacon}
  et~al.}{2015}]{Bacon:2015a}
{Bacon} R.,  et~al., 2015, \mn@doi [\aap] {10.1051/0004-6361/201425419}, \href
  {https://ui.adsabs.harvard.edu/abs/2015A&A...575A..75B} {575, A75}

\bibitem[\protect\citeauthoryear{{Bacon} et~al.,}{{Bacon}
  et~al.}{2017}]{Bacon:2017a}
{Bacon} R.,  et~al., 2017, \mn@doi [\aap] {10.1051/0004-6361/201730833}, \href
  {https://ui.adsabs.harvard.edu/abs/2017A&A...608A...1B} {608, A1}

\bibitem[\protect\citeauthoryear{{Bacon} et~al.,}{{Bacon}
  et~al.}{2021}]{Bacon:2021a}
{Bacon} R.,  et~al., 2021, \mn@doi [\aap] {10.1051/0004-6361/202039887}, \href
  {https://ui.adsabs.harvard.edu/abs/2021A&A...647A.107B} {647, A107}

\bibitem[\protect\citeauthoryear{{Bland} \& {Tully}}{{Bland} \&
  {Tully}}{1988}]{Bland:1988a}
{Bland} J.,  {Tully} B.,  1988, \mn@doi [\nat] {10.1038/334043a0}, \href
  {https://ui.adsabs.harvard.edu/abs/1988Natur.334...43B} {334, 43}

\bibitem[\protect\citeauthoryear{{Boogaard} et~al.,}{{Boogaard}
  et~al.}{2018}]{Boogaard:2018a}
{Boogaard} L.~A.,  et~al., 2018, \mn@doi [\aap] {10.1051/0004-6361/201833136},
  \href {https://ui.adsabs.harvard.edu/abs/2018A&A...619A..27B} {619, A27}

\bibitem[\protect\citeauthoryear{{Bordoloi} et~al.,}{{Bordoloi}
  et~al.}{2011}]{Bordoloi:2011a}
{Bordoloi} R.,  et~al., 2011, \mn@doi [\apj] {10.1088/0004-637X/743/1/10},
  \href {https://ui.adsabs.harvard.edu/abs/2011ApJ...743...10B} {743, 10}

\bibitem[\protect\citeauthoryear{{Bordoloi} et~al.,}{{Bordoloi}
  et~al.}{2014}]{Bordoloi:2014c}
{Bordoloi} R.,  et~al., 2014, \mn@doi [\apj] {10.1088/0004-637X/794/2/130},
  \href {https://ui.adsabs.harvard.edu/abs/2014ApJ...794..130B} {794, 130}

\bibitem[\protect\citeauthoryear{{Borisova} et~al.,}{{Borisova}
  et~al.}{2016}]{Borisova:2016a}
{Borisova} E.,  et~al., 2016, \mn@doi [\apj] {10.3847/0004-637X/831/1/39},
  \href {https://ui.adsabs.harvard.edu/abs/2016ApJ...831...39B} {831, 39}

\bibitem[\protect\citeauthoryear{{Bouch{\'e}}, {Hohensee}, {Vargas},
  {Kacprzak}, {Martin}, {Cooke}  \& {Churchill}}{{Bouch{\'e}}
  et~al.}{2012}]{Bouche:2012a}
{Bouch{\'e}} N.,  {Hohensee} W.,  {Vargas} R.,  {Kacprzak} G.~G.,  {Martin}
  C.~L.,  {Cooke} J.,   {Churchill} C.~W.,  2012, \mn@doi [\mnras]
  {10.1111/j.1365-2966.2012.21114.x}, \href
  {https://ui.adsabs.harvard.edu/abs/2012MNRAS.426..801B} {426, 801}

\bibitem[\protect\citeauthoryear{{Bouch{\'e}}, {Murphy}, {Kacprzak},
  {P{\'e}roux}, {Contini}, {Martin}  \& {Dessauges-Zavadsky}}{{Bouch{\'e}}
  et~al.}{2013}]{Bouche:2013a}
{Bouch{\'e}} N.,  {Murphy} M.~T.,  {Kacprzak} G.~G.,  {P{\'e}roux} C.,
  {Contini} T.,  {Martin} C.~L.,   {Dessauges-Zavadsky} M.,  2013, \mn@doi
  [Science] {10.1126/science.1234209}, \href
  {https://ui.adsabs.harvard.edu/abs/2013Sci...341...50B} {341, 50}

\bibitem[\protect\citeauthoryear{{Bouch{\'e}}, {Carfantan}, {Schroetter},
  {Michel-Dansac}  \& {Contini}}{{Bouch{\'e}} et~al.}{2015}]{Bouche:2015b}
{Bouch{\'e}} N.,  {Carfantan} H.,  {Schroetter} I.,  {Michel-Dansac} L.,
  {Contini} T.,  2015, \mn@doi [\aj] {10.1088/0004-6256/150/3/92}, \href
  {https://ui.adsabs.harvard.edu/abs/2015AJ....150...92B} {150, 92}

\bibitem[\protect\citeauthoryear{{Bouch{\'e}} et~al.,}{{Bouch{\'e}}
  et~al.}{2016}]{Bouche:2016a}
{Bouch{\'e}} N.,  et~al., 2016, \mn@doi [\apj] {10.3847/0004-637X/820/2/121},
  \href {https://ui.adsabs.harvard.edu/abs/2016ApJ...820..121B} {820, 121}

\bibitem[\protect\citeauthoryear{{Bowen}, {Chelouche}, {Jenkins}, {Tripp},
  {Pettini}, {York}  \& {Frye}}{{Bowen} et~al.}{2016}]{Bowen:2016a}
{Bowen} D.~V.,  {Chelouche} D.,  {Jenkins} E.~B.,  {Tripp} T.~M.,  {Pettini}
  M.,  {York} D.~G.,   {Frye} B.~L.,  2016, \mn@doi [\apj]
  {10.3847/0004-637X/826/1/50}, \href
  {https://ui.adsabs.harvard.edu/abs/2016ApJ...826...50B} {826, 50}

\bibitem[\protect\citeauthoryear{{Bruzual} \& {Charlot}}{{Bruzual} \&
  {Charlot}}{2003}]{Bruzual:2003a}
{Bruzual} G.,  {Charlot} S.,  2003, \mn@doi [\mnras]
  {10.1046/j.1365-8711.2003.06897.x}, \href
  {https://ui.adsabs.harvard.edu/abs/2003MNRAS.344.1000B} {344, 1000}

\bibitem[\protect\citeauthoryear{{Burchett}, {Rubin}, {Prochaska}, {Coil},
  {Vaught}  \& {Hennawi}}{{Burchett} et~al.}{2021}]{Burchett:2021a}
{Burchett} J.~N.,  {Rubin} K. H.~R.,  {Prochaska} J.~X.,  {Coil} A.~L.,
  {Vaught} R.~R.,   {Hennawi} J.~F.,  2021, \mn@doi [\apj]
  {10.3847/1538-4357/abd4e0}, \href
  {https://ui.adsabs.harvard.edu/abs/2021ApJ...909..151B} {909, 151}

\bibitem[\protect\citeauthoryear{{Cai} et~al.,}{{Cai} et~al.}{2017}]{Cai:2017a}
{Cai} Z.,  et~al., 2017, \mn@doi [\apj] {10.3847/1538-4357/aa5d14}, \href
  {https://ui.adsabs.harvard.edu/abs/2017ApJ...837...71C} {837, 71}

\bibitem[\protect\citeauthoryear{{Calzetti}, {Armus}, {Bohlin}, {Kinney},
  {Koornneef}  \& {Storchi-Bergmann}}{{Calzetti} et~al.}{2000}]{Calzetti:2000a}
{Calzetti} D.,  {Armus} L.,  {Bohlin} R.~C.,  {Kinney} A.~L.,  {Koornneef} J.,
   {Storchi-Bergmann} T.,  2000, \mn@doi [\apj] {10.1086/308692}, \href
  {https://ui.adsabs.harvard.edu/abs/2000ApJ...533..682C} {533, 682}

\bibitem[\protect\citeauthoryear{{Cappellari}}{{Cappellari}}{2017}]{Cappellari:2017a}
{Cappellari} M.,  2017, \mn@doi [\mnras] {10.1093/mnras/stw3020}, \href
  {https://ui.adsabs.harvard.edu/abs/2017MNRAS.466..798C} {466, 798}

\bibitem[\protect\citeauthoryear{{Carr}, {Scarlata}, {Panagia}  \&
  {Henry}}{{Carr} et~al.}{2018}]{Carr:2018a}
{Carr} C.,  {Scarlata} C.,  {Panagia} N.,   {Henry} A.,  2018, \mn@doi [\apj]
  {10.3847/1538-4357/aac48e}, \href
  {https://ui.adsabs.harvard.edu/abs/2018ApJ...860..143C} {860, 143}

\bibitem[\protect\citeauthoryear{{Chabrier}}{{Chabrier}}{2003}]{Chabrier:2003a}
{Chabrier} G.,  2003, \mn@doi [\pasp] {10.1086/376392}, \href
  {https://ui.adsabs.harvard.edu/abs/2003PASP..115..763C} {115, 763}

\bibitem[\protect\citeauthoryear{{Chen}, {Kennicutt}  \& {Rauch}}{{Chen}
  et~al.}{2005}]{Chen:2005a}
{Chen} H.-W.,  {Kennicutt} Robert~C. J.,   {Rauch} M.,  2005, \mn@doi [\apj]
  {10.1086/427088}, \href
  {https://ui.adsabs.harvard.edu/abs/2005ApJ...620..703C} {620, 703}

\bibitem[\protect\citeauthoryear{{Chen}, {Helsby}, {Gauthier}, {Shectman},
  {Thompson}  \& {Tinker}}{{Chen} et~al.}{2010}]{Chen:2010a}
{Chen} H.-W.,  {Helsby} J.~E.,  {Gauthier} J.-R.,  {Shectman} S.~A.,
  {Thompson} I.~B.,   {Tinker} J.~L.,  2010, \mn@doi [\apj]
  {10.1088/0004-637X/714/2/1521}, \href
  {https://ui.adsabs.harvard.edu/abs/2010ApJ...714.1521C} {714, 1521}

\bibitem[\protect\citeauthoryear{{Chen}, {Boettcher}, {Johnson}, {Zahedy},
  {Rudie}, {Cooksey}, {Rauch}  \& {Mulchaey}}{{Chen} et~al.}{2019}]{Chen:2019a}
{Chen} H.-W.,  {Boettcher} E.,  {Johnson} S.~D.,  {Zahedy} F.~S.,  {Rudie}
  G.~C.,  {Cooksey} K.~L.,  {Rauch} M.,   {Mulchaey} J.~S.,  2019, \mn@doi
  [\apjl] {10.3847/2041-8213/ab25ec}, \href
  {https://ui.adsabs.harvard.edu/abs/2019ApJ...878L..33C} {878, L33}

\bibitem[\protect\citeauthoryear{{Chevallard}, {Charlot}, {Wandelt}  \&
  {Wild}}{{Chevallard} et~al.}{2013}]{Chevallard:2013a}
{Chevallard} J.,  {Charlot} S.,  {Wandelt} B.,   {Wild} V.,  2013, \mn@doi
  [\mnras] {10.1093/mnras/stt523}, \href
  {https://ui.adsabs.harvard.edu/abs/2013MNRAS.432.2061C} {432, 2061}

\bibitem[\protect\citeauthoryear{{Claeyssens} et~al.,}{{Claeyssens}
  et~al.}{2019}]{Claeyssens:2019a}
{Claeyssens} A.,  et~al., 2019, \mn@doi [\mnras] {10.1093/mnras/stz2492}, \href
  {https://ui.adsabs.harvard.edu/abs/2019MNRAS.489.5022C} {489, 5022}

\bibitem[\protect\citeauthoryear{{Daddi} et~al.,}{{Daddi}
  et~al.}{2010}]{Daddi:2010a}
{Daddi} E.,  et~al., 2010, \mn@doi [\apj] {10.1088/0004-637X/713/1/686}, \href
  {https://ui.adsabs.harvard.edu/abs/2010ApJ...713..686D} {713, 686}

\bibitem[\protect\citeauthoryear{{De Cia}, {Ledoux}, {Mattsson}, {Petitjean},
  {Srianand}, {Gavignaud}  \& {Jenkins}}{{De Cia} et~al.}{2016}]{DeCia:2016a}
{De Cia} A.,  {Ledoux} C.,  {Mattsson} L.,  {Petitjean} P.,  {Srianand} R.,
  {Gavignaud} I.,   {Jenkins} E.~B.,  2016, \mn@doi [\aap]
  {10.1051/0004-6361/201527895}, \href
  {https://ui.adsabs.harvard.edu/abs/2016A&A...596A..97D} {596, A97}

\bibitem[\protect\citeauthoryear{{Dekker}, {D'Odorico}, {Kaufer}, {Delabre}  \&
  {Kotzlowski}}{{Dekker} et~al.}{2000}]{Dekker:2000a}
{Dekker} H.,  {D'Odorico} S.,  {Kaufer} A.,  {Delabre} B.,   {Kotzlowski} H.,
  2000, {Design, construction, and performance of UVES, the echelle
  spectrograph for the UT2 Kueyen Telescope at the ESO Paranal Observatory}.
pp 534--545, \mn@doi{10.1117/12.395512}

\bibitem[\protect\citeauthoryear{{Dijkstra} \& {Kramer}}{{Dijkstra} \&
  {Kramer}}{2012}]{Dijkstra:2012a}
{Dijkstra} M.,  {Kramer} R.,  2012, \mn@doi [\mnras]
  {10.1111/j.1365-2966.2012.21131.x}, \href
  {https://ui.adsabs.harvard.edu/abs/2012MNRAS.424.1672D} {424, 1672}

\bibitem[\protect\citeauthoryear{{Epinat} et~al.,}{{Epinat}
  et~al.}{2012}]{Epinat:2012a}
{Epinat} B.,  et~al., 2012, \mn@doi [\aap] {10.1051/0004-6361/201117711}, \href
  {https://ui.adsabs.harvard.edu/abs/2012A&A...539A..92E} {539, A92}

\bibitem[\protect\citeauthoryear{{Epinat} et~al.,}{{Epinat}
  et~al.}{2018}]{Epinat:2018a}
{Epinat} B.,  et~al., 2018, \mn@doi [\aap] {10.1051/0004-6361/201731877}, \href
  {https://ui.adsabs.harvard.edu/abs/2018A&A...609A..40E} {609, A40}

\bibitem[\protect\citeauthoryear{{Erb}, {Quider}, {Henry}  \& {Martin}}{{Erb}
  et~al.}{2012}]{Erb:2012a}
{Erb} D.~K.,  {Quider} A.~M.,  {Henry} A.~L.,   {Martin} C.~L.,  2012, \mn@doi
  [\apj] {10.1088/0004-637X/759/1/26}, \href
  {https://ui.adsabs.harvard.edu/abs/2012ApJ...759...26E} {759, 26}

\bibitem[\protect\citeauthoryear{{Erb}, {Steidel}  \& {Chen}}{{Erb}
  et~al.}{2018}]{Erb:2018a}
{Erb} D.~K.,  {Steidel} C.~C.,   {Chen} Y.,  2018, \mn@doi [\apjl]
  {10.3847/2041-8213/aacff6}, \href
  {https://ui.adsabs.harvard.edu/abs/2018ApJ...862L..10E} {862, L10}

\bibitem[\protect\citeauthoryear{{Feldmeier} et~al.,}{{Feldmeier}
  et~al.}{2013}]{Feldmeier:2013a}
{Feldmeier} J.~J.,  et~al., 2013, \mn@doi [\apj] {10.1088/0004-637X/776/2/75},
  \href {https://ui.adsabs.harvard.edu/abs/2013ApJ...776...75F} {776, 75}

\bibitem[\protect\citeauthoryear{{Feltre} et~al.,}{{Feltre}
  et~al.}{2018}]{Feltre:2018a}
{Feltre} A.,  et~al., 2018, \mn@doi [\aap] {10.1051/0004-6361/201833281}, \href
  {https://ui.adsabs.harvard.edu/abs/2018A&A...617A..62F} {617, A62}

\bibitem[\protect\citeauthoryear{{Finley} et~al.,}{{Finley}
  et~al.}{2017a}]{Finley:2017a}
{Finley} H.,  et~al., 2017a, \mn@doi [\aap] {10.1051/0004-6361/201730428},
  \href {https://ui.adsabs.harvard.edu/abs/2017A&A...605A.118F} {605, A118}

\bibitem[\protect\citeauthoryear{{Finley} et~al.,}{{Finley}
  et~al.}{2017b}]{Finley:2017b}
{Finley} H.,  et~al., 2017b, \mn@doi [\aap] {10.1051/0004-6361/201731499},
  \href {https://ui.adsabs.harvard.edu/abs/2017A&A...608A...7F} {608, A7}

\bibitem[\protect\citeauthoryear{{Francis} et~al.,}{{Francis}
  et~al.}{2001}]{Francis:2001a}
{Francis} P.~J.,  et~al., 2001, \mn@doi [\apj] {10.1086/321417}, \href
  {https://ui.adsabs.harvard.edu/abs/2001ApJ...554.1001F} {554, 1001}

\bibitem[\protect\citeauthoryear{{Freundlich} et~al.,}{{Freundlich}
  et~al.}{2013}]{Freundlich:2013a}
{Freundlich} J.,  et~al., 2013, \mn@doi [\aap] {10.1051/0004-6361/201220981},
  \href {https://ui.adsabs.harvard.edu/abs/2013A&A...553A.130F} {553, A130}

\bibitem[\protect\citeauthoryear{{Freundlich}, {Bouch{\'e}}, {Contini},
  {Daddi}, {Zabl}, {Schroetter}, {Boogaard}  \& {Richard}}{{Freundlich}
  et~al.}{2021}]{Freundlich:2021a}
{Freundlich} J.,  {Bouch{\'e}} N.~F.,  {Contini} T.,  {Daddi} E.,  {Zabl} J.,
  {Schroetter} I.,  {Boogaard} L.,   {Richard} J.,  2021, \mn@doi [\mnras]
  {10.1093/mnras/staa3818}, \href
  {https://ui.adsabs.harvard.edu/abs/2021MNRAS.501.1900F} {501, 1900}

\bibitem[\protect\citeauthoryear{{Heckman}, {Armus}  \& {Miley}}{{Heckman}
  et~al.}{1990}]{Heckman:1990a}
{Heckman} T.~M.,  {Armus} L.,   {Miley} G.~K.,  1990, \mn@doi [\apjs]
  {10.1086/191522}, \href
  {https://ui.adsabs.harvard.edu/abs/1990ApJS...74..833H} {74, 833}

\bibitem[\protect\citeauthoryear{{Ho} \& {Martin}}{{Ho} \&
  {Martin}}{2020}]{Ho:2020a}
{Ho} S.~H.,  {Martin} C.~L.,  2020, \mn@doi [\apj] {10.3847/1538-4357/ab58cd},
  \href {https://ui.adsabs.harvard.edu/abs/2020ApJ...888...14H} {888, 14}

\bibitem[\protect\citeauthoryear{{Ho}, {Martin}, {Kacprzak}  \&
  {Churchill}}{{Ho} et~al.}{2017}]{Ho:2017a}
{Ho} S.~H.,  {Martin} C.~L.,  {Kacprzak} G.~G.,   {Churchill} C.~W.,  2017,
  \mn@doi [\apj] {10.3847/1538-4357/835/2/267}, \href
  {https://ui.adsabs.harvard.edu/abs/2017ApJ...835..267H} {835, 267}

\bibitem[\protect\citeauthoryear{{Hunter}}{{Hunter}}{2007}]{Hunter:2007a}
{Hunter} J.~D.,  2007, \mn@doi [Computing in Science and Engineering]
  {10.1109/MCSE.2007.55}, \href
  {https://ui.adsabs.harvard.edu/abs/2007CSE.....9...90H} {9, 90}

\bibitem[\protect\citeauthoryear{{Johnson} et~al.,}{{Johnson}
  et~al.}{2018}]{Johnson:2018a}
{Johnson} S.~D.,  et~al., 2018, \mn@doi [\apjl] {10.3847/2041-8213/aaf1cf},
  \href {https://ui.adsabs.harvard.edu/abs/2018ApJ...869L...1J} {869, L1}

\bibitem[\protect\citeauthoryear{{Kacprzak}, {Churchill}, {Ceverino},
  {Steidel}, {Klypin}  \& {Murphy}}{{Kacprzak} et~al.}{2010}]{Kacprzak:2010a}
{Kacprzak} G.~G.,  {Churchill} C.~W.,  {Ceverino} D.,  {Steidel} C.~C.,
  {Klypin} A.,   {Murphy} M.~T.,  2010, \mn@doi [\apj]
  {10.1088/0004-637X/711/2/533}, \href
  {https://ui.adsabs.harvard.edu/abs/2010ApJ...711..533K} {711, 533}

\bibitem[\protect\citeauthoryear{{Kacprzak}, {Churchill}, {Barton}  \&
  {Cooke}}{{Kacprzak} et~al.}{2011}]{Kacprzak:2011a}
{Kacprzak} G.~G.,  {Churchill} C.~W.,  {Barton} E.~J.,   {Cooke} J.,  2011,
  \mn@doi [\apj] {10.1088/0004-637X/733/2/105}, \href
  {https://ui.adsabs.harvard.edu/abs/2011ApJ...733..105K} {733, 105}

\bibitem[\protect\citeauthoryear{{Kacprzak}, {Churchill}  \&
  {Nielsen}}{{Kacprzak} et~al.}{2012}]{Kacprzak:2012a}
{Kacprzak} G.~G.,  {Churchill} C.~W.,   {Nielsen} N.~M.,  2012, \mn@doi [\apjl]
  {10.1088/2041-8205/760/1/L7}, \href
  {https://ui.adsabs.harvard.edu/abs/2012ApJ...760L...7K} {760, L7}

\bibitem[\protect\citeauthoryear{{Kacprzak} et~al.,}{{Kacprzak}
  et~al.}{2014}]{Kacprzak:2014a}
{Kacprzak} G.~G.,  et~al., 2014, \mn@doi [\apjl] {10.1088/2041-8205/792/1/L12},
  \href {https://ui.adsabs.harvard.edu/abs/2014ApJ...792L..12K} {792, L12}

\bibitem[\protect\citeauthoryear{{Kamann}, {Wisotzki}  \& {Roth}}{{Kamann}
  et~al.}{2013}]{Kamann:2013a}
{Kamann} S.,  {Wisotzki} L.,   {Roth} M.~M.,  2013, \mn@doi [\aap]
  {10.1051/0004-6361/201220476}, \href
  {https://ui.adsabs.harvard.edu/abs/2013A&A...549A..71K} {549, A71}

\bibitem[\protect\citeauthoryear{{Kassin} et~al.,}{{Kassin}
  et~al.}{2007}]{Kassin:2007a}
{Kassin} S.~A.,  et~al., 2007, \mn@doi [\apjl] {10.1086/517932}, \href
  {https://ui.adsabs.harvard.edu/abs/2007ApJ...660L..35K} {660, L35}

\bibitem[\protect\citeauthoryear{{Kewley} \& {Dopita}}{{Kewley} \&
  {Dopita}}{2002}]{Kewley:2002a}
{Kewley} L.~J.,  {Dopita} M.~A.,  2002, \mn@doi [\apjs] {10.1086/341326}, \href
  {https://ui.adsabs.harvard.edu/abs/2002ApJS..142...35K} {142, 35}

\bibitem[\protect\citeauthoryear{{Kornei}, {Shapley}, {Martin}, {Coil}, {Lotz},
  {Schiminovich}, {Bundy}  \& {Noeske}}{{Kornei} et~al.}{2012}]{Kornei:2012a}
{Kornei} K.~A.,  {Shapley} A.~E.,  {Martin} C.~L.,  {Coil} A.~L.,  {Lotz}
  J.~M.,  {Schiminovich} D.,  {Bundy} K.,   {Noeske} K.~G.,  2012, \mn@doi
  [\apj] {10.1088/0004-637X/758/2/135}, \href
  {https://ui.adsabs.harvard.edu/abs/2012ApJ...758..135K} {758, 135}

\bibitem[\protect\citeauthoryear{{Lan} \& {Fukugita}}{{Lan} \&
  {Fukugita}}{2017}]{Lan:2017a}
{Lan} T.-W.,  {Fukugita} M.,  2017, \mn@doi [\apj] {10.3847/1538-4357/aa93eb},
  \href {https://ui.adsabs.harvard.edu/abs/2017ApJ...850..156L} {850, 156}

\bibitem[\protect\citeauthoryear{{Lan} \& {Mo}}{{Lan} \&
  {Mo}}{2018}]{Lan:2018a}
{Lan} T.-W.,  {Mo} H.,  2018, \mn@doi [\apj] {10.3847/1538-4357/aadc08}, \href
  {https://ui.adsabs.harvard.edu/abs/2018ApJ...866...36L} {866, 36}

\bibitem[\protect\citeauthoryear{{Lan}, {M{\'e}nard}  \& {Zhu}}{{Lan}
  et~al.}{2014}]{Lan:2014a}
{Lan} T.-W.,  {M{\'e}nard} B.,   {Zhu} G.,  2014, \mn@doi [\apj]
  {10.1088/0004-637X/795/1/31}, \href
  {https://ui.adsabs.harvard.edu/abs/2014ApJ...795...31L} {795, 31}

\bibitem[\protect\citeauthoryear{{Lanzetta} \& {Bowen}}{{Lanzetta} \&
  {Bowen}}{1990}]{Lanzetta:1990a}
{Lanzetta} K.~M.,  {Bowen} D.,  1990, \mn@doi [\apj] {10.1086/168922}, \href
  {https://ui.adsabs.harvard.edu/abs/1990ApJ...357..321L} {357, 321}

\bibitem[\protect\citeauthoryear{{Laursen}, {Sommer-Larsen}  \&
  {Andersen}}{{Laursen} et~al.}{2009}]{Laursen:2009a}
{Laursen} P.,  {Sommer-Larsen} J.,   {Andersen} A.~C.,  2009, \mn@doi [\apj]
  {10.1088/0004-637X/704/2/1640}, \href
  {https://ui.adsabs.harvard.edu/abs/2009ApJ...704.1640L} {704, 1640}

\bibitem[\protect\citeauthoryear{{Leclercq} et~al.,}{{Leclercq}
  et~al.}{2017}]{Leclercq:2017a}
{Leclercq} F.,  et~al., 2017, \mn@doi [\aap] {10.1051/0004-6361/201731480},
  \href {https://ui.adsabs.harvard.edu/abs/2017A&A...608A...8L} {608, A8}

\bibitem[\protect\citeauthoryear{{Leclercq} et~al.,}{{Leclercq}
  et~al.}{2020}]{Leclercq:2020a}
{Leclercq} F.,  et~al., 2020, \mn@doi [\aap] {10.1051/0004-6361/201937339},
  \href {https://ui.adsabs.harvard.edu/abs/2020A&A...635A..82L} {635, A82}

\bibitem[\protect\citeauthoryear{{Lopez} et~al.,}{{Lopez}
  et~al.}{2018}]{Lopez:2018a}
{Lopez} S.,  et~al., 2018, \mn@doi [\nat] {10.1038/nature25436}, \href
  {https://ui.adsabs.harvard.edu/abs/2018Natur.554..493L} {554, 493}

\bibitem[\protect\citeauthoryear{{Lopez} et~al.,}{{Lopez}
  et~al.}{2019}]{Lopez:2019a}
{Lopez} S.,  et~al., 2019, \mn@doi [\mnras] {10.1093/mnras/stz3183}, \href
  {https://ui.adsabs.harvard.edu/abs/2019MNRAS.tmp.2763L} {p.~2763}

\bibitem[\protect\citeauthoryear{{Lynds} \& {Sandage}}{{Lynds} \&
  {Sandage}}{1963}]{Lynds:1963a}
{Lynds} C.~R.,  {Sandage} A.~R.,  1963, \mn@doi [\aj] {10.1086/109098}, \href
  {https://ui.adsabs.harvard.edu/abs/1963AJ.....68R.284L} {68, 284}

\bibitem[\protect\citeauthoryear{{Maiolino} et~al.,}{{Maiolino}
  et~al.}{2008}]{Maiolino:2008a}
{Maiolino} R.,  et~al., 2008, \mn@doi [\aap] {10.1051/0004-6361:200809678},
  \href {https://ui.adsabs.harvard.edu/abs/2008A&A...488..463M} {488, 463}

\bibitem[\protect\citeauthoryear{{Martin} \& {Bouch{\'e}}}{{Martin} \&
  {Bouch{\'e}}}{2009}]{Martin:2009a}
{Martin} C.~L.,  {Bouch{\'e}} N.,  2009, \mn@doi [\apj]
  {10.1088/0004-637X/703/2/1394}, \href
  {https://ui.adsabs.harvard.edu/abs/2009ApJ...703.1394M} {703, 1394}

\bibitem[\protect\citeauthoryear{{Martin}, {Shapley}, {Coil}, {Kornei},
  {Bundy}, {Weiner}, {Noeske}  \& {Schiminovich}}{{Martin}
  et~al.}{2012}]{Martin:2012a}
{Martin} C.~L.,  {Shapley} A.~E.,  {Coil} A.~L.,  {Kornei} K.~A.,  {Bundy} K.,
  {Weiner} B.~J.,  {Noeske} K.~G.,   {Schiminovich} D.,  2012, \mn@doi [\apj]
  {10.1088/0004-637X/760/2/127}, \href
  {https://ui.adsabs.harvard.edu/abs/2012ApJ...760..127M} {760, 127}

\bibitem[\protect\citeauthoryear{{Martin}, {Shapley}, {Coil}, {Kornei},
  {Murray}  \& {Pancoast}}{{Martin} et~al.}{2013}]{Martin:2013a}
{Martin} C.~L.,  {Shapley} A.~E.,  {Coil} A.~L.,  {Kornei} K.~A.,  {Murray} N.,
    {Pancoast} A.,  2013, \mn@doi [\apj] {10.1088/0004-637X/770/1/41}, \href
  {https://ui.adsabs.harvard.edu/abs/2013ApJ...770...41M} {770, 41}

\bibitem[\protect\citeauthoryear{{Martin}, {Ho}, {Kacprzak}  \&
  {Churchill}}{{Martin} et~al.}{2019}]{Martin:2019a}
{Martin} C.~L.,  {Ho} S.~H.,  {Kacprzak} G.~G.,   {Churchill} C.~W.,  2019,
  \mn@doi [\apj] {10.3847/1538-4357/ab18ac}, \href
  {https://ui.adsabs.harvard.edu/abs/2019ApJ...878...84M} {878, 84}

\bibitem[\protect\citeauthoryear{{Martini}, {Leroy}, {Mangum}, {Bolatto},
  {Keating}, {Sandstrom}  \& {Walter}}{{Martini} et~al.}{2018}]{Martini:2018a}
{Martini} P.,  {Leroy} A.~K.,  {Mangum} J.~G.,  {Bolatto} A.,  {Keating} K.~M.,
   {Sandstrom} K.,   {Walter} F.,  2018, \mn@doi [\apj]
  {10.3847/1538-4357/aab08e}, \href
  {https://ui.adsabs.harvard.edu/abs/2018ApJ...856...61M} {856, 61}

\bibitem[\protect\citeauthoryear{{McKeith}, {Greve}, {Downes}  \&
  {Prada}}{{McKeith} et~al.}{1995}]{McKeith:1995a}
{McKeith} C.~D.,  {Greve} A.,  {Downes} D.,   {Prada} F.,  1995, \aap, \href
  {https://ui.adsabs.harvard.edu/abs/1995A&A...293..703M} {293, 703}

\bibitem[\protect\citeauthoryear{{Michel-Dansac}, {Blaizot}, {Garel},
  {Verhamme}, {Kimm}  \& {Trebitsch}}{{Michel-Dansac}
  et~al.}{2020}]{Michel-Dansac:2020a}
{Michel-Dansac} L.,  {Blaizot} J.,  {Garel} T.,  {Verhamme} A.,  {Kimm} T.,
  {Trebitsch} M.,  2020, \mn@doi [\aap] {10.1051/0004-6361/201834961}, \href
  {https://ui.adsabs.harvard.edu/abs/2020A&A...635A.154M} {635, A154}

\bibitem[\protect\citeauthoryear{{Mitchell}, {Schaye}  \& {Bower}}{{Mitchell}
  et~al.}{2020}]{Mitchell:2020a}
{Mitchell} P.~D.,  {Schaye} J.,   {Bower} R.~G.,  2020, \mn@doi [\mnras]
  {10.1093/mnras/staa2252}, \href
  {https://ui.adsabs.harvard.edu/abs/2020MNRAS.497.4495M} {497, 4495}

\bibitem[\protect\citeauthoryear{{Moffat}}{{Moffat}}{1969}]{Moffat:1969a}
{Moffat} A.~F.~J.,  1969, \aap, \href
  {https://ui.adsabs.harvard.edu/abs/1969A&A.....3..455M} {3, 455}

\bibitem[\protect\citeauthoryear{{Momose} et~al.,}{{Momose}
  et~al.}{2014}]{Momose:2014a}
{Momose} R.,  et~al., 2014, \mn@doi [\mnras] {10.1093/mnras/stu825}, \href
  {https://ui.adsabs.harvard.edu/abs/2014MNRAS.442..110M} {442, 110}

\bibitem[\protect\citeauthoryear{{Muzahid}, {Kacprzak}, {Churchill},
  {Charlton}, {Nielsen}, {Mathes}  \& {Trujillo-Gomez}}{{Muzahid}
  et~al.}{2015}]{Muzahid:2015a}
{Muzahid} S.,  {Kacprzak} G.~G.,  {Churchill} C.~W.,  {Charlton} J.~C.,
  {Nielsen} N.~M.,  {Mathes} N.~L.,   {Trujillo-Gomez} S.,  2015, \mn@doi
  [\apj] {10.1088/0004-637X/811/2/132}, \href
  {https://ui.adsabs.harvard.edu/abs/2015ApJ...811..132M} {811, 132}

\bibitem[\protect\citeauthoryear{{Nielsen}, {Churchill}, {Kacprzak}  \&
  {Murphy}}{{Nielsen} et~al.}{2013}]{Nielsen:2013a}
{Nielsen} N.~M.,  {Churchill} C.~W.,  {Kacprzak} G.~G.,   {Murphy} M.~T.,
  2013, \mn@doi [\apj] {10.1088/0004-637X/776/2/114}, \href
  {https://ui.adsabs.harvard.edu/abs/2013ApJ...776..114N} {776, 114}

\bibitem[\protect\citeauthoryear{{Nielsen}, {Churchill}, {Kacprzak}, {Murphy}
  \& {Evans}}{{Nielsen} et~al.}{2015}]{Nielsen:2015a}
{Nielsen} N.~M.,  {Churchill} C.~W.,  {Kacprzak} G.~G.,  {Murphy} M.~T.,
  {Evans} J.~L.,  2015, \mn@doi [\apj] {10.1088/0004-637X/812/1/83}, \href
  {https://ui.adsabs.harvard.edu/abs/2015ApJ...812...83N} {812, 83}

\bibitem[\protect\citeauthoryear{{Peng}, {Ho}, {Impey}  \& {Rix}}{{Peng}
  et~al.}{2010}]{Peng:2010a}
{Peng} C.~Y.,  {Ho} L.~C.,  {Impey} C.~D.,   {Rix} H.-W.,  2010, \mn@doi [\aj]
  {10.1088/0004-6256/139/6/2097}, \href
  {https://ui.adsabs.harvard.edu/abs/2010AJ....139.2097P} {139, 2097}

\bibitem[\protect\citeauthoryear{{Piqueras}, {Conseil}, {Shepherd}, {Bacon},
  {Leclercq}  \& {Richard}}{{Piqueras} et~al.}{2017}]{Piqueras:2017a}
{Piqueras} L.,  {Conseil} S.,  {Shepherd} M.,  {Bacon} R.,  {Leclercq} F.,
  {Richard} J.,  2017, arXiv e-prints, \href
  {https://ui.adsabs.harvard.edu/abs/2017arXiv171003554P} {p. arXiv:1710.03554}

\bibitem[\protect\citeauthoryear{{Prescott}, {Dey}  \& {Jannuzi}}{{Prescott}
  et~al.}{2009}]{Prescott:2009a}
{Prescott} M. K.~M.,  {Dey} A.,   {Jannuzi} B.~T.,  2009, \mn@doi [\apj]
  {10.1088/0004-637X/702/1/554}, \href
  {https://ui.adsabs.harvard.edu/abs/2009ApJ...702..554P} {702, 554}

\bibitem[\protect\citeauthoryear{{Prescott}, {Momcheva}, {Brammer}, {Fynbo}  \&
  {M{\o}ller}}{{Prescott} et~al.}{2015}]{Prescott:2015a}
{Prescott} M. K.~M.,  {Momcheva} I.,  {Brammer} G.~B.,  {Fynbo} J. P.~U.,
  {M{\o}ller} P.,  2015, \mn@doi [\apj] {10.1088/0004-637X/802/1/32}, \href
  {https://ui.adsabs.harvard.edu/abs/2015ApJ...802...32P} {802, 32}

\bibitem[\protect\citeauthoryear{{Prochaska}, {Kasen}  \& {Rubin}}{{Prochaska}
  et~al.}{2011}]{Prochaska:2011a}
{Prochaska} J.~X.,  {Kasen} D.,   {Rubin} K.,  2011, \mn@doi [\apj]
  {10.1088/0004-637X/734/1/24}, \href
  {https://ui.adsabs.harvard.edu/abs/2011ApJ...734...24P} {734, 24}

\bibitem[\protect\citeauthoryear{{Quast}, {Baade}  \& {Reimers}}{{Quast}
  et~al.}{2005}]{Quast:2005a}
{Quast} R.,  {Baade} R.,   {Reimers} D.,  2005, \mn@doi [\aap]
  {10.1051/0004-6361:20041601}, \href
  {https://ui.adsabs.harvard.edu/abs/2005A&A...431.1167Q} {431, 1167}

\bibitem[\protect\citeauthoryear{{Rahmani} et~al.,}{{Rahmani}
  et~al.}{2018a}]{Rahmani:2018a}
{Rahmani} H.,  et~al., 2018a, \mn@doi [\mnras] {10.1093/mnras/stx2726}, \href
  {https://ui.adsabs.harvard.edu/abs/2018MNRAS.474..254R} {474, 254}

\bibitem[\protect\citeauthoryear{{Rahmani} et~al.,}{{Rahmani}
  et~al.}{2018b}]{Rahmani:2018b}
{Rahmani} H.,  et~al., 2018b, \mn@doi [\mnras] {10.1093/mnras/sty2216}, \href
  {https://ui.adsabs.harvard.edu/abs/2018MNRAS.480.5046R} {480, 5046}

\bibitem[\protect\citeauthoryear{{Rickards Vaught}, {Rubin}, {Arrigoni
  Battaia}, {Prochaska}  \& {Hennawi}}{{Rickards Vaught}
  et~al.}{2019}]{RickardsVaught:2019a}
{Rickards Vaught} R.~J.,  {Rubin} K. H.~R.,  {Arrigoni Battaia} F.,
  {Prochaska} J.~X.,   {Hennawi} J.~F.,  2019, \mn@doi [\apj]
  {10.3847/1538-4357/ab211f}, \href
  {https://ui.adsabs.harvard.edu/abs/2019ApJ...879....7R} {879, 7}

\bibitem[\protect\citeauthoryear{{Rubin}, {Weiner}, {Koo}, {Martin},
  {Prochaska}, {Coil}  \& {Newman}}{{Rubin} et~al.}{2010}]{Rubin:2010b}
{Rubin} K. H.~R.,  {Weiner} B.~J.,  {Koo} D.~C.,  {Martin} C.~L.,  {Prochaska}
  J.~X.,  {Coil} A.~L.,   {Newman} J.~A.,  2010, \mn@doi [\apj]
  {10.1088/0004-637X/719/2/1503}, \href
  {https://ui.adsabs.harvard.edu/abs/2010ApJ...719.1503R} {719, 1503}

\bibitem[\protect\citeauthoryear{{Rubin}, {Prochaska}, {M{\'e}nard}, {Murray},
  {Kasen}, {Koo}  \& {Phillips}}{{Rubin} et~al.}{2011}]{Rubin:2011a}
{Rubin} K. H.~R.,  {Prochaska} J.~X.,  {M{\'e}nard} B.,  {Murray} N.,  {Kasen}
  D.,  {Koo} D.~C.,   {Phillips} A.~C.,  2011, \mn@doi [\apj]
  {10.1088/0004-637X/728/1/55}, \href
  {https://ui.adsabs.harvard.edu/abs/2011ApJ...728...55R} {728, 55}

\bibitem[\protect\citeauthoryear{{Rubin}, {Prochaska}, {Koo}, {Phillips},
  {Martin}  \& {Winstrom}}{{Rubin} et~al.}{2014}]{Rubin:2014a}
{Rubin} K. H.~R.,  {Prochaska} J.~X.,  {Koo} D.~C.,  {Phillips} A.~C.,
  {Martin} C.~L.,   {Winstrom} L.~O.,  2014, \mn@doi [\apj]
  {10.1088/0004-637X/794/2/156}, \href
  {https://ui.adsabs.harvard.edu/abs/2014ApJ...794..156R} {794, 156}

\bibitem[\protect\citeauthoryear{{Rupke} et~al.,}{{Rupke}
  et~al.}{2019}]{Rupke:2019a}
{Rupke} D. S.~N.,  et~al., 2019, \mn@doi [\nat] {10.1038/s41586-019-1686-1},
  \href {https://ui.adsabs.harvard.edu/abs/2019Natur.574..643R} {574, 643}

\bibitem[\protect\citeauthoryear{{Saintonge} et~al.,}{{Saintonge}
  et~al.}{2013}]{Saintonge:2013a}
{Saintonge} A.,  et~al., 2013, \mn@doi [\apj] {10.1088/0004-637X/778/1/2},
  \href {https://ui.adsabs.harvard.edu/abs/2013ApJ...778....2S} {778, 2}

\bibitem[\protect\citeauthoryear{{Scarlata} \& {Panagia}}{{Scarlata} \&
  {Panagia}}{2015}]{Scarlata:2015a}
{Scarlata} C.,  {Panagia} N.,  2015, \mn@doi [\apj]
  {10.1088/0004-637X/801/1/43}, \href
  {https://ui.adsabs.harvard.edu/abs/2015ApJ...801...43S} {801, 43}

\bibitem[\protect\citeauthoryear{{Schlegel}, {Finkbeiner}  \&
  {Davis}}{{Schlegel} et~al.}{1998}]{Schlegel:1998a}
{Schlegel} D.~J.,  {Finkbeiner} D.~P.,   {Davis} M.,  1998, \mn@doi [\apj]
  {10.1086/305772}, \href
  {https://ui.adsabs.harvard.edu/abs/1998ApJ...500..525S} {500, 525}

\bibitem[\protect\citeauthoryear{{Schmidt} et~al.,}{{Schmidt}
  et~al.}{2019}]{Schmidt:2019a}
{Schmidt} K.~B.,  et~al., 2019, \mn@doi [\aap] {10.1051/0004-6361/201935857},
  \href {https://ui.adsabs.harvard.edu/abs/2019A&A...628A..91S} {628, A91}

\bibitem[\protect\citeauthoryear{{Schroetter}, {Bouch{\'e}}, {P{\'e}roux},
  {Murphy}, {Contini}  \& {Finley}}{{Schroetter}
  et~al.}{2015}]{Schroetter:2015a}
{Schroetter} I.,  {Bouch{\'e}} N.,  {P{\'e}roux} C.,  {Murphy} M.~T.,
  {Contini} T.,   {Finley} H.,  2015, \mn@doi [\apj]
  {10.1088/0004-637X/804/2/83}, \href
  {https://ui.adsabs.harvard.edu/abs/2015ApJ...804...83S} {804, 83}

\bibitem[\protect\citeauthoryear{{Schroetter} et~al.,}{{Schroetter}
  et~al.}{2016}]{Schroetter:2016a}
{Schroetter} I.,  et~al., 2016, \mn@doi [\apj] {10.3847/1538-4357/833/1/39},
  \href {https://ui.adsabs.harvard.edu/abs/2016ApJ...833...39S} {833, 39}

\bibitem[\protect\citeauthoryear{{Schroetter} et~al.,}{{Schroetter}
  et~al.}{2019}]{Schroetter:2019a}
{Schroetter} I.,  et~al., 2019, \mn@doi [\mnras] {10.1093/mnras/stz2822}, \href
  {https://ui.adsabs.harvard.edu/abs/2019MNRAS.490.4368S} {490, 4368}

\bibitem[\protect\citeauthoryear{{Schroetter} et~al.,}{{Schroetter}
  et~al.}{2020}]{Schroetter:2020a}
{Schroetter} I.,  et~al., 2020, arXiv e-prints, \href
  {https://ui.adsabs.harvard.edu/abs/2020arXiv201204935S} {p. arXiv:2012.04935}

\bibitem[\protect\citeauthoryear{{Sharp} \& {Bland-Hawthorn}}{{Sharp} \&
  {Bland-Hawthorn}}{2010}]{Sharp:2010a}
{Sharp} R.~G.,  {Bland-Hawthorn} J.,  2010, \mn@doi [\apj]
  {10.1088/0004-637X/711/2/818}, \href
  {https://ui.adsabs.harvard.edu/abs/2010ApJ...711..818S} {711, 818}

\bibitem[\protect\citeauthoryear{{Sobolev}}{{Sobolev}}{1960}]{Sobolev:1960b}
{Sobolev} V.~V.,  1960, {Moving envelopes of stars}

\bibitem[\protect\citeauthoryear{{Soto}, {Lilly}, {Bacon}, {Richard}  \&
  {Conseil}}{{Soto} et~al.}{2016}]{Soto:2016b}
{Soto} K.~T.,  {Lilly} S.~J.,  {Bacon} R.,  {Richard} J.,   {Conseil} S.,
  2016, {ZAP: Zurich Atmosphere Purge} (\mn@eprint {ascl} {1602.003})

\bibitem[\protect\citeauthoryear{{Steidel}, {Adelberger}, {Shapley}, {Pettini},
  {Dickinson}  \& {Giavalisco}}{{Steidel} et~al.}{2000}]{Steidel:2000a}
{Steidel} C.~C.,  {Adelberger} K.~L.,  {Shapley} A.~E.,  {Pettini} M.,
  {Dickinson} M.,   {Giavalisco} M.,  2000, \mn@doi [\apj] {10.1086/308568},
  \href {https://ui.adsabs.harvard.edu/abs/2000ApJ...532..170S} {532, 170}

\bibitem[\protect\citeauthoryear{{Steidel}, {Kollmeier}, {Shapley},
  {Churchill}, {Dickinson}  \& {Pettini}}{{Steidel}
  et~al.}{2002}]{Steidel:2002a}
{Steidel} C.~C.,  {Kollmeier} J.~A.,  {Shapley} A.~E.,  {Churchill} C.~W.,
  {Dickinson} M.,   {Pettini} M.,  2002, \mn@doi [\apj] {10.1086/339792}, \href
  {https://ui.adsabs.harvard.edu/abs/2002ApJ...570..526S} {570, 526}

\bibitem[\protect\citeauthoryear{{Steidel}, {Bogosavljevi{\'c}}, {Shapley},
  {Kollmeier}, {Reddy}, {Erb}  \& {Pettini}}{{Steidel}
  et~al.}{2011}]{Steidel:2011a}
{Steidel} C.~C.,  {Bogosavljevi{\'c}} M.,  {Shapley} A.~E.,  {Kollmeier} J.~A.,
   {Reddy} N.~A.,  {Erb} D.~K.,   {Pettini} M.,  2011, \mn@doi [\apj]
  {10.1088/0004-637X/736/2/160}, \href
  {https://ui.adsabs.harvard.edu/abs/2011ApJ...736..160S} {736, 160}

\bibitem[\protect\citeauthoryear{{Sugahara}, {Ouchi}, {Lin}, {Martin}, {Ono},
  {Harikane}, {Shibuya}  \& {Yan}}{{Sugahara} et~al.}{2017}]{Sugahara:2017a}
{Sugahara} Y.,  {Ouchi} M.,  {Lin} L.,  {Martin} C.~L.,  {Ono} Y.,  {Harikane}
  Y.,  {Shibuya} T.,   {Yan} R.,  2017, \mn@doi [\apj]
  {10.3847/1538-4357/aa956d}, \href
  {https://ui.adsabs.harvard.edu/abs/2017ApJ...850...51S} {850, 51}

\bibitem[\protect\citeauthoryear{{Tacconi} et~al.,}{{Tacconi}
  et~al.}{2010}]{Tacconi:2010a}
{Tacconi} L.~J.,  et~al., 2010, \mn@doi [\nat] {10.1038/nature08773}, \href
  {https://ui.adsabs.harvard.edu/abs/2010Natur.463..781T} {463, 781}

\bibitem[\protect\citeauthoryear{{Tacconi} et~al.,}{{Tacconi}
  et~al.}{2013}]{Tacconi:2013a}
{Tacconi} L.~J.,  et~al., 2013, \mn@doi [\apj] {10.1088/0004-637X/768/1/74},
  \href {https://ui.adsabs.harvard.edu/abs/2013ApJ...768...74T} {768, 74}

\bibitem[\protect\citeauthoryear{{Tumlinson}, {Peeples}  \& {Werk}}{{Tumlinson}
  et~al.}{2017}]{Tumlinson:2017a}
{Tumlinson} J.,  {Peeples} M.~S.,   {Werk} J.~K.,  2017, \mn@doi [\araa]
  {10.1146/annurev-astro-091916-055240}, \href
  {https://ui.adsabs.harvard.edu/abs/2017ARA&A..55..389T} {55, 389}

\bibitem[\protect\citeauthoryear{{Vazdekis}, {Koleva}, {Ricciardelli},
  {R{\"o}ck}  \& {Falc{\'o}n-Barroso}}{{Vazdekis}
  et~al.}{2016}]{Vazdekis:2016a}
{Vazdekis} A.,  {Koleva} M.,  {Ricciardelli} E.,  {R{\"o}ck} B.,
  {Falc{\'o}n-Barroso} J.,  2016, \mn@doi [\mnras] {10.1093/mnras/stw2231},
  \href {https://ui.adsabs.harvard.edu/abs/2016MNRAS.463.3409V} {463, 3409}

\bibitem[\protect\citeauthoryear{{Verhamme}, {Schaerer}  \&
  {Maselli}}{{Verhamme} et~al.}{2006}]{Verhamme:2006a}
{Verhamme} A.,  {Schaerer} D.,   {Maselli} A.,  2006, \mn@doi [\aap]
  {10.1051/0004-6361:20065554}, \href
  {https://ui.adsabs.harvard.edu/abs/2006A&A...460..397V} {460, 397}

\bibitem[\protect\citeauthoryear{{Weilbacher}, {Streicher}  \&
  {Palsa}}{{Weilbacher} et~al.}{2016}]{Weilbacher:2016a}
{Weilbacher} P.~M.,  {Streicher} O.,   {Palsa} R.,  2016, {MUSE-DRP: MUSE Data
  Reduction Pipeline} (\mn@eprint {ascl} {1610.004})

\bibitem[\protect\citeauthoryear{{Weilbacher} et~al.,}{{Weilbacher}
  et~al.}{2020}]{Weilbacher:2020a}
{Weilbacher} P.~M.,  et~al., 2020, \mn@doi [\aap]
  {10.1051/0004-6361/202037855}, \href
  {https://ui.adsabs.harvard.edu/abs/2020A&A...641A..28W} {641, A28}

\bibitem[\protect\citeauthoryear{{Weiner} et~al.,}{{Weiner}
  et~al.}{2009}]{Weiner:2009a}
{Weiner} B.~J.,  et~al., 2009, \mn@doi [\apj] {10.1088/0004-637X/692/1/187},
  \href {https://ui.adsabs.harvard.edu/abs/2009ApJ...692..187W} {692, 187}

\bibitem[\protect\citeauthoryear{{Wendt} \& {Molaro}}{{Wendt} \&
  {Molaro}}{2012}]{Wendt:2012a}
{Wendt} M.,  {Molaro} P.,  2012, \mn@doi [\aap] {10.1051/0004-6361/201218862},
  \href {https://ui.adsabs.harvard.edu/abs/2012A&A...541A..69W} {541, A69}

\bibitem[\protect\citeauthoryear{{Wendt}, {Bouch{\'e}}, {Zabl}, {Schroetter}
  \& {Muzahid}}{{Wendt} et~al.}{2021}]{Wendt:2021a}
{Wendt} M.,  {Bouch{\'e}} N.~F.,  {Zabl} J.,  {Schroetter} I.,   {Muzahid} S.,
  2021, \mn@doi [\mnras] {10.1093/mnras/stab049}, \href
  {https://ui.adsabs.harvard.edu/abs/2021MNRAS.502.3733W} {502, 3733}

\bibitem[\protect\citeauthoryear{{Whitaker}, {van Dokkum}, {Brammer}  \&
  {Franx}}{{Whitaker} et~al.}{2012}]{Whitaker:2012a}
{Whitaker} K.~E.,  {van Dokkum} P.~G.,  {Brammer} G.,   {Franx} M.,  2012,
  \mn@doi [\apjl] {10.1088/2041-8205/754/2/L29}, \href
  {https://ui.adsabs.harvard.edu/abs/2012ApJ...754L..29W} {754, L29}

\bibitem[\protect\citeauthoryear{{Wisotzki} et~al.,}{{Wisotzki}
  et~al.}{2016}]{Wisotzki:2016a}
{Wisotzki} L.,  et~al., 2016, \mn@doi [\aap] {10.1051/0004-6361/201527384},
  \href {https://ui.adsabs.harvard.edu/abs/2016A&A...587A..98W} {587, A98}

\bibitem[\protect\citeauthoryear{{Wisotzki} et~al.,}{{Wisotzki}
  et~al.}{2018}]{Wisotzki:2018a}
{Wisotzki} L.,  et~al., 2018, \mn@doi [\nat] {10.1038/s41586-018-0564-6}, \href
  {https://ui.adsabs.harvard.edu/abs/2018Natur.562..229W} {562, 229}

\bibitem[\protect\citeauthoryear{{Yang}, {Zabludoff}, {Tremonti}, {Eisenstein}
  \& {Dav{\'e}}}{{Yang} et~al.}{2009}]{Yang:2009a}
{Yang} Y.,  {Zabludoff} A.,  {Tremonti} C.,  {Eisenstein} D.,   {Dav{\'e}} R.,
  2009, \mn@doi [\apj] {10.1088/0004-637X/693/2/1579}, \href
  {https://ui.adsabs.harvard.edu/abs/2009ApJ...693.1579Y} {693, 1579}

\bibitem[\protect\citeauthoryear{{Yip}, {Szalay}, {Wyse}, {Dobos},
  {Budav{\'a}ri}  \& {Csabai}}{{Yip} et~al.}{2010}]{Yip:2010a}
{Yip} C.-W.,  {Szalay} A.~S.,  {Wyse} R. F.~G.,  {Dobos} L.,  {Budav{\'a}ri}
  T.,   {Csabai} I.,  2010, \mn@doi [\apj] {10.1088/0004-637X/709/2/780}, \href
  {https://ui.adsabs.harvard.edu/abs/2010ApJ...709..780Y} {709, 780}

\bibitem[\protect\citeauthoryear{{Zabl}, {Freudling}, {M{\o}ller},
  {Milvang-Jensen}, {Nilsson}, {Fynbo}, {Le F{\`e}vre}  \& {Tasca}}{{Zabl}
  et~al.}{2016}]{Zabl:2016a}
{Zabl} J.,  {Freudling} W.,  {M{\o}ller} P.,  {Milvang-Jensen} B.,  {Nilsson}
  K.~K.,  {Fynbo} J.~P.~U.,  {Le F{\`e}vre} O.,   {Tasca} L.~A.~M.,  2016,
  \mn@doi [\aap] {10.1051/0004-6361/201526378}, \href
  {https://ui.adsabs.harvard.edu/abs/2016A&A...590A..66Z} {590, A66}

\bibitem[\protect\citeauthoryear{{Zabl} et~al.,}{{Zabl}
  et~al.}{2019}]{Zabl:2019a}
{Zabl} J.,  et~al., 2019, \mn@doi [\mnras] {10.1093/mnras/stz392}, \href
  {https://ui.adsabs.harvard.edu/abs/2019MNRAS.485.1961Z} {485, 1961}

\bibitem[\protect\citeauthoryear{{Zabl} et~al.,}{{Zabl}
  et~al.}{2020}]{Zabl:2020a}
{Zabl} J.,  et~al., 2020, \mn@doi [\mnras] {10.1093/mnras/stz3607}, \href
  {https://ui.adsabs.harvard.edu/abs/2020MNRAS.492.4576Z} {492, 4576}

\bibitem[\protect\citeauthoryear{{van der Walt}, {Colbert}  \&
  {Varoquaux}}{{van der Walt} et~al.}{2011}]{vanderWalt:2011a}
{van der Walt} S.,  {Colbert} S.~C.,   {Varoquaux} G.,  2011, \mn@doi
  [Computing in Science and Engineering] {10.1109/MCSE.2011.37}, \href
  {https://ui.adsabs.harvard.edu/abs/2011CSE....13b..22V} {13, 22}

\makeatother
\end{thebibliography}



\appendix

\section{Galaxy properties}

\begin{figure}
	\centering
	\includegraphics[width=\columnwidth]{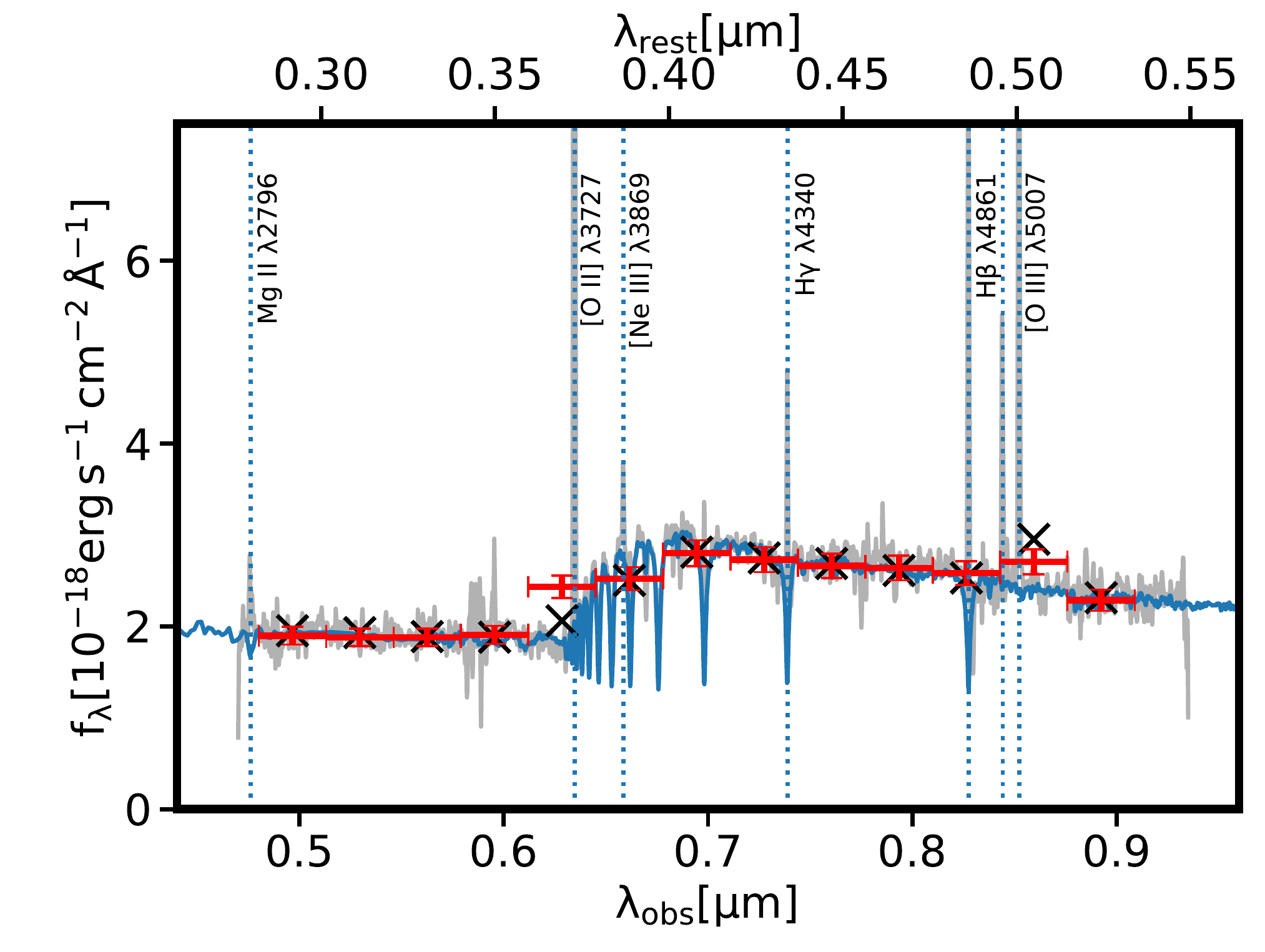}
	\caption{SED fit for the \nameprime{} galaxy. The blue curve shows
  the best-fit SED model, while the grey curve is the observed
  spectrum (smoothed with a Gaussian of width $\sigma=1.5\,\mathrm{pixel}$). The SED
  fit was done to photometry in 13 medium band filters represented by the red horizontal error bars.
  While the red error bars are plotted at the measured flux density, the black
  crosses show the flux density predicted by the best-fit SED model, which includes the contribution from emission lines.
 }
	\label{fig:sed_prime}
\end{figure}

\subsection{Stellar masses}
\label{appendix:mstar}

We determined the stellar masses of the galaxies using our custom SED fitting code \coniecto{} \citep{Zabl:2016a}.
Identical to earlier papers in the \mfl{} series, we fit BC03 \citep{Bruzual:2003a} models to photometry determined from 13 pseudo-medium filters optimized to cover the full MUSE wavelength range.
 The width of these filters is shown by the red horizontal error bars in \Fig{fig:sed_prime}.
As in \citetalias{Zabl:2020a}{}, we use here a delayed-tau star-formation history.

The  the stellar mass obtained from the SED fit (shown in \Fig{fig:sed_prime}) for the \nameprime{}  galaxies is $\log(M_\star/M_\odot) = 	\mainXpropXsedXmass$. The \namesecond{} galaxy has  $\log(M_\star/M_\odot)=\secondXpropXsedXmass$. This means that the \namesecond{} galaxy has less than 1/5 the mass of the \nameprime{} galaxy and can be considered as a minor companion to the \nameprime{} galaxy.

\subsection{Morphology}
\label{appendix:morpho}

\begin{figure*}
	\begin{tabular}{ccc}
	\OII{} & \OIII{} & Ratio / Color map \\
		\includegraphics[width=0.25\textwidth]{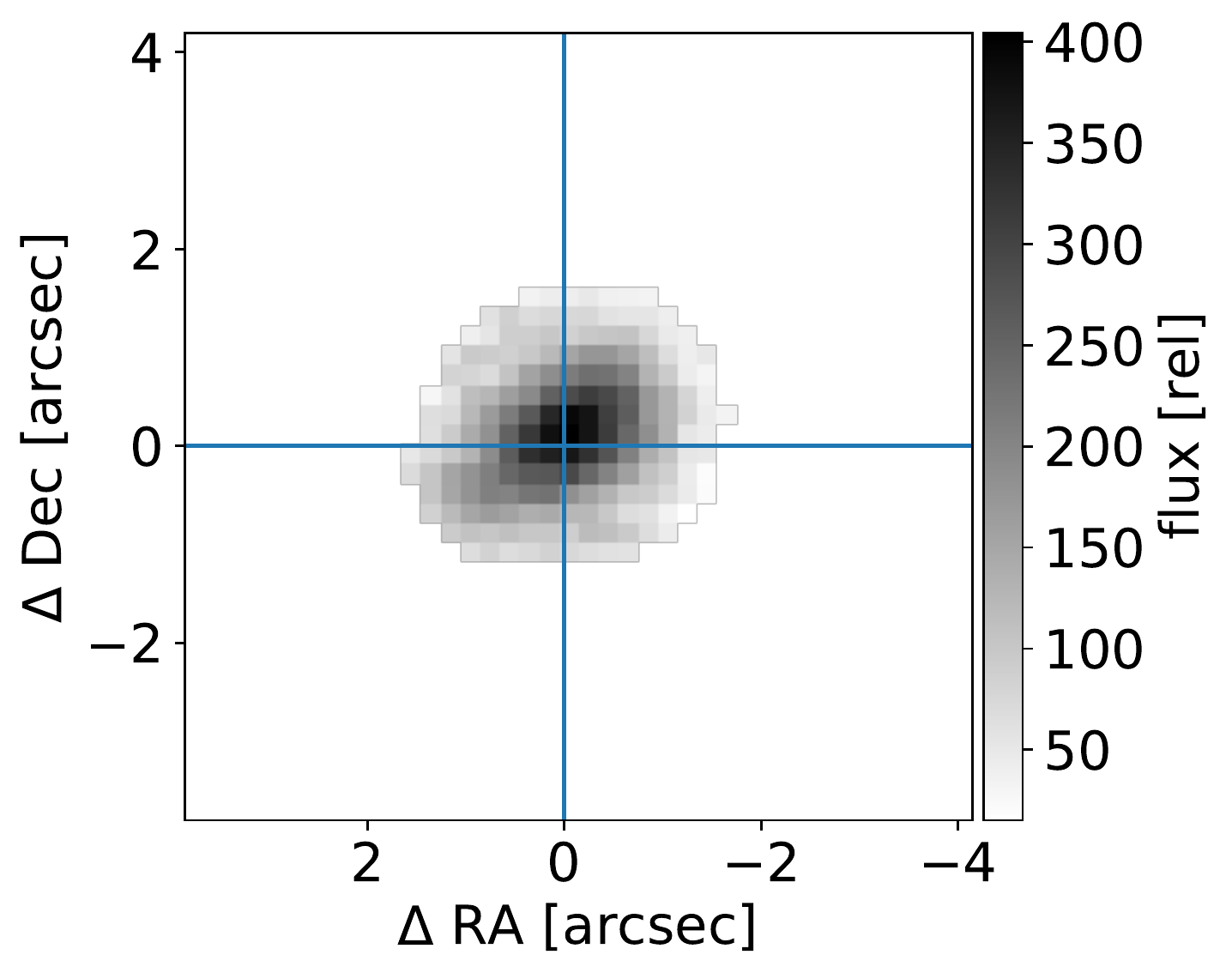} & \includegraphics[width=0.25\textwidth]{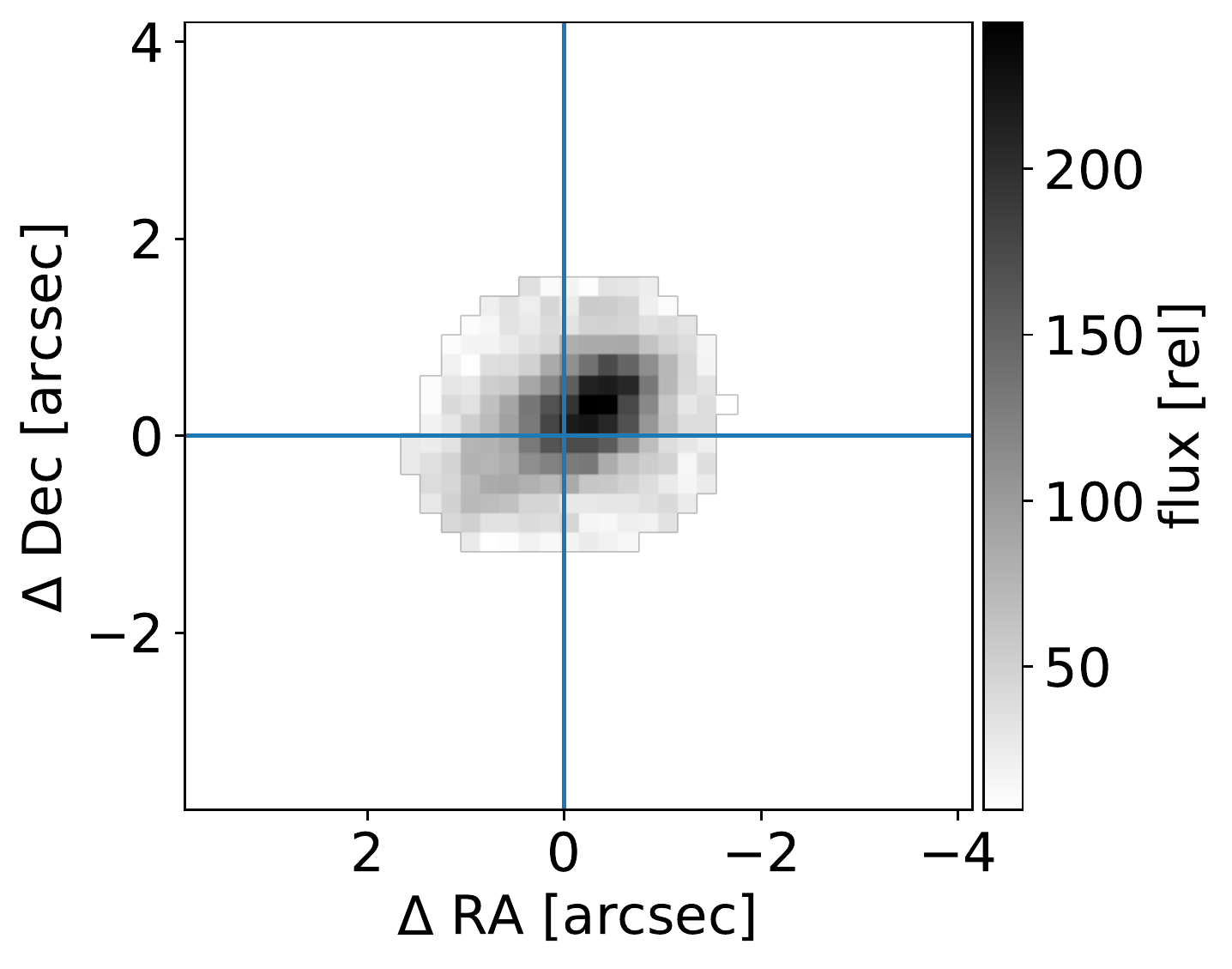}  &  \includegraphics[width=0.25\textwidth]{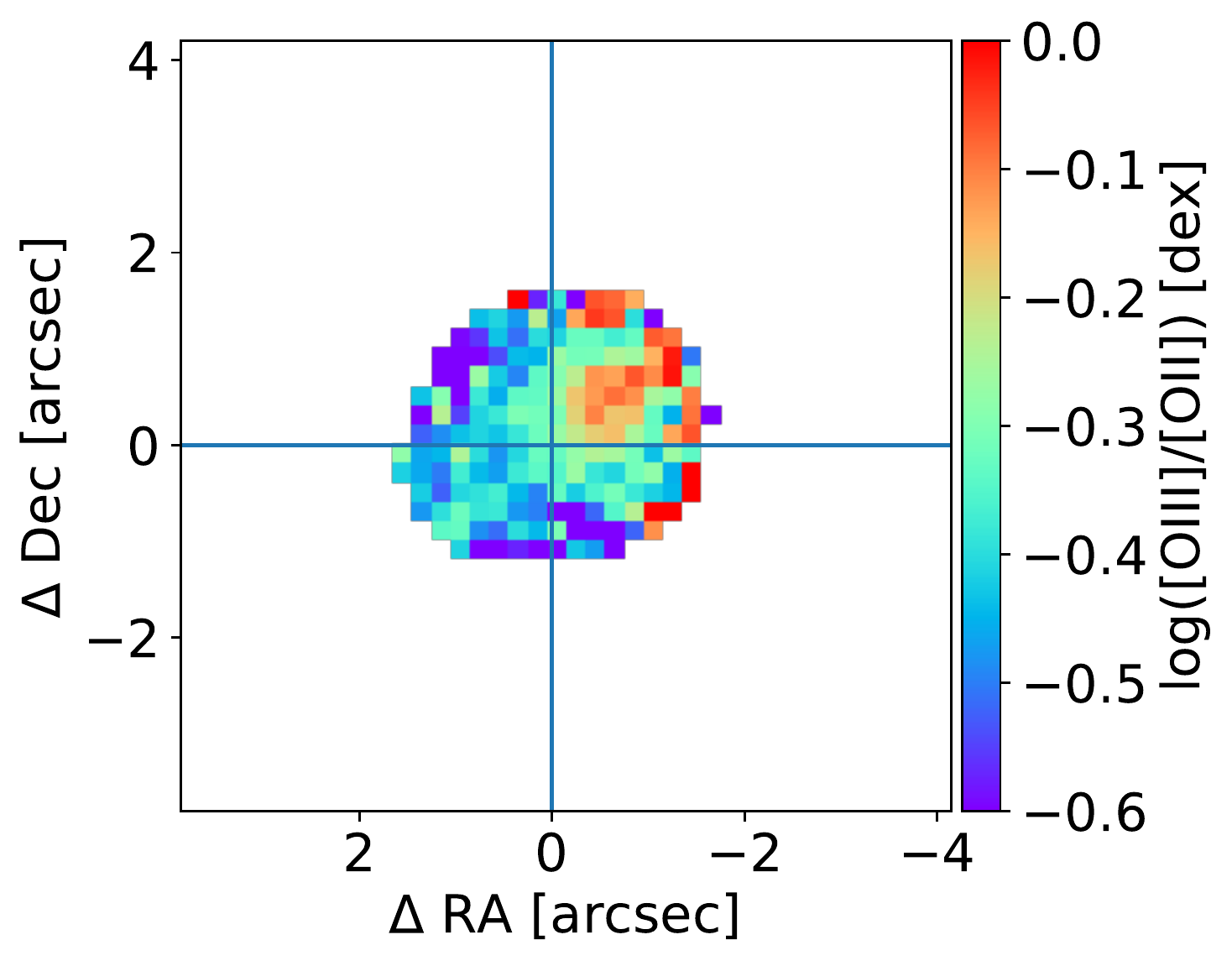} \\
		\includegraphics[width=0.25\textwidth]{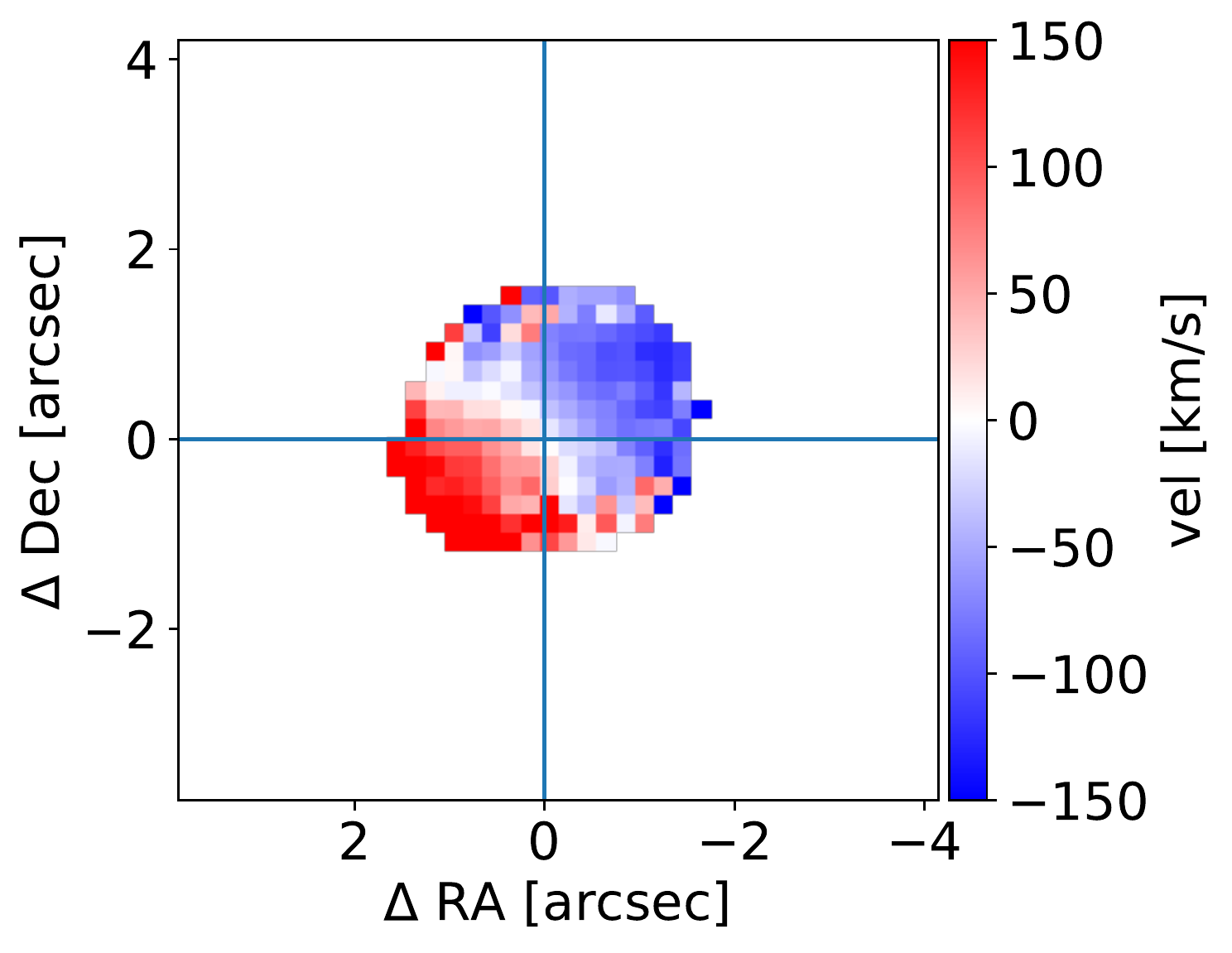} & \includegraphics[width=0.25\textwidth]{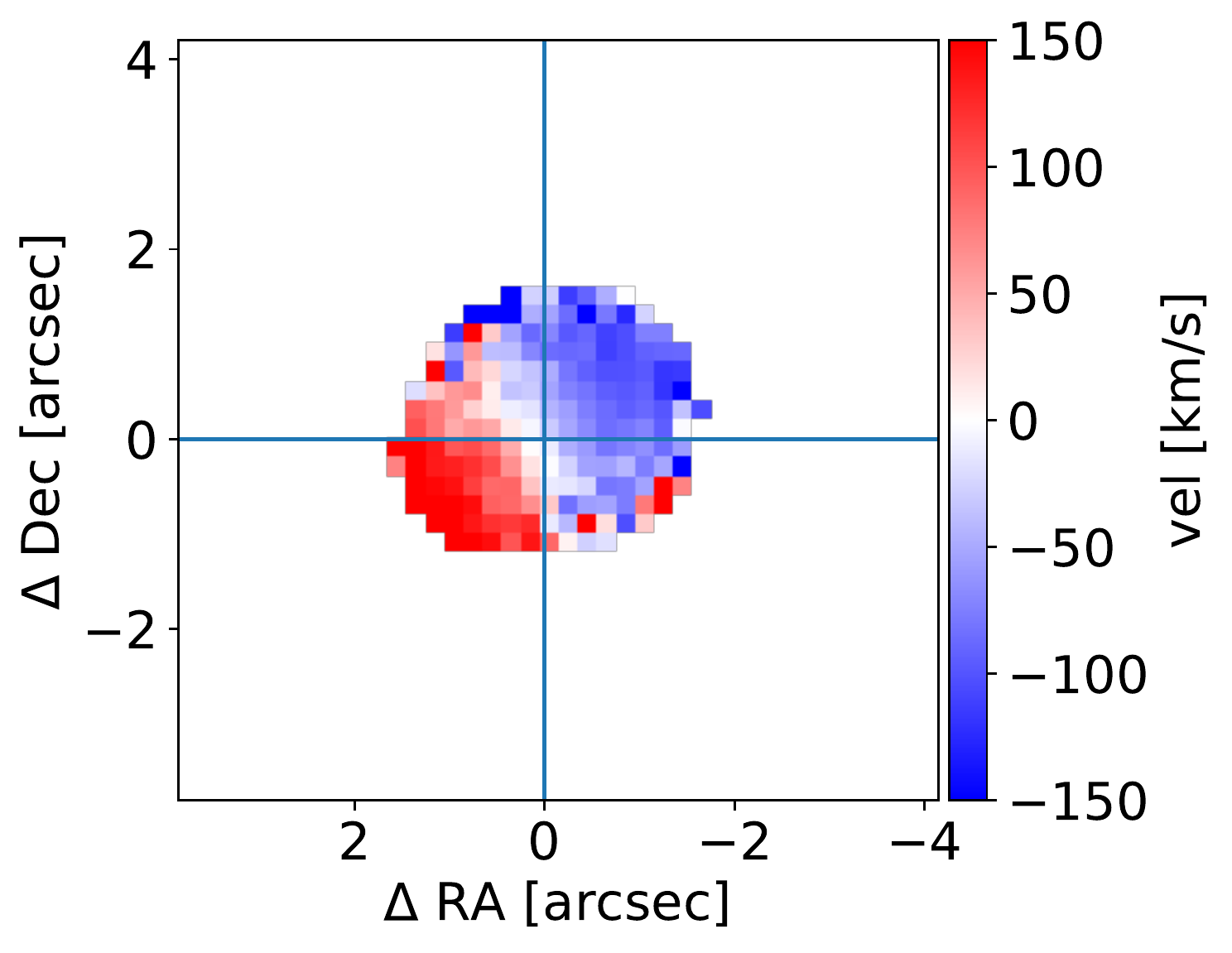} &
		\hspace{-6Ex}\includegraphics[width=0.201\textwidth]{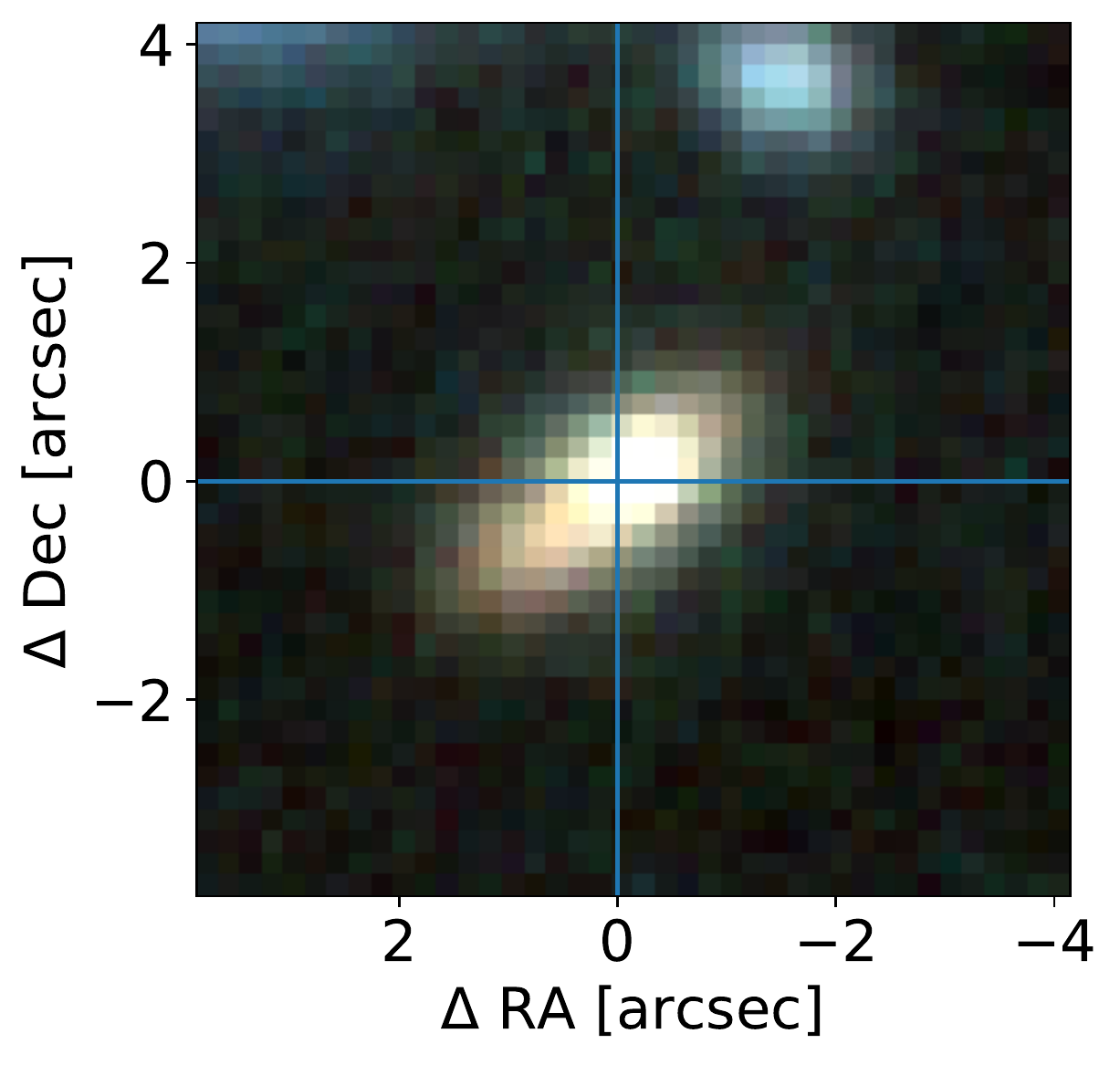}
		\\
	\end{tabular}
	\caption{\label{fig:flux_and_kinematics} Flux and kinematic maps. The blue
    cross is in all six panels at the same position. This position indicates the
    location we define as the formal centre of the galaxy throughout the paper.
	  \textbf{Left and Centre:} The \OII{} and \OIII{} flux (top) and velocity (bottom) maps have been determined by fitting the respective emission lines in each spaxel using the \camel{} code. To increase the S/N the cube has been sligtly smoothed in the spatial direction with a FWHM of 1~pix before fitting. \task{What is the zero velocity of these cubes} \textbf{Upper right:} Ratio between the \OIII{} and \OII{} flux maps. The ratio appears to increase towards the north-west.  \task{The cross might currently not be at exactly correct position}
	  \textbf{Lower right:} RGB image using  i', r', and V filters. The continuum has an extension towards the south-east, which is only weakly visible in the emission lines. The blue galaxy near the top is a background galaxy (\namebonepseven).
	  All figures in this panel are based on the data cube created from the best-seeing data only.
	   }
\end{figure*}

In order to characterize the morphology of the \nameprime{} galaxy, we use a combination of broad- and narrow-band maps.
\Fig{fig:flux_and_kinematics}
shows an \OII{} flux map (upper left), an \OIII{} flux map (upper center), and a color composite created from three broadband filters (lower right), using 
the `best-seeing' cube from Section~\ref{sec:obs}.

Looking first at the lower right image in \Fig{fig:flux_and_kinematics}, one can see that 
  the galaxy is redder towards the south-east and bluer towards the north-west.
This redder region could have an older stellar population or more dust.
Unfortunately, the S/N of $\Hg{}$ is not sufficient to allow for a spatially resolved dust map from $\Hg{}/\Hb{}$ (\Ha{} is not covered).

From \Fig{fig:flux_and_kinematics}, a more striking variation is  between the spatial distribution of $\OII{}\,\lambda\lambda 3727,3729$ (top left) and $\OIII{}\,\lambda 5007$ (top middle), where  \OIII{}   is enhanced compared to \OII{}   towards the north-west. The \OIII{}/\OII{} ratio map (upper right panel of \Fig{fig:flux_and_kinematics}) makes this apparent. This difference can be caused by variations in the ionization parameter, $q$, and/or the gas metallicity  \citep[e.g.][]{Kewley:2002a}.
Together, the  \OIII{}/\OII{} map and the color map indicate that this galaxy has a large clump of enhanced SFR, slightly offset ($\delta x=0.5$~kpc) from the galaxy center.~\footnote{Alternatively, the galaxy  could be undergoing a minor merger.}

\begin{figure}
\includegraphics[width=\columnwidth]{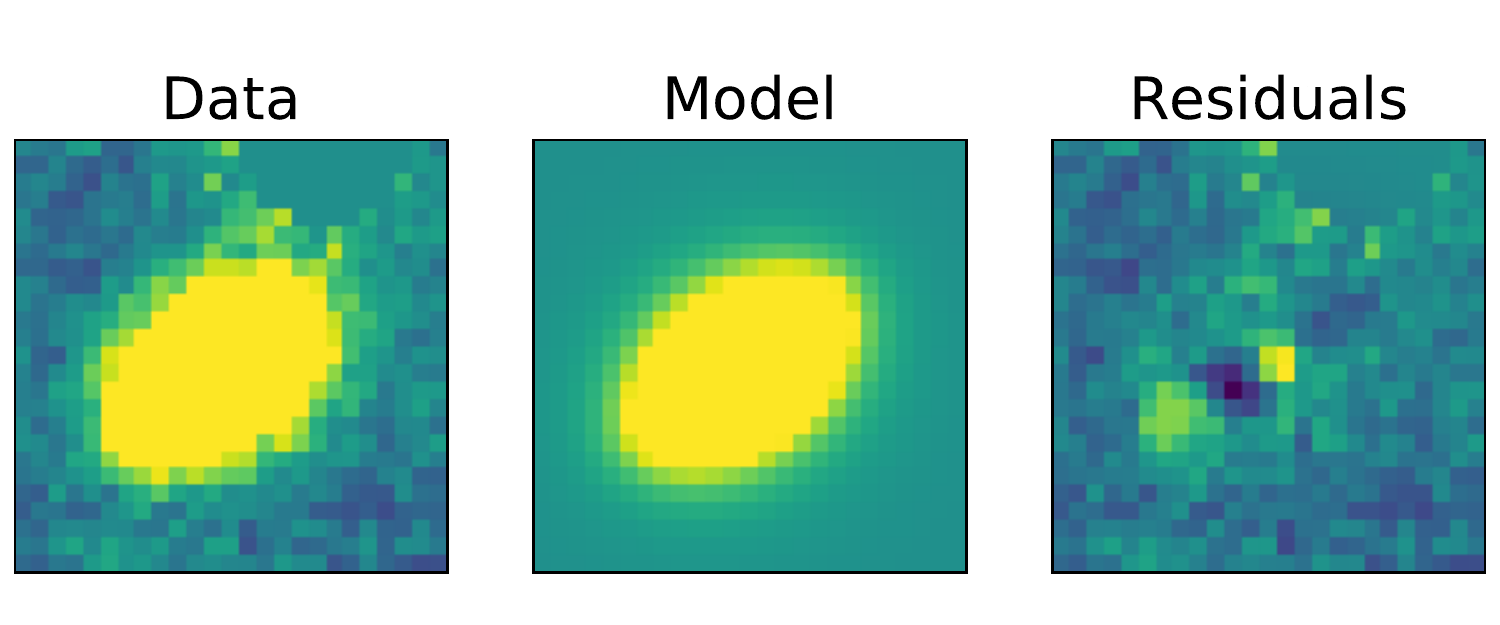}
\caption{\label{fig:galfit_rp} Comparison between data and \galfit{} model for a broadband image ($r'$) of the \nameprime{} galaxy. All the data, the model, and the residuals (data-model) are scaled identically. The cutouts have a size of $5\arcsec \times 5\arcsec$.}
\end{figure}

Because the \nameprime{} galaxy contains this significant  clump of SFR and because  \OII\ is also extended towards the minor axis, the inclination from \OII{} will be biased. We thus revert to fitting the continuum to  estimate the morphological parameters of the \nameprime{} galaxy.
 \Fig{fig:galfit_rp} shows the best-fit single-\sersic{} model obtained with \galfit{} \citep{Peng:2010a} to an $r'$ broadband image created from the MUSE cube and using a Moffat PSF with the parameters as measured from the quasar. This best-fit model has a \sersic{} index of $\mainXpropXgalfitXn$, an axis ratio of $\mainXpropXgalfitXbtoa{}$ ($\incl \approx \mainXpropXgalfitXincl \deg$), and a half-light radius $\rhalf=4.6\;\kpc$.  \task{Jarle complained that here only teh uncertainties are stated} \task{Galfit was not made based on best-seeing cube shown in the figure; currently not yet using this PA and centre from this fit, but from old [OII] fit. Need to maybe update.}

\subsection{Kinematics}
\label{appendix:kinematics}

The bottom left and middle panels of 
Fig.~\ref{fig:flux_and_kinematics}
show the \OII{} and \OIII{} velocity fields determined using the \camel{} code \citep{Epinat:2012a}. The rotation field is consistent between both \OII{} and \OIII{} and looks undisturbed.

As discussed in Appendix \ref{appendix:morpho}, the line flux distribution deviates from a \sersic{} profile due to the presence of a SFR clump, and thus prevents us from perform a simultaneous fit of the morphology and kinematics with \gpk{}. We thus first perform a two component (two S{\'e}rsics, one representing the clump) model fit with  \galfit{} \citep{Peng:2010a}
 on the \OIII{} NB image, where the PSF resolution is higher than \OII{}.
\OIII{} is also preferred as the \OIII{} SB profile extends little beyond the continuum (see \Fig{fig:sb_profiles_pa}) and the fit gives hence a good representation of the star-forming ISM.
Subsequently, we use this intrinsic two component flux profile, with the two components independently scaled to \OII{}, and the inclination inferred from the continuum fit (Appendix \ref{appendix:morpho}) as input when we perform the 3D kinematic modelling using  \gpk{} \citep{Bouche:2015b}.

\subsection{Line fluxes and metallicity}
\label{appendix:linefluxes}

\begin{table}
\caption{\label{tab:linefluxes} Line fluxes as measured with \ppxf{}.}
\begin{center}
\begin{tabular}{crrr}
\hline
Ion & $\lambda$ (obs) &  $\lambda$ (rest) & line flux \\
 & [$\textnormal{\AA}$] &  [$\textnormal{\AA}$] & [$10^{-18}\uerglf{}$]\\

\hline
\OII{} & 6344 & 3727 & $71.0{\scriptstyle \pm0.8}$ \\
\OII{} & 6349 & 3730 & $104.4{\scriptstyle \pm1.6}$ \\
H I & 6420 & 3772 & $1.2{\scriptstyle \pm0.5}$ \\
H I & 6466 & 3799 & $1.9{\scriptstyle \pm0.4}$ \\
H I & 6530 & 3836 & $9.0{\scriptstyle \pm0.3}$ \\
\NeIII{} & 6587 & 3870 & $8.7{\scriptstyle \pm0.4}$ \\
H I & 6621 & 3890 & $7.6{\scriptstyle \pm0.4}$ \\
H$\delta$ & 6984 & 4103 & $14.3{\scriptstyle \pm0.5}$ \\
H$\gamma$ & 7390 & 4342 & $26.6{\scriptstyle \pm0.5}$ \\
H$\beta$ & 8277 & 4863 & $62.0{\scriptstyle \pm0.7}$ \\
\OIII{} & 8443 & 4960 & $24.9{\scriptstyle \pm0.3}$ \\
\OIII{} & 8525 & 5008 & $76.3{\scriptstyle \pm0.5}$ \\

\hline
\end{tabular}
\end{center}
\end{table}

\begin{figure*}
\centering
\includegraphics[width=0.8\textwidth]{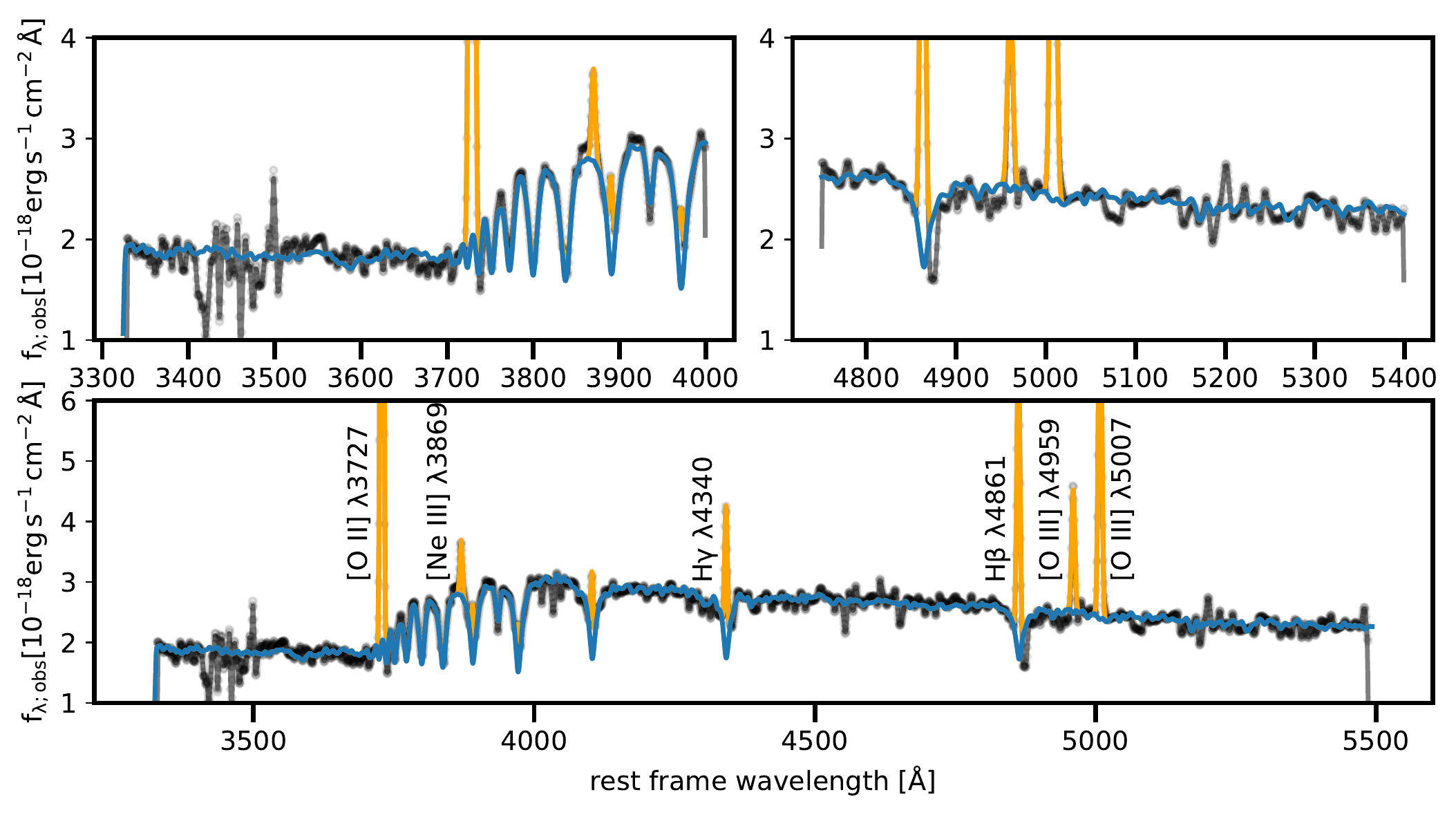}
\caption{\label{fig:ppxf} Nebular and stellar decomposition using \ppxf{} for the \nameprime{} galaxy. The data (black) are fit by a combination of BC03 stellar population models (blue) and emission lines (orange). The two upper panels are zooms into the lower panel.}
\end{figure*}

We measured line fluxes using the \ppxf{} code, which allows to simultaneously fit a stellar continuum model and the emission lines. For the continuum we allowed an arbitrary linear combination of single age BC03 stellar populations models, convolved to the wavelength-dependent spectral resolution of MUSE.
All fit emission lines  were assumed to be Gaussians with the intrinsic velocity dispersion shared between the different transitions.
While we imposed on $\OII{}\,\lambda3729/\OII{}\,\lambda3727$ and  $\OIII\,\lambda4959 /\OIII\,\lambda5007$ constraints from atomic physics,\footnote{$0.28 < \OII{}\,\lambda3729/\OII{}\,\lambda3727 < 1.47$  and  $\OIII\,\lambda4959 / \OIII\,\lambda5007 = 0.33$.} the ratio between all others lines was allowed to freely vary. The resulting fit is shown in \Fig{fig:ppxf} and the fluxes are listed in \Tab{tab:linefluxes}.

\begin{figure}
  \centering
    \includegraphics[width=1.0\columnwidth]{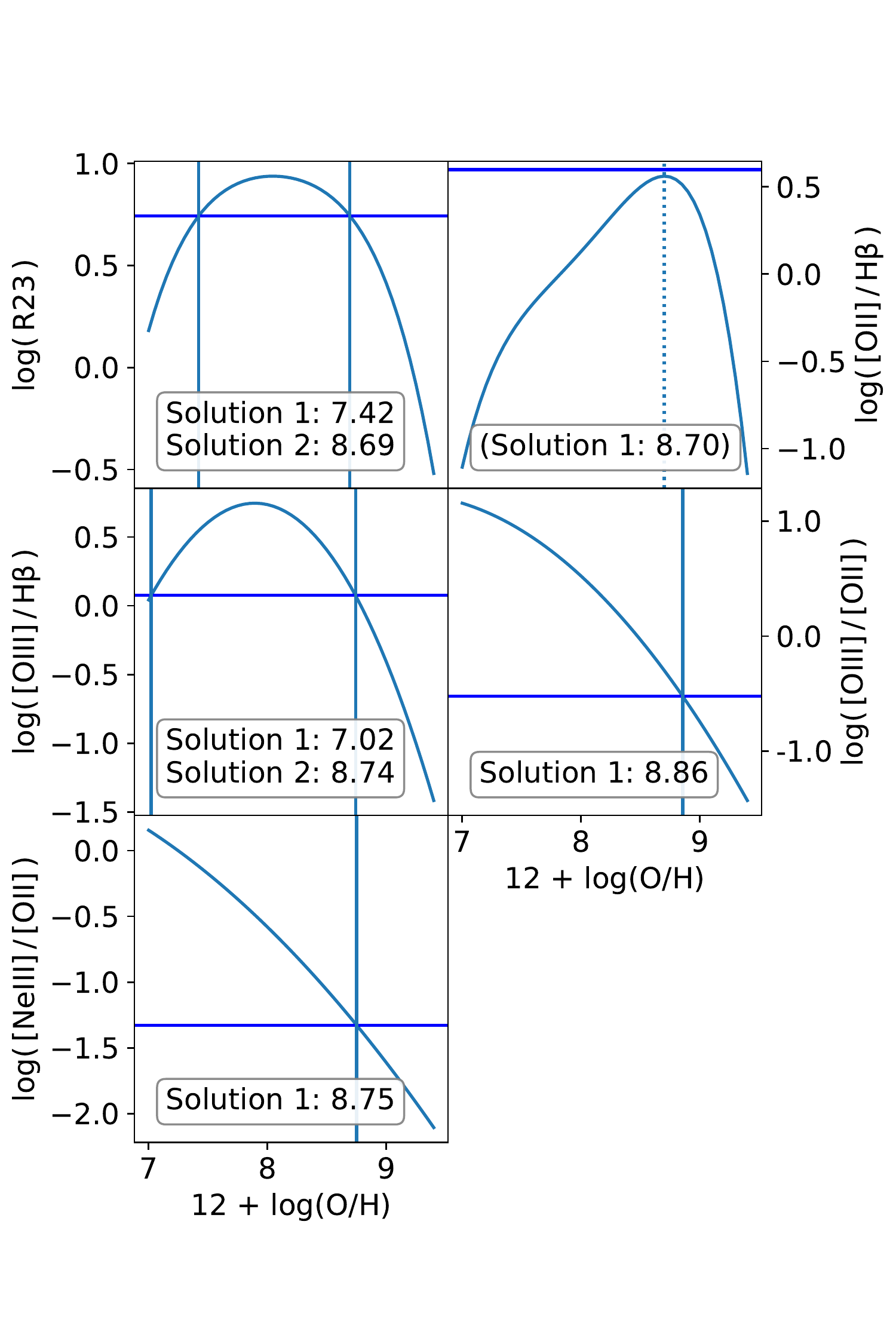} 
  \caption{\label{fig:metallicity_maiolino} Oxygen gas-phase metallicity
    estimates using the strong-line calibrations from \citep{Maiolino:2008a}.
   }	
\end{figure}

The measured fluxes, after de-reddening assuming a \citet{Calzetti:2000a} extinction curve and an $\ebv{}=0.29$ as estimated from the stellar mass (see \Tab{tab:prop_gals}), allow us to use several of the common strong-line gas-phase
metallicity indicators, except those involving $\NII{}$, which is not covered by
our MUSE data. Using the calibrations from \citet{Maiolino:2008a}, we
have estimated the oxygen abundance, 12+log(O/H), with five different indicators using ratios
between the \OII{}, \OIII{}, $\Hb$, and \NeIII{} lines (see \Fig{fig:metallicity_maiolino}).
The combination of the five indicators prefers a $\mathrm{12+log(O/H)}\approx 8.7\mbox{--}8.8$, indicating approximately solar abundance.  This means that
the \nameprime{} galaxy is a typical galaxy on the mass-metallicity relation at
$z=0.7$.
We note that the agreement between the different indicators could be improved by assuming a lower $\ebv\approx0.2$. This could indicate that the $\ebv$ estimate as obtained from the $\ebv$-$M_*$ relation might be a slight overestimate.

\section{\MgII{} SB map for wide filter}
\label{app:wide_mgii_sb_map}
\begin{figure*}
  \includegraphics[width=0.8\columnwidth]{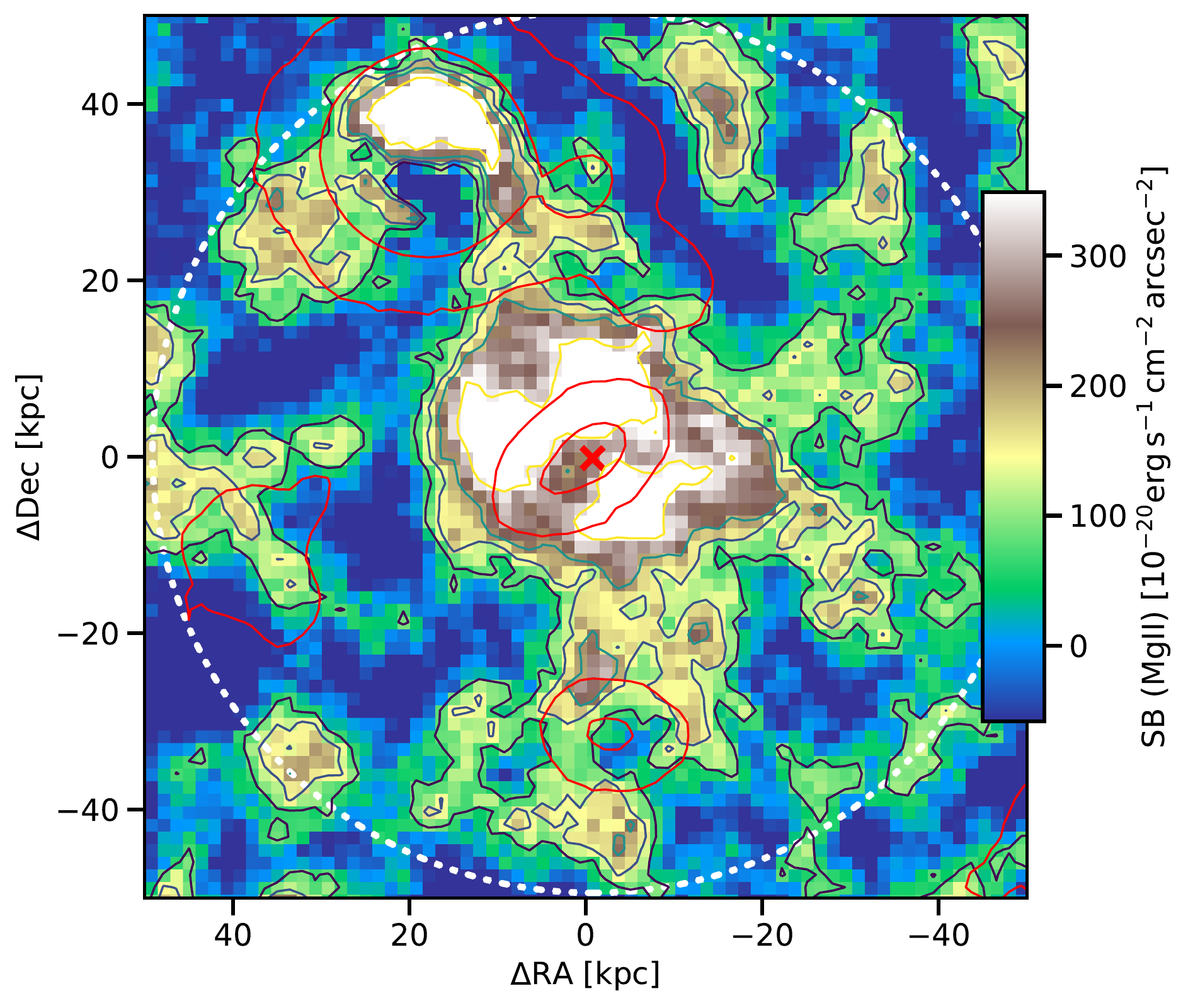}
  \includegraphics[width=0.8\columnwidth]{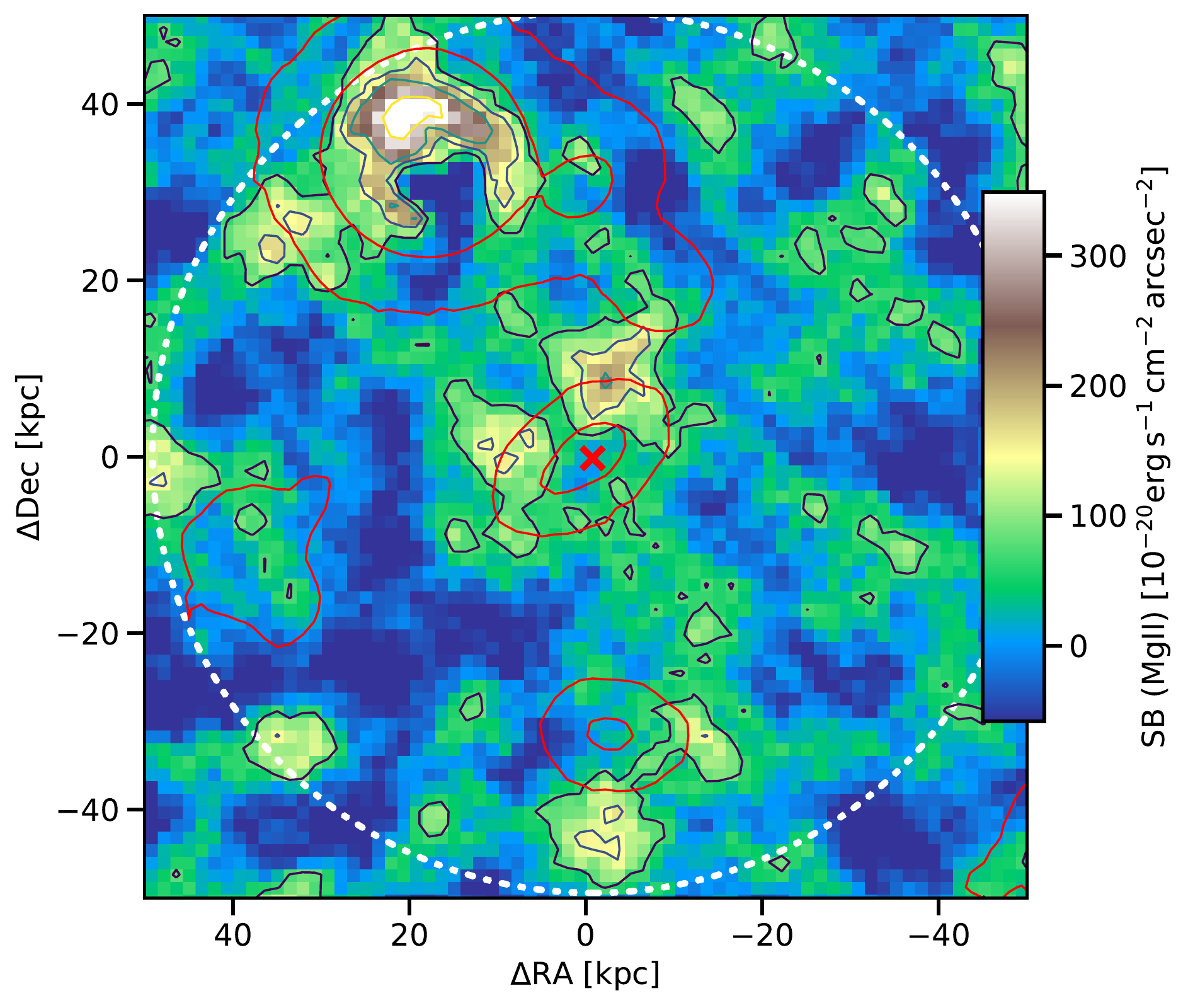}
\caption{\label{fig:wide_mgii_sb_map}
\textbf{Left:} \MgII{} SB emission map as in \Fig{fig:main_maps} (right), but using instead of the double NB filter ($2\times600\,\kms$; blue in \Fig{fig:mgii_spec_full_extract}) a broad single filter ($2270\,\kms$; yellow in \Fig{fig:mgii_spec_full_extract}). \textbf{Right:} The difference between the wide-filter SB map and the map based on the double NB as in \Fig{fig:main_maps} (right) is shown.
The contour levels are in both panels at the same SB levels as in  \Fig{fig:main_maps} (right).
The flux at the position of the quasar is likely a residual from the quasar PSF subtraction and not real emission.}
\end{figure*}

We investigated whether using a broad NB filter ($2270\,\kms$; yellow in \Fig{fig:mgii_spec_full_extract}) instead of the double NB filter ($2\times600\,\kms$; blue in \Fig{fig:mgii_spec_full_extract}) does have a substantial impact on the \MgII{} morphology of the halo.  The left panel of \Fig{fig:wide_mgii_sb_map} shows the SB map for the wide filter, while the right panel shows the difference between this map and the map for the double NB filter, as shown in \Fig{fig:main_maps} (right). We can conclude that the morphology does not substantially change when using a wider filter. As a narrower filter has lower noise and less problems with continuum residuals from other sources in the field, we therefore decided to use the double NB filter also for extraction of the radial SB profiles in \Fig{fig:sb_profiles_pa}.

\section{Absence of rotation in outflow regions}
\label{sec:app:rotouter}
\begin{figure*}
  \includegraphics[height=4.5cm]{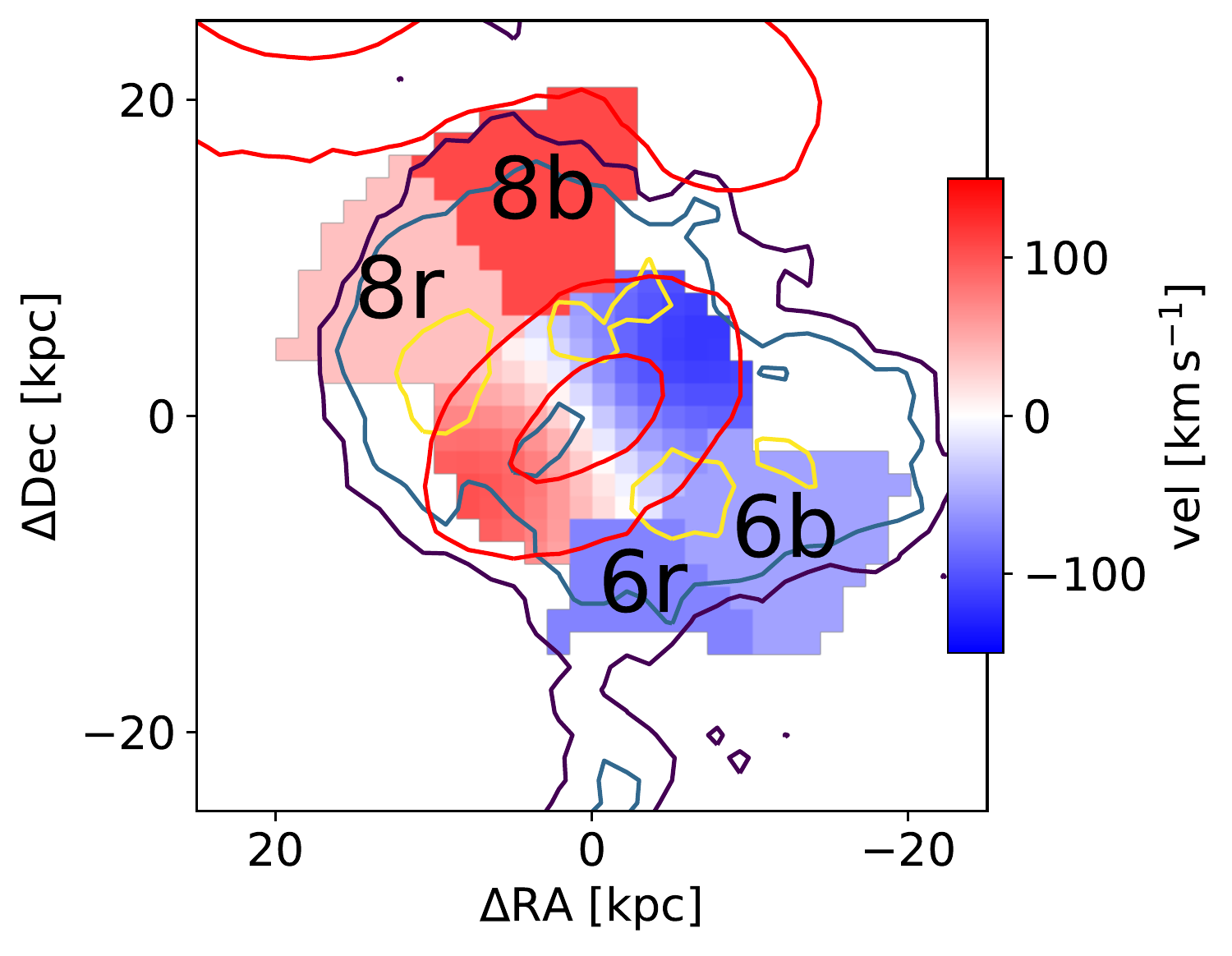}
  \includegraphics[height=4.5cm]{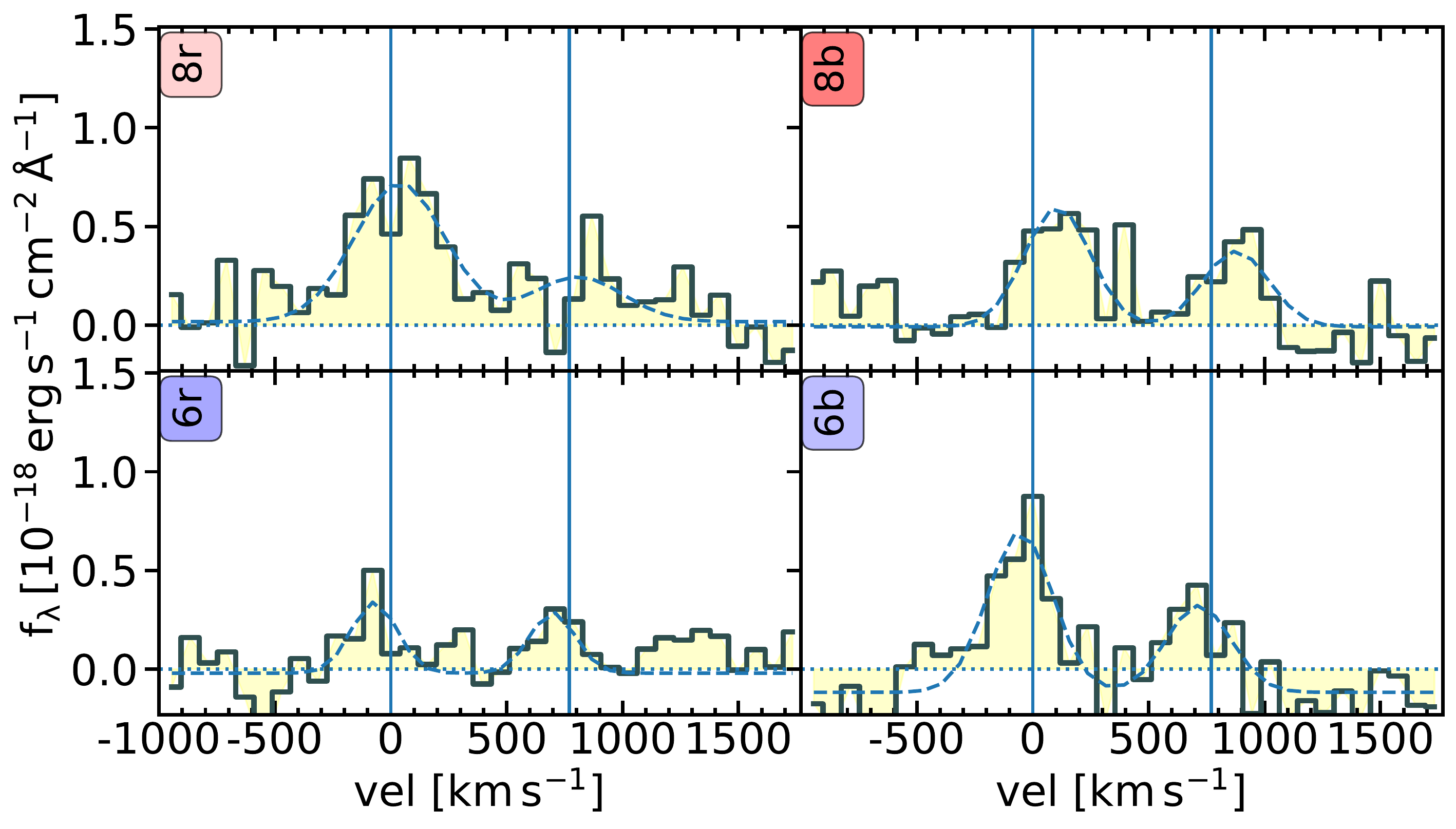}
  \caption{\label{fig:kin_split_outer} \MgII{} spectra extracted from the
    numbered regions shown in the left panel are shown in the right panel. These regions
    split the two outer minor axis regions from \Fig{fig:quadrant_spectra} into
    the part where the extrapolation of the galaxy rotation field would be
    blueshifted (``b'') and the part where it would be redshifted (``r'').
    A fit to the $\MgII{}$ doublet in each of the four regions is shown as
    blue-dashed line. The $\vlos$ from this fit sets the color-scale of the
    regions in the map. The galaxy rotation field measured from \OII{} with \gpk{} is shown
  in the centre of the map. }
\end{figure*}

In \Sec{section:halo:kinematics} we found that the kinematics in the two outer
minor-axis regions (6 \& 8 in \Fig{fig:quadrant_spectra}) are consistent with
the expectation from a biconical outflow, with redshifted emission in region 8
and blueshifted emission in region 6. However, as the emission flux in the outer
part of the cone-like regions is not completely symmetric w.r.t. the galaxy's
minor axis, the kinematic signature could in principle also be caused by a
(very) extended, clumpy, co-rotating disk: In region 8 the \MgII{} emission is
slightly brighter on the redshifted side of the galaxy rotation field, while in
region 6 it is somewhat brighter on the blue-shifted side.
To rule out rotation as explanation for the kinematics in these regions, we split
regions 6 \& 8 into two halves each, split by the galaxy's minor axis. The
regions and the spectra extracted from these regions are shown in
\Fig{fig:kin_split_outer}, where the ``b'' and ``r'' regions are on the blue-
and redshifted side of the galaxy rotation field, respectively. We find that in
both cases ``b'' and ``r'' have the same velocity sign, which strongly supports
the outflow scenario. Furthermore, this result tentatively indicates that the
outflow does not carry high angular momentum.


\bsp	
\label{lastpage}
\end{document}